\documentclass[bibyear]{aa}  
\usepackage{graphicx}
\usepackage{mathtools}
\usepackage{ifpdf}
\usepackage{txfonts}
\usepackage{color}
\usepackage{natbib}
 \usepackage{lscape}
\DeclareRobustCommand{\ion}[2]{%
\relax\ifmmode
\ifx\testbx\f@series
{\mathbf{#1\,\mathsc{#2}}}\else
{\mathrm{#1\,\mathsc{#2}}}\fi
\else\textup{#1\,{\mdseries\textsc{#2}}}%
\fi}

\begin{document}
   \title{Ionized gas kinematics of galaxies in the CALIFA survey \thanks{Based on observations collected at the Centro Astronómico Hispano Alemán (CAHA) at Calar Alto, operated jointly by the Max-Planck Institut f\"ur Astronomie and the Instituto de Astrof\'{\i}sica de Andaluc\'{\i}a (CSIC)} I }
\subtitle{Velocity fields, kinematic parameters of the dominant component, and presence of kinematically distinct gaseous systems \thanks{Figures
in Appendix C are only available in electronic form via
http://www.edpsciences.org}}


   \author{B. Garc\'{\i}a-Lorenzo\inst{1,2}
          \and
          I. M\'arquez\inst{3}
          \and
          J.K. Barrera-Ballesteros\inst{1,2}
          \and
          J. Masegosa\inst{3}
          \and
          B. Husemann\inst{4,5}
          \and
          J. Falc\'on-Barroso\inst{1,2}
          \and
          M. Lyubenova\inst{6}
          \and
          S. F. S\'anchez\inst{7}
          \and
          J. Walcher\inst{5}
          \and
          D. Mast\inst{3,8}
          \and
          R. Garc\'{\i}a-Benito\inst{3}
          \and
          J. M\'endez-Abreu\inst{9,1,2}
          \and
          G. van de Ven\inst{6}
          \and
          K. Spekkens\inst{10}
          \and
          L. Holmes\inst{10}
          \and
          A. Monreal-Ibero\inst{5,11}
          \and
          A. del Olmo\inst{3}
          \and
          B. Ziegler\inst{12}
          \and
          J. Bland-Hawthorn\inst{13} 
          \and
          P. S\'anchez-Bl\'azquez\inst{14}
          \and
          J. Iglesias-P\'aramo\inst{3,8}
          \and
          J.A.L. Aguerri\inst{1,2}
          \and
          P. Papaderos\inst{15}
          \and
          J.M. Gomes\inst{15}
          \and
          R.A. Marino\inst{16}
          \and
          R.M. Gonz\'alez Delgado\inst{3}
          \and
          C. Cortijo-Ferrero\inst{3}
          \and
          A.R. L\'opez-S\'anchez\inst{17,18}
          \and
          S. Bekerait\.e\inst{5}
          \and
          L. Wisotzki\inst{5}
          \and
          D. Bomans\inst{19}
          \and
          the CALIFA team}

   \institute{Instituto de Astrof\'{\i}sica de Canarias, C/Via Lactea S/N, 38200-La Laguna, Tenerife, Spain\\
              \email{bgarcia@iac.es}
         \and
             Dept. Astrof\'{\i}sica, Universidad de La Laguna, C/ Astrof\'{\i}sico Francisco S\'anchez, E-38205-La Laguna, Tenerife, Spain
         \and
          Instituto de Astrof\'{\i}sica de Andaluc\'{\i}a (CSIC), Glorieta de la Astronom\'{\i}a S/N, 18008 Granada, Spain
         \and
European Southern Observatory, Karl-Schwarzschild-Str. 2, D-85748 Garching b. Muenchen, Germany
         \and
Leibniz-Institut f\"ur Astrophysik Potsdam (AIP), An der Sternwarte 16, 14482, Potsdam, Germany
         \and
Max Planck Institute for Astronomy, K\"onigstuhl 17, 69117 Heidelberg, Germany
         \and
Instituto de Astronom\'{\i}a, Universidad Nacional Aut\'onoma de Mexico, A.P.70-264,04510, M\'axico, D.F.
         \and
Centro Astron\'omico Hispano Alem\'an de Calar Alto (CSIC-MPG), C/ Jes\'us Durb\'an Rem\'on 2-2, 4004 Almer\'{\i}a, Spain
         \and
School of Physics and Astronomy, University of St Andrews, North Haugh, St Andrews, KY16 9SS, UK
         \and
Department of Physics, Royal Military College of Canada, PO Box 17000, Station Forces, Kingston, Ontario, K7K 7B4, Canada 
         \and
GEPI Observatoire de Paris, CNRS, Universit\'e Paris Diderot, Place Jules Janssen, 92190 Meudon, France
         \and
University of Vienna, Department of Astrophysics, T\"urkenschanzstr. 17, 1180 Vienna, Austria
         \and
Sydney Institute for Astronomy, School of Physics A28, University of Sydney, NSW 2006, Australia
         \and
Departamento de F\'{\i}sica Te\'orica, Universidad Aut\'onoma de Madrid, 28049 Madrid, Spain
         \and
Centro de Astrof\'{\i}sica and Faculdade de Ci\^encias, Universidade do Porto, Rua das Estrelas, 4150-762 Porto, Portugal
         \and
CEI Campus Moncloa, UCM-UPM, Departamento de Astrof\'{\i}sica
y CC. de la Atm\'osfera, Facultad de CC. F\'{\i}sicas, Universidad
Complutense de Madrid, Avda. Complutense s/n, 28040 Madrid, Spain
         \and
Australian Astronomical Observatory, PO Box 915, North Ryde, NSW 1670, Australia
         \and
Department of Physics and Astronomy, Macquarie University
, NSW 2109, Australia
         \and
Astronomical Institute of the Ruhr-University Bochum Universitaetsstr. 150, 44801 Bochum, Germany
}

   \date{Received Month day, 2014; accepted month day, 2014}

 
  \abstract
  {Ionized gas kinematics provide important clues to the dynamical structure of galaxies and hold constraints to the processes driving their evolution.}
   { This work provides an overall characterization of the kinematic behavior of the ionized gas of the galaxies included in the Calar Alto Legacy Integral field Area (CALIFA), offering kinematic clues to potential users of the CALIFA survey for including kinematical criteria in their selection of targets for specific studies. From the first 200 galaxies observed by CALIFA, we present the 2D kinematic view of the 177 galaxies satisfaying a gas content/detection threshold.}
   {After removing the stellar contribution, we used the cross-correlation technique to obtain the radial velocity of the dominant gaseous component for different emission lines (namely, [\ion{O}{ii}]~$\lambda\lambda3726,3729$, [\ion{O}{iii}]~$\lambda\lambda4959,5007$, H$\alpha$+[\ion{N}{ii}]~$\lambda\lambda6548,6584$, and [SII]$\lambda\lambda6716,6730$). 
The main kinematic parameters measured on the plane of the sky were directly derived from the radial velocities with no assumptions on the internal prevailing motions. Evidence of the presence of several gaseous components with different kinematics were detected by using [\ion{O}{iii}]~$\lambda\lambda4959,5007$ emission line profiles. }
   {At the velocity resolution of CALIFA, most objects in the sample show regular velocity fields, although the ionized-gas kinematics are rarely consistent with simple coplanar circular motions. Thirty-five percent of the objects present evidence of a displacement between the photometric and kinematic centers larger than the original spaxel radii. Only 17\% of the objects in the sample exhibit kinematic lopsidedness when comparing receding and approaching sides of the velocity fields, but most of them are interacting galaxies exhibiting nuclear activity (AGN or LINER). Early-type (E+S0) galaxies in the sample present clear photometric-kinematic misaligments.
 There is evidence of asymmetries in the emission line profiles 
in 117 out of the 177 analyzed galaxies,
suggesting the presence of kinematically distinct gaseous components located at different distances from the nucleus. The kinematic decoupling between the dominant and secondary component/s suggested by the observed asymmetries in the profiles can be characterized by a limited set of parameters.}
   {This work constitutes 
the first determination of the ionized gas kinematics of the galaxies observed in the CALIFA survey. The derived velocity fields, 
the reported kinematic distortions/peculiarities and the identification of the presence of several gaseous components in different regions of the objects might
be used as additional criteria for selecting galaxies for specific studies.}

\authorrunning{Garc\'{\i}a-Lorenzo et al.}
\titlerunning{Ionized gas kinematics of CALIFA galaxies}

   \maketitle
%

\section{Introduction}
\label{sec1}




Galaxies in the Local Universe are the result of billions of years of
cosmic evolution. The study of their statistical and detailed properties 
are therefore expected to 
explain the evolution of galaxies. In particular, the analysis of the
gas kinematics and their interplay with other galaxy components
  (essentially the stellar components)
allows us to infer
the dynamical structure of galaxies and to constrain the
processes leading to their formation and evolution \citep[see, e.g.,][]{2010A&A...521A..63L}. 
These 
processes include the role of 
interactions during the galaxy lifetime \citep[see, e.g.,][]{2012ApJ...761L...6M,2010ApJ...711L..61M,2004AJ....127.1344A}, the 
relative mass contribution of luminous and dark matter \citep[see][for a review]{2001ARA&A..39..137S}, the presence
of supermassive black holes and their relationship with the large-scale properties of the host galaxies \citep[see][for a review]{2001ApJ...547..140M}, the 
origin of kinematically decoupled components \cite[see][for a review]{1999IAUS..186..149B}
and of disk heating \citep[][ and references therein]{2001ASPC..230..221M}, the presence of pressure-supported ionized gas in bulges \citep{1999MNRAS.307..433C,1995ApJ...448L..13B}, the fraction of quiescent and/or disturbed galaxies \citep[e.g.,][]{2010ApJ...714L.108S}, or even the fraction of galaxies supported by rotation \citep[e.g.,][]{2006ApJ...653.1027W}. 

The advent of integral field spectrographs made it possible for 
the first time to obtain 2D kinematics of the gas and stars and to overcome
the difficulties associated with the interpretation of 1D
velocity curves. The kinematics derived with the SAURON spectrograph 
(with a field of view of 33 x 41 arcsec$^2$) have been found to be rarely consistent
with simple coplanar circular motions in early-type  galaxies \citep{2006MNRAS.366.1151S}; in late types objects, gas and stars are decoupled in the inner
central region \citep{2006MNRAS.367...46G}. The ATLAS 3D project \citep{2011MNRAS.413..813C}, aimed at enlarging the sample, also uses SAURON. The DiskMass Survey
\citep{2010ApJ...716..198B} uses the SparsePak \citep{2004PASP..116..565B,2005ApJS..156..311B} and PPak \citep{2004AN....325..151V, 2006PASP..118..129K} integral field units with
1arcmin FoVs and jointly analyses gas and stellar kynematics for a
sample of 30 nearly face-on spiral galaxies; among other results they
report that galaxy disks are submaximal and that disks with a fainter
central surface brightness in bluer and less luminous galaxies of
later morphological types are kinematically colder with respect to
their rotational velocities \citep{2013A&A...557A.130M}. 

All these issues will greatly benefit from a survey like the Calar
Alto Legacy Integral Field Area survey\footnote{http://califa.caha.es}
\citep[][S12 hereafter]{2012A&A...538A...8S}. CALIFA is devoted to the
comparative measurements of ionized gas together with other
spectroscopic properties. It is an Integral Field Spectroscopy (IFS)
survey that is acquiring spatially resolved spectroscopic information
of a diameter-selected sample of $\sim$600 galaxies (0.005 $<$ z $<$
0.03), with the same FoV as in the DiskMass survey. These
galaxies cover the color-magnitude diagram with a large enough number
of objects per color/mag bin to enable statistical studies as a
function of galaxy type.

Studies based on gas kinematics, like the one presented in
this paper (the stellar kinematics will be presented in a forthcoming
paper by Falc\'on Barroso et al., 
in preparation), will be mainly devoted to the
analysis of gas velocity fields, and provide a first estimation of
the regularity of the assumed gravitational potential. For disk
galaxies, departures from regular rotation can be easily traced, and
asymmetries can be interpreted in terms 

of the observed
morphology. The importance of disk heating can be estimated by
measuring the emission lines widths. Strong velocity gradients may be
evidence of the presence of shocks; several line components will be
eventual tracers of several kinematical components. 
This is the first of a series of works devoted to disentangling the ionized gas
behavior through a comprehensive and homogeneous characterization of
the main kinematic properties of the galaxies observed by the CALIFA
survey. This work tries to promote the inclusion of kinematic
criteria in the selection of galaxies from CALIFA survey for specific
works.

The structure of this article is as follows. In Sect. \ref{sec2}, we
summarize the observations, data reduction and main properties of the
sample. In Sect. \ref{medir} we describe the procedure to derive the
ionized gas velocity fields, the adopted methods to estimate some
kinematic parameters directly from the observed radial velocities and
the process to detect the presence of kinematically distinct gaseous
components from the observed emission line profiles. In
Sect. \ref{result}, we explore some dependences of the estimated
kinematic indicators with galaxy types. Finally, a summary of the
results is given in Sect. \ref{summary}.

\section{Observations and data reduction}
\label{sec2}
\subsection{CALIFA survey}
The galaxies presented in this work are included in the CALIFA survey
mother sample (S12),
that comprises
939 galaxies
selected from the SDSS DR7 \citep{2009ApJS..182..543A}. The main selection
criteria to build this sample are the angular isophotal size (45$^{\prime\prime}$ $< D_{25} <$ 80$^{\prime\prime}$, where $D_{25}$ is the isophotal diameter in the SDSS $r$-band) and the proximity of these galaxies (0.005 $< z < $ 0.03). These criteria are such that the
selected objects represent a wide range of galactic properties such as
morphological types, luminosities, stellar masses and colors. Further
details on the selection criteria effects and a detailed
characterization on the CALIFA mother sample are explained in S12 \citep[see also][]{2014arXiv1407.2939W}. CALIFA has already released the
 first set of fully reduced, quality tested, and scientifically useful data cubes for 100 galaxies to the astronomical community \citep[see][H13 hereafter]{2013A&A...549A..87H}. The data are available at http://califa.caha.es/DR1/ webpage.

\subsection{Observations and data reduction}
\label{observa}

Detailed information of the observational strategy and data reduction
are presented in both H13 and S12. In this section we summarize the
most important aspects regarding the observational setup and data
reduction of the CALIFA galaxies.

Observations were carried out using the PPak (PMAS fiber Package)
fiber bundle \citep{2006PASP..118..129K} of the Potsdam Multi-Aperture
Spectrophotometer \citep[][PMAS]{2005PASP..117..620R} at the 3.5m telescope of the Calar Alto Observatory (Almer\'{\i}a, Spain). Its
main component consists of 331 fibers each with a diameter of 2\farcs7,
concentrated in a single hexagonal bundle covering a field-of-view of
74\arcsec$\times$64\arcsec, with a filling factor of $\sim$60\% and a fiber-to-fiber pitch of $\sim$3.6 arcsec \citep{2006PASP..118..129K}. In
order to cover the complete FoV and sample well the PSF, a dithering scheme of three pointings has been adopted. 
	
The objects analyzed in this work have been observed in two spectral setups namely V500 and V1200.
The V500 (V1200) setup has a nominal resolution of $\lambda/\Delta\lambda\sim 850$ ($\lambda/\Delta\lambda\sim 1650$) at $\sim
5000$ \AA\ (at $\sim 4500$ \AA) and its nominal wavelength range is 3745--7300 \AA \ (3400--4750 \AA). 
The exposure time is fixed for all the observed objects. For
the V500 setup a single exposure of 900 s per pointing of the
dithering scheme is taken while for the V1200 setup 3 or 2 exposures
of 600\,s or 900\,s, respectively, are obtained per pointing.

The data reduction is performed by a pipeline designed specifically
for the CALIFA survey. The detailed reduction process is explained in S12,
while improvements on this pipeline are presented in H13 (current
pipeline version: V1.3c). The usual reduction tasks per pointing include cosmic
rays rejection, optimal extraction, flexure correction, wavelength and
flux calibration, and sky subtraction. Finally, all three pointing are
combined using a flux-conserving inverse-distance weighting scheme
(see S12 for details) to reconstruct a spatially resampled data cube, with a 1''$\times1''$ sampling (see details in S12). The final FITS data cubes include science data, propagated error vectors, masks, and error weighting factors as described in H13. 
\subsection{The sample}
\label{sample}

The CALIFA project reached 200 galaxies observed with the two setups V500 and V1200 (see Sect. \ref{observa} and S12 for details on setups) in January 2013.
From these 200 objects, we selected those galaxies with a minimum ionized gas contect/detection established as follows: detection of the H$\alpha$ emission line with signal-to-noise (S/N hereafter) $\geq$ 20 (after stellar background subtraction) in at least the number of spaxels ($\sim10$) subtending the effective angular resolution ($\sim$3.7 arcsec, see H13) of the final 1 arcsec/pixel scale resampled data cube (S12). Given the large number of spaxels to analyse and the variety of profiles, the S/N of each emission line was just
estimated from the peak of the line and the standard deviation of a region nearby to the emission line after the stellar continuum subtraction. We note that this quick estimation of S/N depends to first order on the spectral resolution. However, the adopted criterion is intended to select galaxies with blocks of contiguous spaxels to afford a kinematic view of the ionized gas, but any other ionized gas detection criteria could be adopted for specific studies \citep[see, e.g.,][]{2013A&A...555L...1P,2013A&A...558A..43S}. The number of objects satisfying the established minimum gas detection criterion is 177. 

\begin{figure}
  \resizebox{\hsize}{!}{\includegraphics[angle=90]{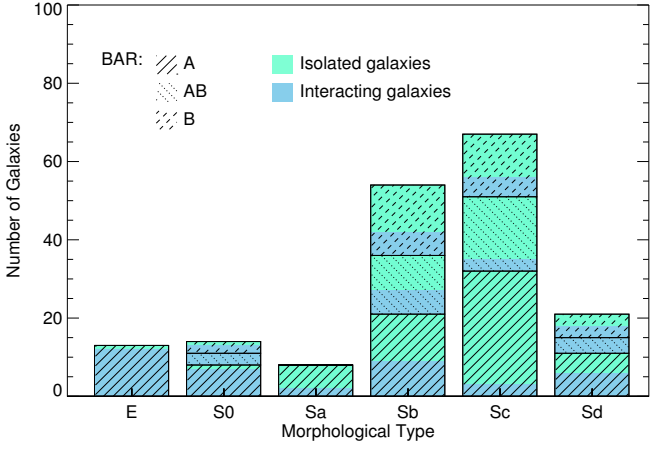}}
   \caption{Distribution by morphological type of the galaxies in the sample. We divide the galaxies into ellipticals (E), lenticulars (S0), and spirals (Sa, Sb, Sc, and Sd). The fraction of non-barred spirals (A), weakly barred spirals (AB), strongly barred spirals (B) is marked (see the top-left corner legend in the plot). Colors indicate the fraction of interacting and isolated galaxies in each morphological division.} 
              \label{tipomorfo}%
    \end{figure}

\begin{figure}
  \resizebox{\hsize}{!}{\includegraphics{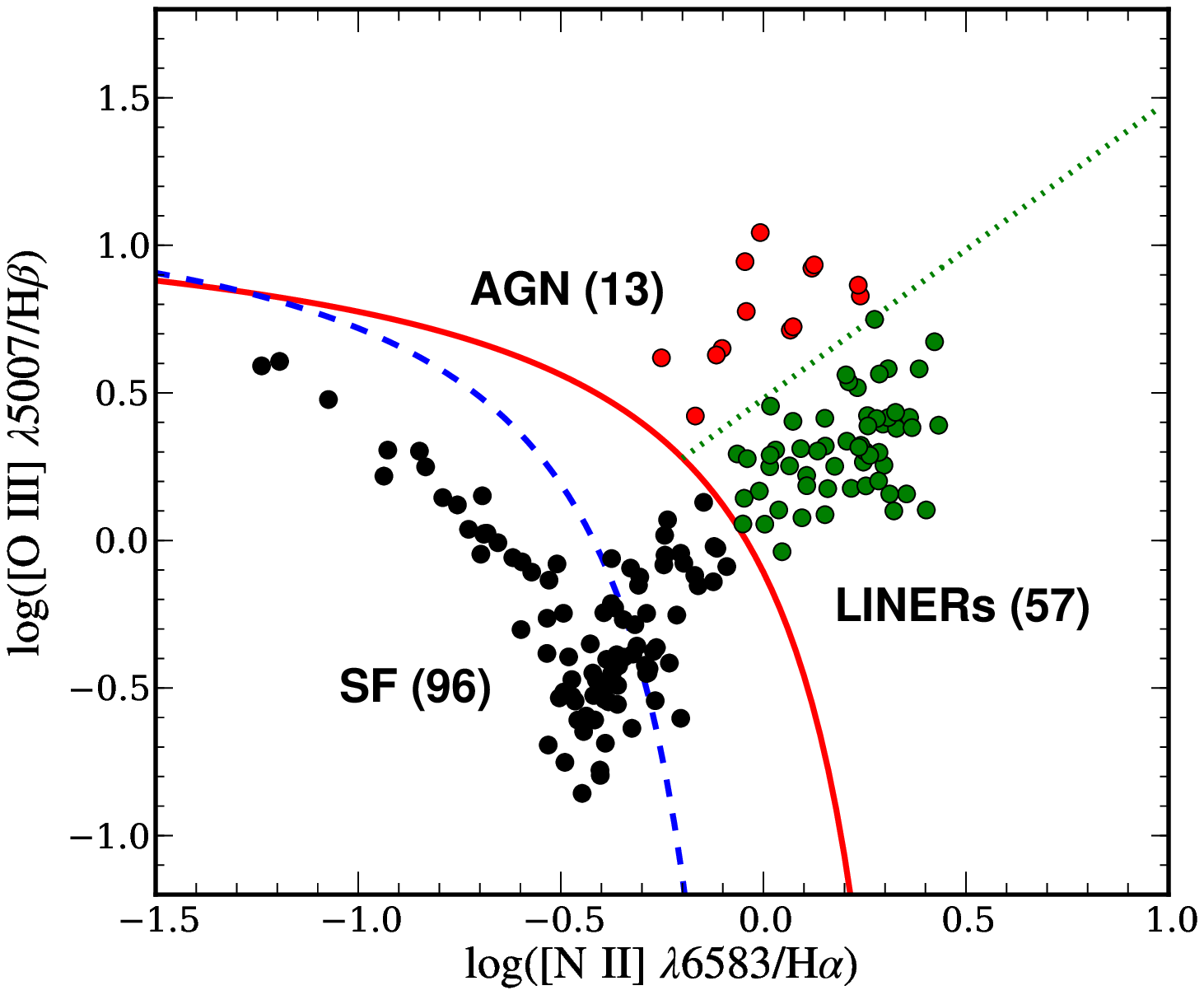}}
   \caption{Emission-line diagnostic diagram for the optical nucleus spectrum of each galaxy in the sample. Only objects for which all required emission lines have signal-to-noise larger than three are shown (166 galaxies). The demarcation lines of \citet{2001ApJS..132...37K} (red curve), \citet{2003MNRAS.346.1055K} (dashed-blue curve), and \citet{2010MNRAS.403.1036C} (dotted-green straight line) are used to classify the galaxies into star forming (SF), active galactic nuclei (AGN), and LINER-type galaxies, which are denoted with black, red, and green symbols, respectively. } 
              \label{BPT}%
    \end{figure}

This sample of 177 objects spans all morphological types, as Fig. \ref{tipomorfo} shows; the morphological type was inferred by combining the independent visual classifications of several members of the CALIFA collaboration \citep[see][]{2014arXiv1407.2939W}. Because of the CALIFA selection criteria (S12), a significant number ($\sim$34\%) of the galaxies in this sample have high inclinations (ellipticity $\ge$ 0.6). Inclinations were inferred from the outer regions of the SDSS r-band images of the galaxies in the sample using the standard task {\it ellipse} of IRAF\footnote{IRAF is distributed by the National Optical Astronomy Observatory, which is operated by the Association of Universities for Research in Astronomy (AURA) under cooperative agreement with the National Science Foundation}. The outer isophotes for most of the galaxies in the sample cover a external region of the disk free of the spiral arms; however, we cannot rule out that some values for the inclination could be contaminated by other structures or asymmetries due to external disturbances. In addition, we have not taken into account the fact that disks are not infinitely thin. Assuming an ellipticity of 0.83 ( for UGC10297) as our minimum disk intrinsic thickness, only six galaxies ($\sim$3\%) in our sample have errors in the inclination larger than 5 degrees due to this effect. The CALIFA mother sample \citep[S12;][]{2014arXiv1407.2939W} includes galaxies from quite different environments. We visually inspected SDSS r-band images of CALIFA galaxies aiming to search for signatures of interactions (e.g., tidal tails) and nearby companions (positions and redshifts taken from NED\footnote{Nasa/IPAC Extragalactic database. http://ned.ipac.caltech.edu/} ) with spatially projected separation $\leq$ 100 kpc ($\sim$twice the physical radius of the extendest object of the sample) and systemic velocity difference $\leq$ 1000 km/s \citep{2013MNRAS.436.1765M}. These selection criteria basically identify ongoing one-to-one galaxy interactions, mergers showing tidal features and compact groups of galaxies, that we refer as interacting objects hereafter. The sample analyzed in this work includes $\sim$40\% of galaxies identified as in interaction following these criteria. The remaining objects in the sample will be referred as isolated galaxies, including: (1) objects with a low probability of environmental effects over a Hubble time \citep[e.g.,][]{1986A&A...154..343V}, (2) minor mergers of large mass ratios (M1/M2) which has non-visible effects on the morphology of the most massive galaxy (M1), and (3) galaxies in the boundaries of groups. Different nuclear ionization mechanisms (nuclear types, hereafter), 
including pure star formation (SF), low-ionization nuclear
emission-line regions (LINERS) and active galactic nuclei (AGN) are also represented in this sample of 177 CALIFA galaxies. Nuclear types were inferred from the location of each object in different diagnostic diagrams \citep{1981PASP...93....5B,1987ApJS...63..295V}: fluxes were obtained by fitting Gaussians to the emission lines (after removing the stellar continuum) in the nuclear spectrum (data cube central spaxel, see H13) of each galaxy.  Only the 166 galaxies in our sample with a S/N$>$3 in the required emission lines for diagnostic diagrams have a nuclear type classification. Figure \ref{BPT} shows the  emission-line diagnostic diagram which considers the [\ion{O}{iii}]/H$\beta$ versus [\ion{N}{ii}]/H$\alpha$ ratios for these objects. The remaining galaxies not satisfying the S/N threshold  have assigned 'indef' as
their nuclear type. 

In Appendix A (Table \ref{tabprop}), we list the galaxies in the sample together with their primary morphological characteristics, interaction/companion status and nuclear type obtained within the CALIFA collaboration. Table \ref{tabprop} also includes the systemic velocity and photometric major position angles of each object obtained from NED. 
 
\section{Measuring the ionized gas kinematics}
\label{medir}

\begin{figure}
  \resizebox{\hsize}{!}{\includegraphics{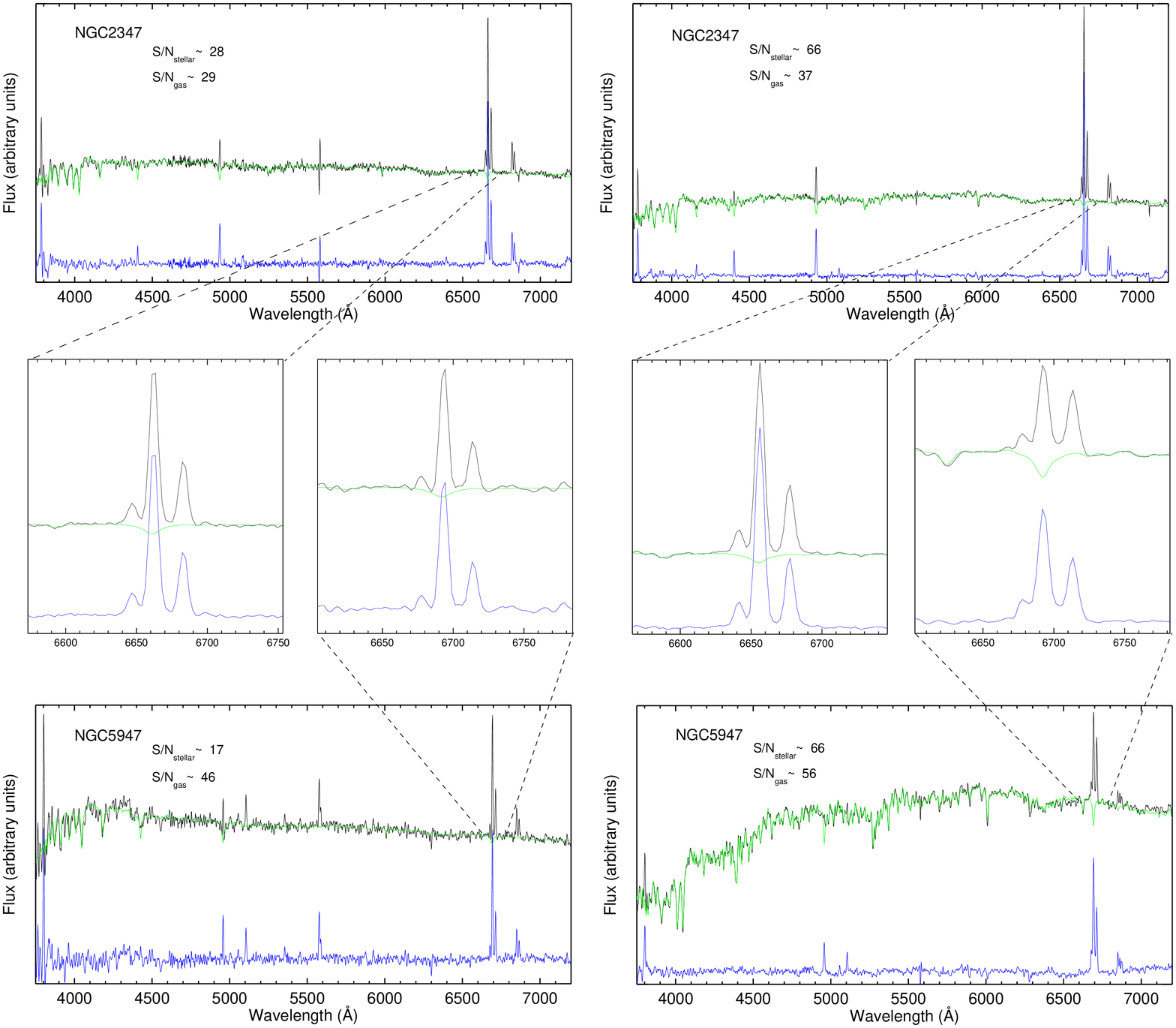}}
   \caption{Spectra of different S/N (in the stellar continuum and ionized gas) of \object{NGC~2347} and \object{NGC~5947} in the full spectral range (top and botton panels) and in a range around H$\alpha$+[NII]. Black spectra correspond to the data and green to the bestfit stellar continuum. Residual spectra, including emission lines, are in blue. } 
              \label{continuum_sub}%
    \end{figure}

The ionized-gas kinematics presented in this paper are measured from pure
emission-line data cubes resulting from the subtraction of the best stellar
continuum fit to the original CALIFA data (see some examples in Fig. \ref{continuum_sub}). The  bestfit stellar continuum is obtained
as a product of the stellar kinematics analysis of the data as described in
Falc\'on-Barroso et al. (in preparation). The wavelength range of CALIFA observations
includes many bright emission lines (e.g., [\ion{O}{ii}]~$\lambda\lambda3726,3729$ for V1200 mode and H$\alpha$+[\ion{N}{ii}]~$\lambda\lambda6548,6584$ or [\ion{O}{iii}]~$\lambda\lambda4959,5007$ for V500 mode), which were masked during the fitting of the stellar continuum. The stellar kinematic results are not computed for each spaxel of the data cube, but on Voronoi bins \citep[see][]{2003MNRAS.342..345C} ensuring a minimun signal-to-noise in the continuum of 20 per bin. Only spaxels with S/N $\geq$ 6 in the stellar continuum were used for binning (see details in Falc\'on-Barroso et al. in preparation). In order to extract the emission-line spectrum for each spaxel in the data cube, the individual spectra belonging to a given Voronoi bin were subtracted from the best stellar continuum fit for that bin. In the process we used a low order polynomial to ensure a flat continuum in the final emission-line data cube. Voronoi bins comes from the merging of a number of neighbouring spaxels in general not larger than the number of spaxels subtending the effective angular resolution of the data (except in the outer regions of the objects). Then, we do not expect strong variations of the stellar populations within each Voronoi bin. 

\subsection{Radial velocities}
\label{radialvel}

  In order to obtain an integrated kinematic view of the ionized gas of
the CALIFA galaxies, we have adopted the cross-correlation (CC hereafter) technique following the procedure proposed in \citet[][GL13 hereafter]{2013MNRAS.429.2903G}. We note that the CC technique globally compares a problem spectrum with a reference spectrum, quantitatively measuring their similarities. This
technique allows the dominant radial velocity of components integrated along the line of sight to be derived for systems with a variety of morphologies. Radial velocities are measured on the pure emission-line data
cubes without adopting any optimal binning. We employ a Gaussian profile to
model the upper part of the peak of the CC 
function. Estimated uncertainties in the location of the maximum of the
CC 
function are $\sim10$ km~s$^{-1}$
from the covariance matrix of the standard errors in the fitting
parameters for both V1200 and V500 configurations. Estimations of velocity dispersions for the ionized gas
were also obtained from the full-width-half-maximum of the
CC 
function for each spectrum at each selected spectral
range. However, any analysis of velocity dispersions goes beyond the
 scope of this work.

\begin{figure*}
   \resizebox{\hsize}{!}{\includegraphics[width=\textwidth]{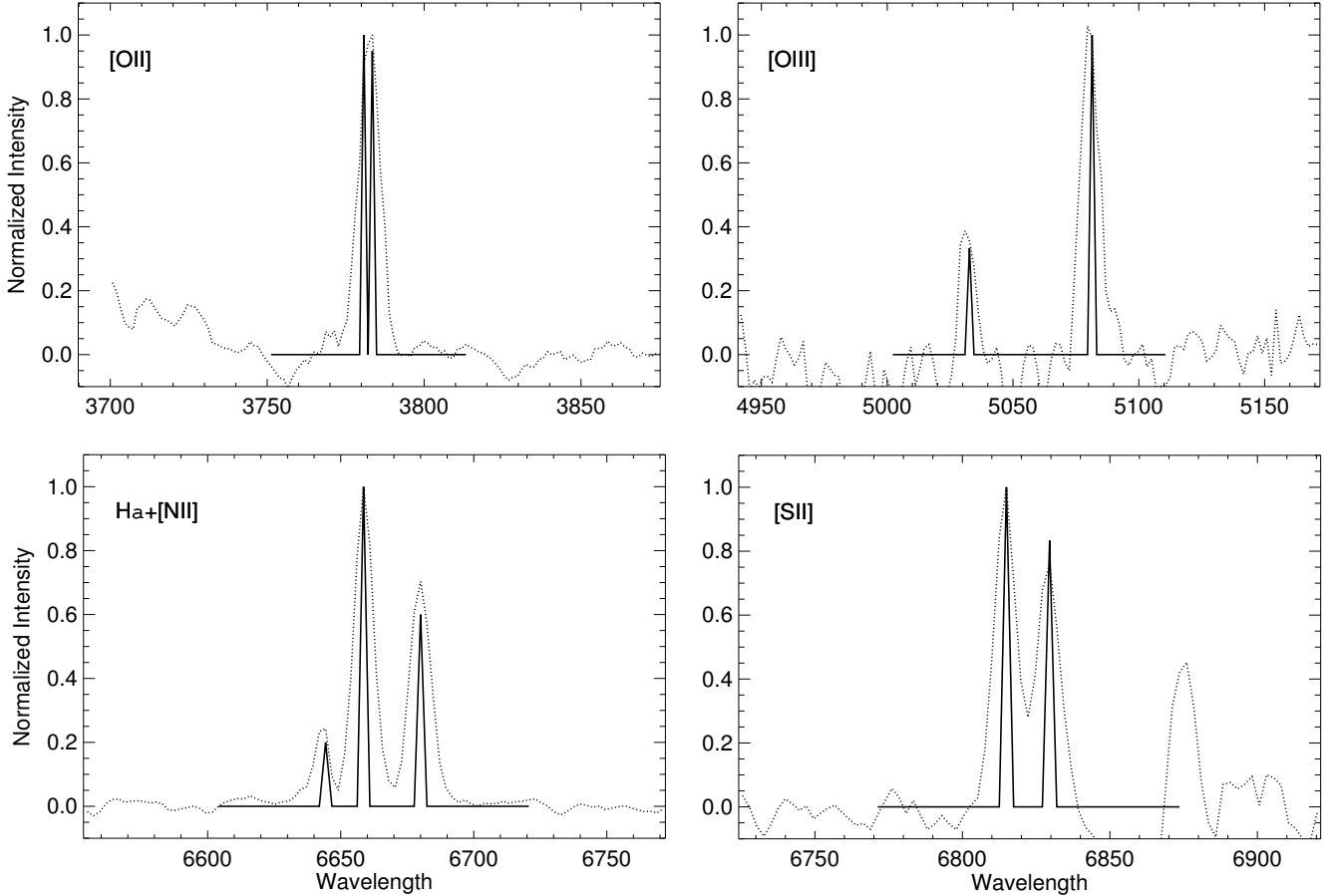}}
   \caption{Templates (solid line) generated including the appropiate number of Gaussians to simulate the emission lines in the four spectral ranges selected to apply the CC technique. For each galaxy Gaussians are centered at the corresponding wavelength according to their redshift in NED. Intensity ratio between emission lines were fixed according to atomic parameters. For those intensities unlinked to other lines, we checked the typical ratios across the objects (see text and Table \ref{tab1}). Dotted lines correspond to the central spectrum of \object{NGC~2347}.}

              \label{templates}%
    \end{figure*}

The 2D distribution of ionized gas was obtained by
integrating the signal in the spectral ranges selected to apply the
CC 
technique. Noise in the selected spectral ranges affects the recovered emission line distributions and
therefore these maps could slightly differ from those obtained by fitting
Gaussians to emission lines in the spectra.  Four spectral ranges (in
rest-frame wavelength) including the bright emission lines [\ion{O}{ii}]~$\lambda3726,3729$, 
[\ion{O}{iii}]~$\lambda\lambda4959,5007$, H$\alpha$+[\ion{N}{ii}]~$\lambda\lambda6548,6584$ and [SII]$\lambda\lambda6716,6730$, were selected
to infer the integrated ionized gas kinematics of the galaxies in our
sample (see Table \ref{tab1}). The reference spectrum needed for the
application of the CC 
technique was generated including
as many Gaussians as single emission lines are expected in each
spectral range selected to apply the technique (see Fig. \ref{templates}). The noise 
contribution was not included in the templates to avoid degradation of the
signal-to-noise ratio when the CC function is computed. The widths of
the Gaussians were selected to be negligible compared to the problem
spectra in order to keep the spectral resolution
in the resultant CC 
function. The wavelength (or
velocity) position of the required Gaussians to generate the template
spectrum corresponds to the center of the emission lines at the
redshift of the galaxies obtained from NED. Different
weights (in flux) were given to the Gaussians to account for fixed
intensity ratio between emission lines according to atomic
parameters. Because CC technique also account for differences/similarities between reference and problem spectra, those parameters unfixed by atomic values could have an impact on the uncertainties when determining radial velocities. 
Integrating the signal in narrow band filters (two pixels wide) centered on emission lines, we estimated the intensity ratios of those emission lines not linked by atomic parameters. The typical intensity ratio adopted to generate the reference spectra are indicated in Table \ref{tab1}.

\begin{table*}
\caption{\label{tab1} Spectral rest-frame bands selected for applying the CC 
technique to derive the radial velocities of the dominant ionized gas component.}
\begin{tabular}{lccc}
\hline\hline
CALIFA mode & Spectral range (\AA ) & Emission lines & Intensity ratio \\
\hline
V1200       & 3696-3759 \AA         & [\ion{O}{ii}]~$\lambda3726$+[\ion{O}{ii}]~$\lambda3729$ & [\ion{O}{ii}]~$\lambda3729$/[\ion{O}{ii}]~$\lambda3726\sim 0.38-1.3$ \\
V500        & 4929-5037 \AA         & [\ion{O}{iii}]~$\lambda4959$+[\ion{O}{iii}]~$\lambda5007$ & [\ion{O}{iii}]~$\lambda5007$/[\ion{O}{iii}]~$\lambda4959$=3\\
V500        & 6508-6623 \AA         & [\ion{N}{ii}]~$\lambda6548$+H$\alpha$+[\ion{N}{ii}]~$\lambda6584$ & H$\alpha$/[\ion{N}{ii}]~$\lambda6584\sim0.5-1.5$ \\
            &                       &                                                 & [\ion{N}{ii}]~$\lambda6584$/[\ion{N}{ii}]~$\lambda6548$=3 \\
V500        & 6686-6761 \AA         & [SII]$\lambda\lambda6716,6730$ & [SII]$\lambda6716$/[SII]$\lambda6730\sim0.7-1.2$ \\
\hline
\end{tabular}
\end{table*}

Experiments using a set of different templates (with varying 
velocity, dispersion and relative intensity of the input Gaussians to generate the templates) were performed for the different selected spectral ranges to assess for the uncertainties due to template selection. The selection of unfixed intensity ratios, such as 
[\ion{O}{ii}]~$\lambda\lambda3726,3729$ or [\ion{S}{ii}]~$\lambda\lambda6716,6730$, may have an important influence, increasing when lines are significantly blended (e.g., in the case of [\ion{O}{ii}]~$\lambda\lambda3726,3729$). To mitigate this limitation, we checked the average intensity ratios for each object in the sample, adapting the input ratio consequently. In same particular cases, the intensity ratios were even adapted in regions of a few spaxels. 
 In general, the uncertanties associated with the template selection is $\sim10$ km s$^{-1}$; in the worst cases ([\ion{O}{ii}]~$\lambda\lambda3726,3729$)), the selection of the template may 
contribute to increasing the radial velocity uncertainties up to 40 km s$^{-1}$, but this is only the case for a few spaxels. However, it is important to point out here that these uncertainties refer to individual spaxels, while we are exploring global trends in the sense of coherent motions retrieved from hundreds of contiguous spaxels.  

In order to assess for statistical uncertainties, we employed a Monte-Carlo simulation, deriving the different measurements from many realizations of the input data by adding the formal noise of the original data cubes (see H13 for error data cubes details). The procedure was repeated five hundred times, resulting in five hundred different input data cubes for each object in our sample. For each simulation we determined radial velocities only for those emission lines with an estimated S/N $\geq$ 6. In this way, we do not have the same final number of simulations per spectrum but we only consider those with a minimum number of estimations of one hundred, applying the Shapiro-Wilk test to check the normality of the distributions. Finally, we take the average as the measurement of the radial velocity. From the standard deviation we calculate the confidence interval at the 95\% confidence level, which indicate the reliability of the measured radial velocities. In general, although these errors vary from spectrum to spectrum and from object to object, radial velocity measurements are accurate (95\% confidence) to within 40 km s$^{-1}$ for [\ion{O}{ii}], 15 km s$^{-1}$ for [\ion{O}{iii}], 22 km s$^{-1}$ for H$\alpha$+[\ion{N}{ii}], and 30 km s$^{-1}$ for [SII] (see the standard deviation 2D-distributions in Appendix C).

The error in the estimated radial velocity for each spectrum is actually a combination of the three previous sources of uncertainties: CC peak centroid, template selection and statistical uncertainty, being the last the dominant in most of the spectra.

The emission line
profiles observed in galaxies arise mainly from one gaseous
component, although the presence of different gaseous systems with
different kinematics (
as happens in the central region of galaxies
with a certain degree of activity) can give rise to complex emission
line profiles, showing blue/red wings, shoulders or double peaks \citep[see, e.g.,][and references therein]{2011ApJ...741...50W}. These complex profiles are
commonly identified by visual inspection and their analysis requires a
kinematic deprojection in components \citep[see, e.g.,][]{1996ApJ...463..509A}. 
Additionally, the CC 
function was also used to infer the presence of complex emission line
profiles in the analyzed spectra by tracing line bisectors (see Sect. \ref{asym_medir}). 
 Again the Monte-Carlo simulation approach is used to estimate statistical uncertainties in the calculation of bisectors. From the many realizations, we compute average bisectors for each spectrum and calculated their standard deviations at the different bisector levels. These standard deviations are then used to compute the confidence interval (95\% confidence level) which is taken as uncertainties in tracing the bisectors. In general, bisector measurements are accurate within 20 km s$^{-1}$ (95\% confidence) up to 40\% bisector level (from the line peak to 40\% of the peak intensity) for [\ion{O}{iii}]~$\lambda5007$ emission line with a S/N$\geq$10. For spectra with an estimated S/N$\geq$20, the same accuracy is obtained for all the bisector levels.



\subsection{Direct estimation of kinematic parameters}
\label{estima}

Most of our sample 
galaxies 
with extended emission line distributions show a global receding-approaching velocity field resembling that of rotating systems or, at least, ordered motions (see Appendix C). Nevertheless, 
departures from circular/ordered motions are 
evident in many objects in the form of 
clear kinematic distortions in the velocity fields. For the sake of uniformity in the analysis of the entire sample, a simple approach has been adopted 
to estimate the kinematic parameters observed in the plane of the sky directly from the measured radial velocities. Hence we avoid using
any kinematical model or any assumption on internal dynamics or projection effects, and follow 
 the procedures described in Sect.s \ref{vsys_medir}, \ref{k_medir}, and \ref{pa_medir}. The adopted approach allows a rapid determination of 
the frequency of kinematic distortions detected in the target galaxies in the plane of the sky. Kinematic models are being applied for specific studies within the CALIFA collaboration \citep[see, e.g.,][]{holmes13}. 

Since enough gas is not always present, the distance from the center over which the kinematic parameters are estimated depends on each object, its surface brightness and its ionized gas content. Moreover, the reader should take into account that the sample includes many edge-on galaxies (ellipticity $\ge$ 0.6) and the estimation of the kinematic parameters for these objects are affected by projection effects or/and dust obscuration. Table \ref{tabpropkin} in appendix A includes the maximum radius (distance from the center) used to estimate the kinematic parameters for each galaxy. In general, the errors in the estimated parameters are taken as the standard deviation of the parameters in that distance range.

\subsubsection{Systemic velocity}
\label{vsys_medir}

The bulk motion of a galaxy is provided by its systemic velocity, typically adopted as the radial velocity at the kinematic center. At the CALIFA spatial resolution (see S12 and H13), the photometric and dynamical centers should agree in position for most of the objects (see Sect. \ref{k_medir}). Hence, the systemic velocity (V$_{sys}$ hereafter) for the objects in our sample have been estimated by averaging the radial velocities obtained in an aperture of 3.7 arcsec in radius (corresponding to twice the full-width-half-maximum of the CALIFA point-spread-function, see H13 for details) around the location of the optical nucleus (central spaxel of the data cube, see H13 for details). The systemic velocities were corrected from observed to heliocentric values using the correction in the header of each CALIFA data cube (see Table 4 in H13 for keyword). The standard deviation of the radial velocity measurements is taken as an indicator of the uncertainty in the determination of the systemic velocity. Large standard deviations indicate a large variation of the velocities in the central region of the objects. We note that the adopted aperture size corresponds to quite different physical sizes on objects at different redshifts. Indeed, an aperture of 3.7 arcsec on the galaxy at the lowest redshift in our sample (\object{NGC~3057}) corresponds to 376 parsec, while it covers 
almost 2.3 kpc for 
the largest redshift (\object{NGC~6166NED01}). Values of V$_{sys}$ have been obtained from [\ion{O}{ii}] (V$_{sys}^{[OII]}$), [\ion{O}{iii}] (V$_{sys}^{[OIII]}$), H$\alpha$+[\ion{N}{ii}] (V$_{sys}^{H\alpha}$),  and [SII] (V$_{sys}^{[SII]}$). In appendix A (Table \ref{tabpropasy}) we list these values for each object. Only one 
galaxy (\object{NGC~3158}) has undetectable emission in the central 3.7 arcsec around the optical nucleus location. Indeed the emission in \object{NGC~3158} (V$_{sys}$=6989 km s$^{-1}$ from NED) is concentrated in a region $\sim20$ arcsec northeast from its optical nucleus and could correspond to a small companion at 7171 km s$^{-1}$.  

\subsubsection{Kinematic center}
\label{k_medir}

For an ideal purely rotating disk galaxy, the rotation velocity varies with radius from the kinematic center, which coincides with the galactic center, and it is settled by the radial distribution of mass within the galaxy. The kinematic center of such a galaxy has a zero rotation velocity, and it is the location of the largest velocity gradient in the galaxy. Based on this idea and in order to 
estimate the location of the kinematic center (KC hereafter) of the sample galaxies, 
the average directional derivative of the H$\alpha$+[\ion{N}{ii}] velocity field was computed by calculating the average absolute difference of the obtained velocity for each spectrum with the velocity of the surrounding regions. The resulting image (velocity gradient image hereafter) emphasizes those regions in the velocity field where the data are changing rapidly. Therefore, for galaxies showing regular motions, the peak of the average directional derivative image (velocity gradient peak hereafter) should indicate the KC \citep{2008A&A...488..117K, 1997ApJ...490..227A}. This procedure works quite well for regular velocity fields, with uncertainties smaller than the spaxel size. For complex kinematics, more than a single velocity gradient peak can arise in the gradient image, indicating several locations where radial velocities are changing rapidly. Indeed, in many cases the velocity gradient distribution shows the optical nucleus surrounded by a ring-like or a bar-like structure of large velocity gradient values. The presence of several velocity gradient peaks in the velocity field of a galaxy is then identified as a clear departure from pure rotation. 

\begin{figure*}
   \resizebox{\hsize}{!}{\includegraphics[width=\textwidth]{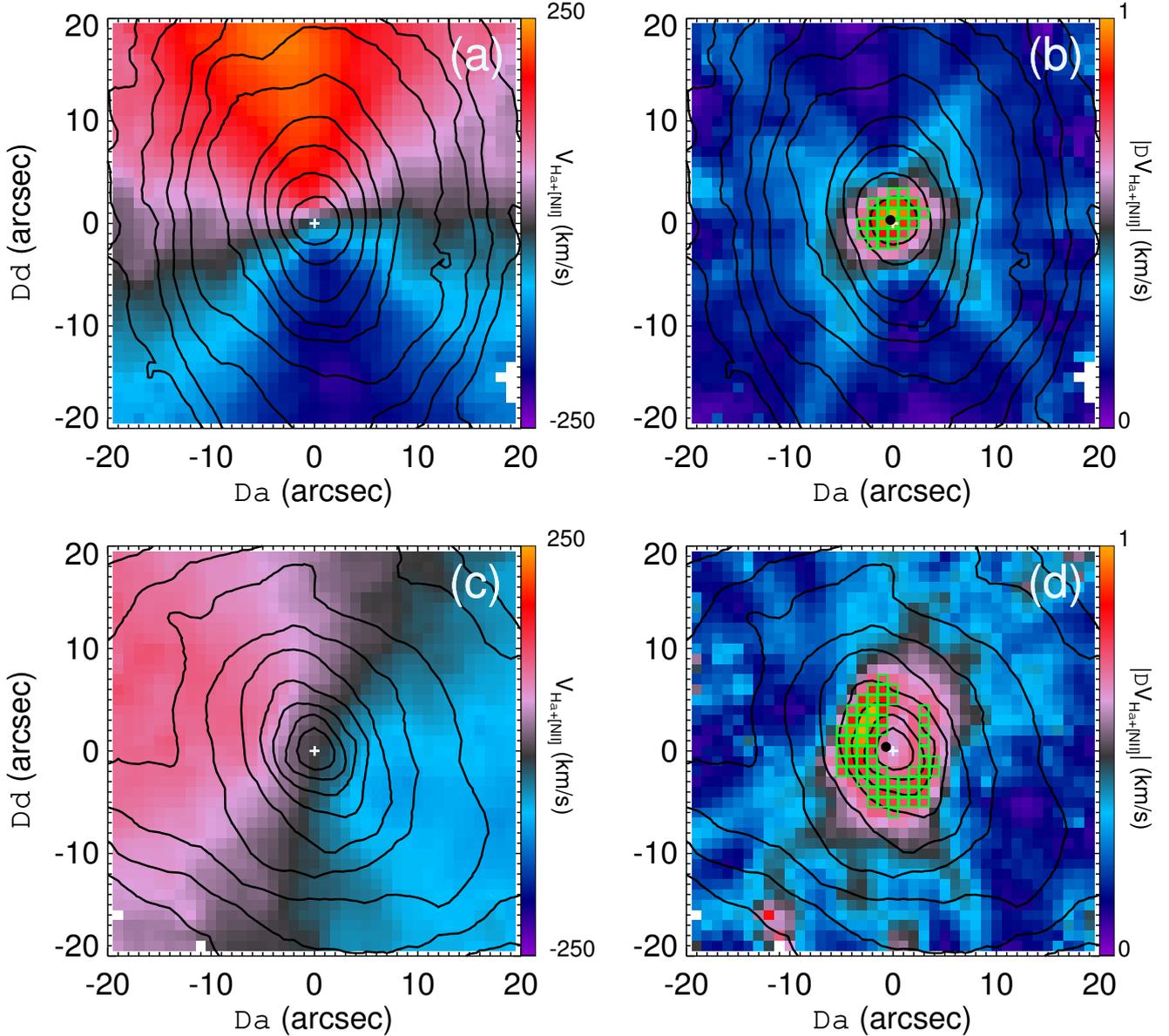}}
   \caption{Examples of the estimation of the kinematic center (KC) position for \object{NGC~2347}, a galaxy with a single peak in the velocity gradient image (panels a and b), and \object{NGC~5947}, an object with a ring structure in the central region of the velocity gradient distribution (panels c and d). Left panels show the ionized gas velocity field (from H$\alpha$+[\ion{N}{ii}] following the procedure in \ref{radialvel}) in the inner region of \object{NGC~2347} (panel a), and \object{NGC~5947} (panel c).  Right panels correspond to the average derivative of the H$\alpha$+[\ion{N}{ii}] velocity field for \object{NGC~2347} (panel b), and \object{NGC~5947} (panel d) normalized to their velocity gradient peaks. Green open squares indicate the velocity gradient pixels used to estimate the KC location through the average of their positions weigthed by their gradient values. Contours from the stellar continuum are overlapped. The white plus sign marks the location of the optical nucleus (peak of the stellar continuum) and the black circle indicates the estimated KC position.}

              \label{idiff}%
    \end{figure*}

In order to estimate the position of the KC, a region of 10$\times$10 arcsec$^2$ is selected around the largest velocity gradient peak. Then, we select those positions with a velocity gradient larger than the average velocity gradient inside this box. Finally, the KC is estimated from the weighted average location of the selected positions defining the peaks/structures of the velocity gradient map, using as weights the velocity gradient value at each location (see Fig. \ref{idiff}a, and \ref{idiff}b). Using the KC produces a more symmetric pseudo-rotation curve (see Sect. \ref{pa_medir}) than the optical nucleus or any of the locations with a large velocity gradient in the velocity field. This is the reason why we have adopted this procedure to estimate the location of the KC instead of just selecting the position of the largest velocity gradient in the maps. Uncertainties in the location of KCs are estimated assuming different radii --- from the spaxel size to the PSF size (H13) --- to define the surroundings of each spaxel when deriving the gradient images. Obviously, the detection of structures in the velocity gradient images is limited to the spatial resolution of the CALIFA data cubes.


The velocity at the KC (V$_{KC}$ hereafter) was estimated following the same procedure as for systemic velocities (see Sect. \ref{vsys_medir}), but with the aperture centered on the KC (see Table \ref{tabpropkin} in Appendix A).
 
\subsubsection{Position angle of kinematic axes }
\label{pa_medir}

\begin{figure*}
   \resizebox{\hsize}{!}{\includegraphics[width=\textwidth]{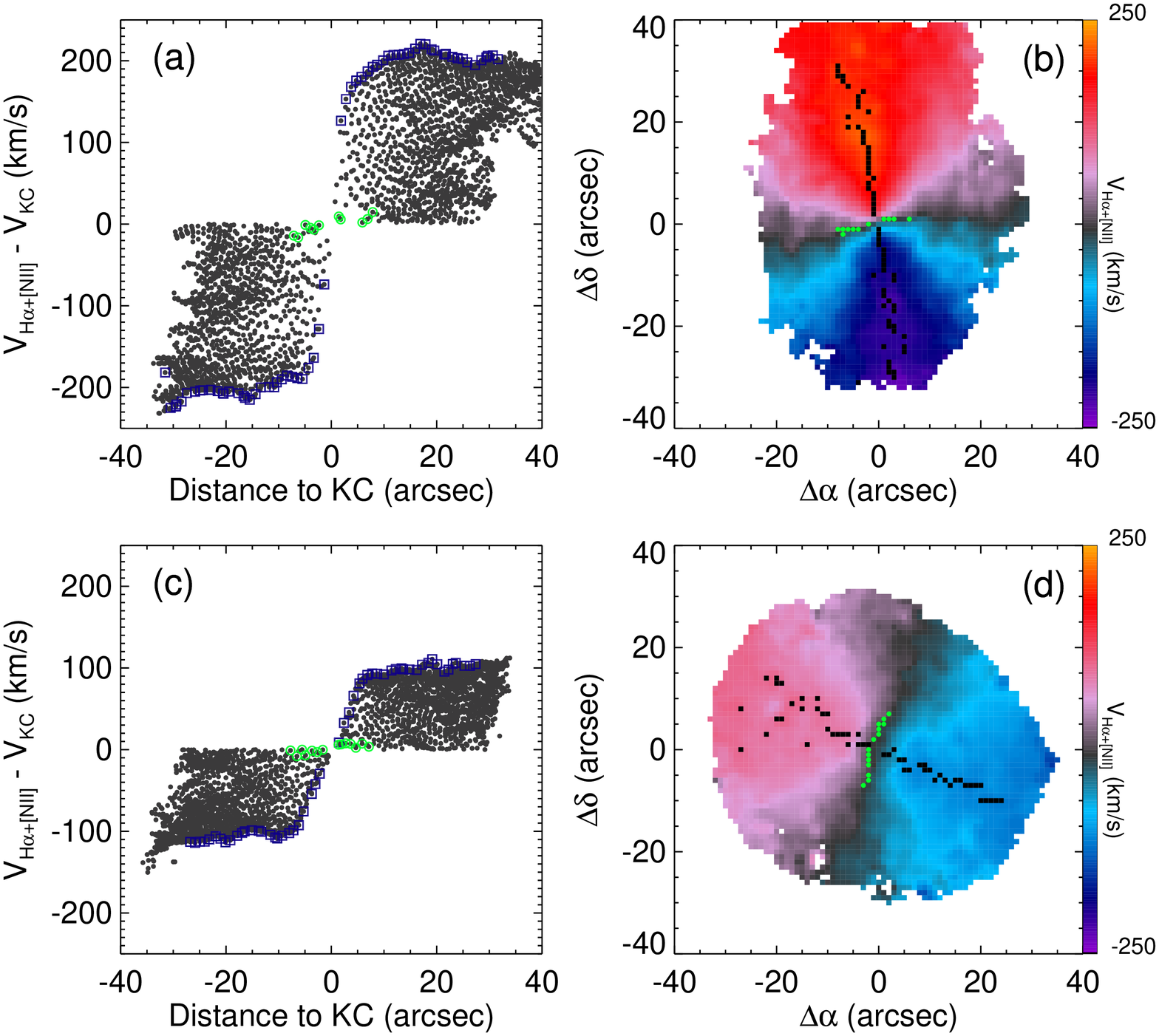}}
   \caption{(Left-panels) Distance to the  KC
(in arcsec) versus the observed velocity for each spatial element of the CALIFA data cube for (a) \object{NGC~2347}, and (c) \object{NGC~5947}. The blue squares indicate those spaxels with the largest difference in velocity respect to the radial velocity of the KC at each radius, tracing a pseudo-rotation curve and defining a kinematic major axis. Green circles correspond to those spaxels with the lowest difference in velocity to the KC selected to estimate the PA$_{minor}$.
(Right-panels) The ionized gas velocity field (from H$\alpha$+[\ion{N}{ii}] following the procedure in \ref{medir}) of (b) \object{NGC~2347}, and (d) \object{NGC~5947}. 
Filled squares mark the spaxels with the largest absolute velocity difference with the KC
velocity at each radius -- the same than those marked in blue open squares in (a) and (c)--. These spaxels trace the largest velocity gradient in the velocity field providing a direct estimation of the kinematic major axis position angle. Filled green circles correspond to those spaxels with a similar velocity than the KC -- the same than green open circles in (a) and (c) -- and tracing the kinematic minor axis.}
              \label{radial-velocity}%
    \end{figure*}

 The position angle of the kinematic major axis provides the mean orientation of the ionized gas velocity field. It is usually defined as the angle between the north and the receding side of the velocity field \citep[e.g.,][]{2009ApJ...692.1623H,1997MNRAS.292..349S}. For a rotating disk, the dependence of the kinematic major axis on galactocentric distance is negligible. 
The average orientation of the observed velocity fields (PA$_{kin}$ hereafter) can be directly estimated from the polar position of the spaxels defining the kinematic line of nodes \citep{1992ApJ...387..503N,1987MNRAS.228..595B} inferred as follows: (1) we plot the radial velocity of each spectrum/spaxel in a distance-velocity diagram, in which the origin (reference point) is taken as the KC position (see Appendices A and C); (2) we select those spectra/spaxels with the largest velocity differences and uncertainties smaller than the typical velocity error (22 km/s) with respect to V$_{KC}$ as a function of radius (see Fig. \ref{radial-velocity}a,c), which trace the observed pseudo-rotation curve; (3) we locate the selected spaxels on the velocity field (see Fig. \ref{radial-velocity}b,d) to trace the kinematic line of nodes; and (4) we average the polar coordinates (respect to the KC) of the selected spaxels to obtain PA$_{kin}$. 
This simplistic approach to trace the kinematic major axis additionally allows us to determine the degree of symmetry of the velocity field by comparing 
mean position angle 
 from the receding side (PA$_{kin,rec}$) with that 
from the approaching side (PA$_{kin,app}$). In a similar way, we can estimate a mean position angle for the kinematic minor axis (PA$_{minor}$ hereafter) by selecting those spectra with the lowest velocity differences to V$_{KC}$ at any radius (see Fig. \ref{radial-velocity}a,c) and locating them on the velocity field (see Fig. \ref{radial-velocity}b,d). Only those spectra with a velocity difference smaller than the typical error for H$\alpha$+[\ion{N}{ii}] velocities (22 km s$^{-1}$) were considered to trace the PA$_{minor}$. In a pure rotating disk galaxy, the kinematic minor and major axes are everywhere perpendicular \citep{1998gaas.book.....B}. Therefore, the comparison of mean position angles for both axes also provides a parameter to account for kinematic distortions in the velocity field and departures from rotation.

 For a rotating disk galaxy, this procedure traces the kinematic major and minor axes. For a distorted velocity field, the selected spaxels from the pseudo-rotation curve in the distance-velocity diagram may not be necessarily 
aligned and defining a clear direction on the velocity field. The reported position angles for kinematic major and minor axes in this work (see Table \ref{tabpropkin} in Appendix A) actually corresponds to the average of the polar coordinates of spaxels selected from the position-velocity diagram relative to the adopted KC (see Sect. \ref{k_results}). The standard deviation ($\delta$PA hereafter) will provide the degree of alignment of these positions and then the agreement (or not) of the traced kinematic line of nodes with the classical idea of kinematic axis. It is important to note here that a rotating disk showing a variation of the inclination with galactocentric distance (tilted rings) will show a curved kinematic major axis, instead of a straight axis. Indeed, the axis curvature is related to the galactocentric variation of the disk inclination. The variation of PA$_{kin}$ as a function of a tilt follows the relation:
\begin{equation}
\label{inclination}
tan \ \Sigma[i\pm\Delta i] = cos(i \pm\Delta i)/cos(i) * tan \ \Sigma[i]
\end{equation}

where $\Sigma$[i] is the position angle of the major kinematic axis of a flat rotating disk that is seen at an angle i and $\Sigma$[i$\pm\Delta$i] is the corresponding position angle when the disk is tilted by $\Delta$i. For these rotating systems, the adopted PA$_{kin}$ approach will result in larger standard deviations than for flat disks. Following Eq. \ref{inclination}, a linear variation of 30$^{\circ}$ in $\Delta$i from the center to the outer parts of a tilted disk could increase the $\delta$PA$_{kin}$ up to $\sim$20$^{\circ}$ for high inclined objects (i$\geq$60$^{\circ}$). When i$\leq50^{\circ}$, $\delta$PA$_{kin}$ will be smaller than 10$^{\circ}$ for a similar $\Delta$i galactocentric variation.

The accuracy in the estimation of the kinematic PA following the procedure described is a complex function of the actual position of the spatial elements of the CALIFA data cube set by the image reconstruction (see S12 for details), the uncertainties in determining the optical nucleus (taken at the data cube central spaxel, see H13 for details), the errors in the location of the KC, and the radial velocities uncertainties. As a reference, the accuracy in determining a defined position angle from a set of spatial elements in the CALIFA data cube is smaller than 0.5 degrees. Through the five hundred velocity fields for each object resulting from the Monte-Carlo simulations, we account for statistical uncertainties approaching the kinematic PAs. In general, PA$_{kin}$ and PA$_{minor}$ and their standard deviations are accurate (95\% confidence) within 2 degrees. Regardless, the standard deviation of the positions averaged to estimate PA$_{kin}$ is taken as the uncertainty of this parameter.



 


 

\subsection{Presence of kinematically distinct gaseous components}
\label{asym_medir}

The presence of double/multiple gaseous components with different kinematics in galaxies is evident from the shape of the emission line profiles in their spectra, showing asymmetries, shoulders or double peaks. These features have been interpreted as due to rotating gaseous disks, outflows/inflows or dual active galactic nuclei \citep[see, e.g.,][and references therein]{2012ApJ...745...67F}. The [\ion{O}{iii}]~$\lambda5007$ line is usually selected to look for double/multiple gaseous components in the spectra of galaxies, since it is the brightest unblended emission line in the optical range for typical spectral resolutions (including CALIFA V500 data). Moreover, only faint stellar features are present in the [\ion{O}{iii}] spectral region and hence, [\ion{O}{iii}] is little affected by uncertainties in the subtraction of the stellar component, the opposite that in the case of e.g., H$\beta$. A systematic search of double-peaked emission-line profiles can be done by tracing the bisector of a single emission line (e.g., [\ion{O}{iii}]~$\lambda5007$) or the bisector of the CC
peak function (obtained when comparing a problem spectrum with a reference created using a defined shape for the lines in the spectral range) and studying the shape of these bisectors, in particular the deviation from the central position/velocity for different bisector levels (GL13).

  The spectral resolution of CALIFA is not the best to identify dynamically distinct gaseous components, but the identification of asymmetries in the emission line profiles will indicate where such multiple components may exist. It should also be noted that, because of the limited spatial resolution, beam-smearing of the velocity gradient could translate into non-Gaussian line structure. In any case, the strength of these asymmetries could have an impact when measuring emission line fluxes if these asymmetries are not taken into account (see Appendix B). The searching for asymmetries in the emission line profiles of the galaxies in the analyzed sample has been performed by analyzing the bisector shape of the CC
peak function obtained when applying the CC 
 technique to the [\ion{O}{iii}]~$\lambda\lambda4959,5007$ spectral range (from 4929 to 5036 \AA \ in rest frame). The [\ion{O}{iii}]~$\lambda\lambda4959,5007$ spectral range was selected instead of the single [\ion{O}{iii}]~$\lambda5007$ emission line profile to mitigate the influence of any observational or instrumental signature (e.g., cosmic ray) not properly removed during the data reduction process and affecting a single emission line, since the CC
 technique will smooth out these features 
providing an average profile shape. The reference spectrum (template) was generated following the procedure described in Sect. \ref{radialvel}. It is important to note here that the level of noise in the problem spectrum affects the detection of asymmetries (the template is generated without noise). In appendix B we analyze the limits in the detection of double/multiple gaseous components in CALIFA V500 data cubes through a single-Gaussian model as a function of the signal-to-noise of the profile. Based on the results in appendix B, the searching of double/multiple gaseous components through the asymmetries in the [\ion{O}{iii}] emission profiles should be restricted to observed spectra with an estimated S/N ([\ion{O}{iii}]~$\lambda5007$) $\geq$ 30 and over a 10\% intensity of the peak (10\% bisector level). However, the Monte Carlo simulations in appendix B do not include the effects of the stellar subtraction that could be playing a role in the observed [\ion{O}{iii}] profiles, mainly in those with lower S/N. To mitigate the impact of uncertainties in the stellar subtraction on the detection of multiple gaseous components, we only will consider those spectra with a minimum S/N of 40. We establish that a profile is actually asymmetric only when the absolute differences between the central velocity provided by the CC 
function and the velocity at two bisector levels (at least) are larger than the limits at each level established in Appendix B (equations \ref{eqc1} and \ref{eqc4}).

 

\section{Results}
\label{result}

Appendix A provides some structural parameters obtained from NED (V$_{sys}^{NED}$, and PA$_{NED}$) and from measurements within the CALIFA collaboration (Table \ref{tabprop}). Appendix A also includes the kinematic parameters (see Sect. \ref{estima}) estimated from the H$\alpha$+[\ion{N}{ii}] velocity field (kinematic center positions, velocities of kinematic center and position angles of kinematic axes) of each object in the sample (Table \ref{tabpropkin}). The ionized gas velocity fields derived from [\ion{O}{ii}]~$\lambda\lambda3726,3729$ (V1200 mode), [\ion{O}{iii}]~$\lambda\lambda4959,5007$, H$\alpha$+[\ion{N}{ii}]~$\lambda\lambda6548,6584$, and [SII]$\lambda\lambda6716,6730$ (V500 mode) emission lines for the 177 galaxies with a minimum gas content/detection observed up to January 2013 in the CALIFA survey in both V1200 and V500 configurations are shown in Appendix C. In the following, we present the general kinematic properties of this sample. 

\subsection{Ionized gas distribution and velocity fields} 

For a general kinematic study of the ionized gas of galaxies in the CALIFA Survey, only spectra in the CALIFA data cubes with a S/N $\geq$ 6 in both the stellar continuum and ionized gas are considered (see Sect. 3). With these criteria, H$\alpha$+[\ion{N}{ii}] emission is detected in at least 5\% of the spatial elements (about 4420 in the resampled data cube, see S12 for details) on all galaxies in our sample. The simultaneous detection of [\ion{O}{ii}], [\ion{O}{iii}], H$\alpha$+[\ion{N}{ii}], and [SII] emission lines was positive for 152 objects (86\%). The galaxies in the sample show a large variety of ionized gas 2D distributions and, in general, their velocity fields show a global pattern of receding and approaching velocities (see appendix C).


\subsubsection{Systemic velocity}
\label{vsys_results}

\begin{figure}
   \resizebox{\hsize}{!}{\includegraphics{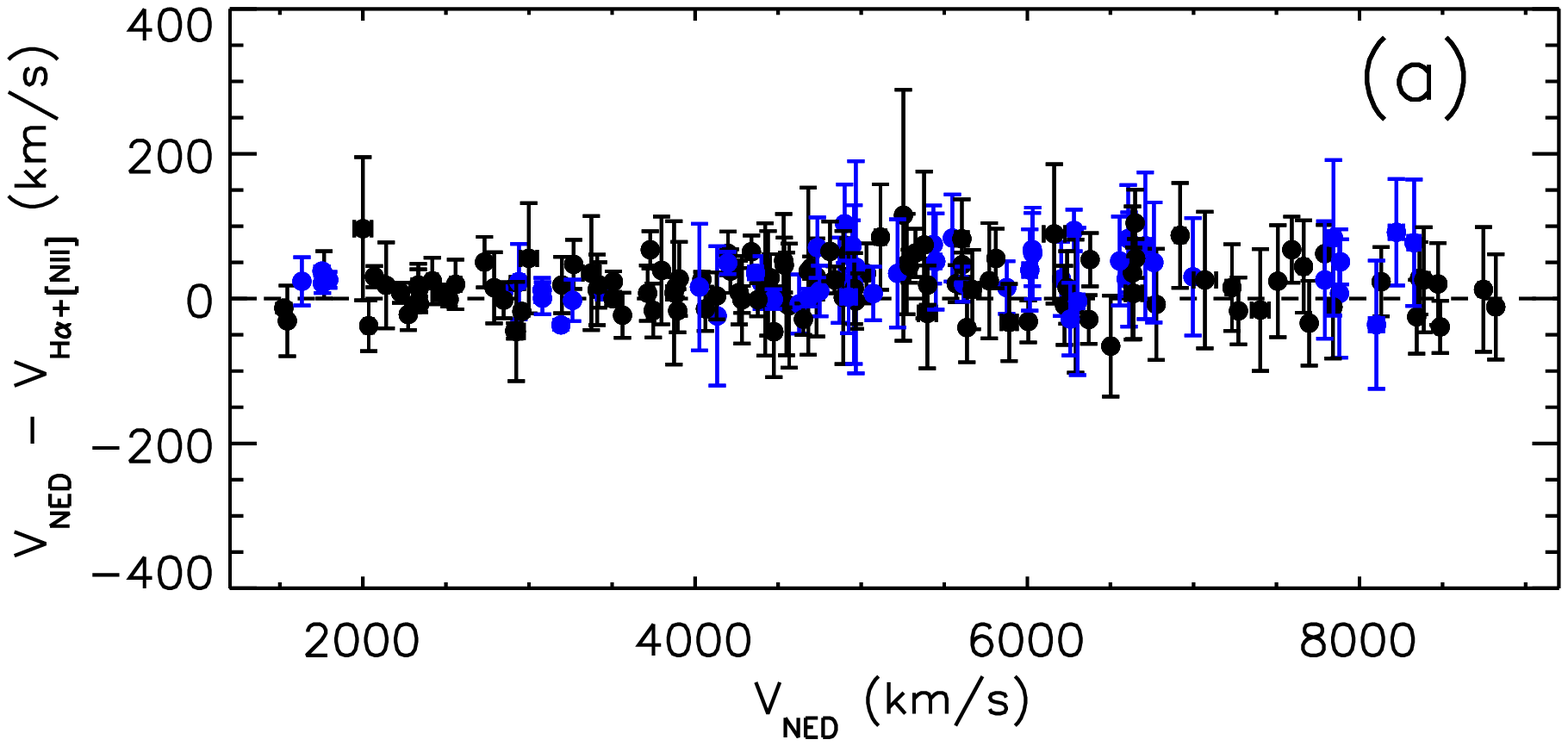}}
   \resizebox{\hsize}{!}{\includegraphics{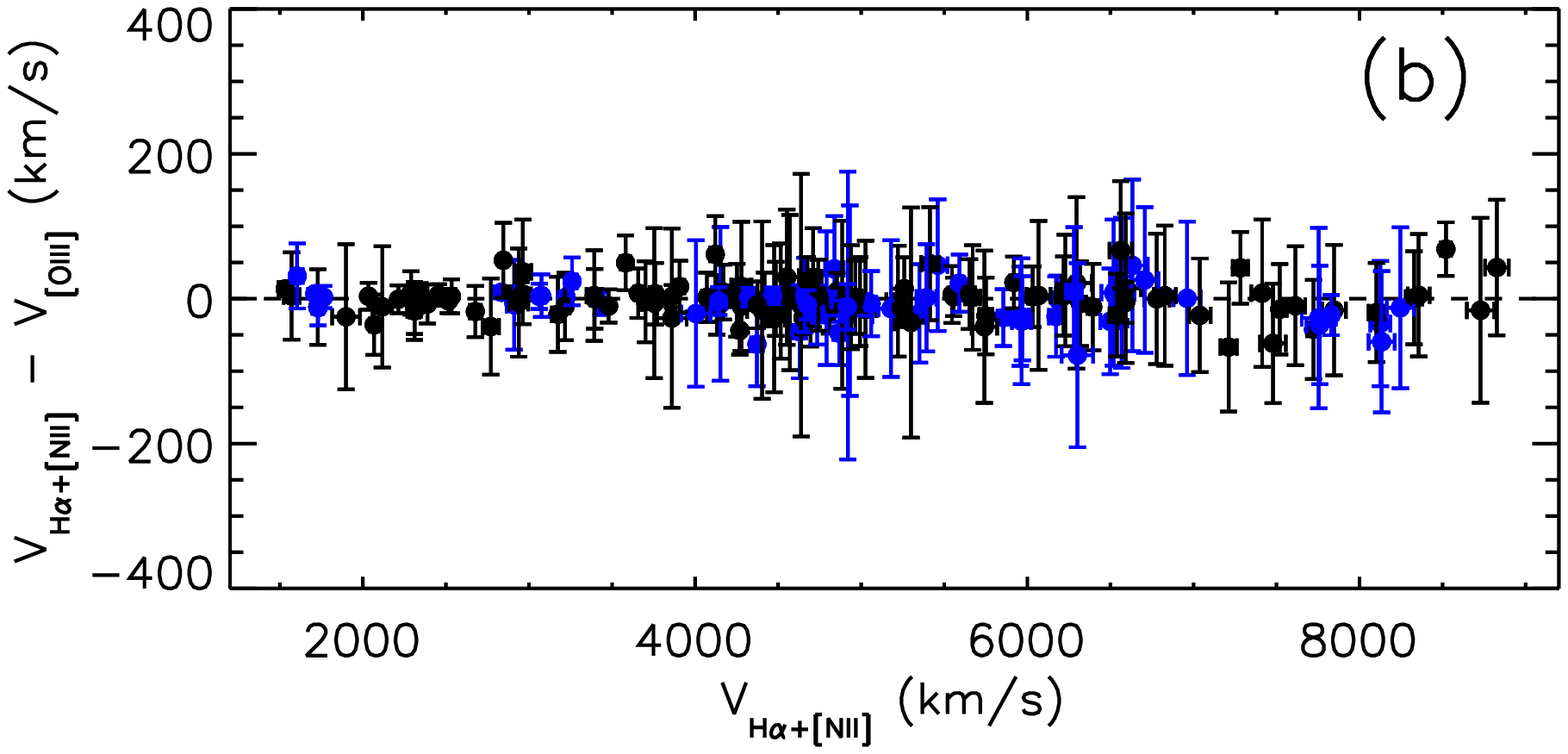}}
   \caption{(a) Comparison of the systemic velocities obtained from NED and those estimated from the H$\alpha$+[\ion{N}{ii}] radial velocities (see Sect. \ref{vsys_medir}) for the objects in the sample. Error bars in the horizontal axis correspond to the velocity uncertainties provided by NED. These velocity uncertainties and the standard deviation of the averaged radial H$\alpha$+[\ion{N}{ii}] velocities of each object were combined to calculate the error bars for the vertical axis. (b) Comparison of the systemic velocities derived from H$\alpha$+[\ion{N}{ii}] and [\ion{O}{iii}] emission lines for the objects in the sample. Error bars in the horizontal axis correspond to the standard deviation of the averaged radial velocities to estimate V$_{sys}^{H\alpha}$ for each object. Error bars for the vertical axis were calculated from the standard deviation of H$\alpha$+[\ion{N}{ii}] and [\ion{O}{iii}] radial velocities. In both panels, blue dots correspond to objects identified as galaxies in interaction (see Sect. \ref{sample} and Table \ref{tabprop} in Appendix A)}
              \label{vsys}%
    \end{figure}

The derived V$_{sys}^{H\alpha}$ and systemic velocities taken from NED (V$_{sys}^{NED}$ hereafter) are in good agreement (see Fig. \ref{vsys}a). The weighted mean of V$_{sys}^{NED}$-V$_{sys}^{H\alpha}$ is 8.6 km s$^{-1}$, using as weights the error bars in Fig. \ref{vsys}, which were derived from the standard deviation of the radial measurements and the published velocity uncertainties for V$_{sys}^{NED}$. For 175 of the 176 objects in the sample with H$\alpha$+[\ion{N}{ii}] in the central region, the discrepancies between V$_{sys}^{NED}$ and V$_{sys}^{H\alpha}$ can be attributed to differences in the procedures to determine them. Indeed, velocities in NED come not only from ionized gas but also from stellar or \ion{H}{i} observations, velocities that can be significantly different. V$_{sys}^{NED}$ could correspond to the velocity at the optical nucleus or to the brightest zone of each object, which could be far of the nucleus. Indeed, the large difference of 115 km s$^{-1}$ in V$_{sys}^{NED}$-V$_{sys}^{H\alpha}$, corresponding to \object{NGC~0160}, is well explained if V$_{sys}^{NED}$ comes from the brighter emission knots at the southwest of its nucleus. Only the derived V$_{sys}^{H\alpha}$ for \object{NGC~6166NED01} presents a large discrepancy (larger than 1200 km s$^{-1}$) with its V$_{sys}^{NED}$. The SDSS r-band image for this object shows several peaks and the center of the CALIFA data cube is located at the brightest, leaving the others at the southwest. These knots actually correspond to at least three objects: \object{NGC~6166A} at V$_{sys}^{NED}$=9271 km s$^{-1}$; \object{NGC~6166B} at V$_{sys}^{NED}$=8104 km s$^{-1}$; and \object{NGC~6166C} at V$_{sys}^{NED}$=9850 km s$^{-1}$. At the southwest of the CALIFA data cube center, H$\alpha$+[\ion{N}{ii}] velocities (with an average V$_{H\alpha+[NII]}$ of 9238 km s$^{-1}$) are in agreement with the NED values for \object{NGC~6166A}. The average velocity of the brightest knot (CALIFA data cube center) is 8048 km s$^{-1}$, in agreement with V$_{sys}^{NED}$ for \object{NGC~6166B}.


According to the adopted signal-to-noise threshold, [\ion{O}{iii}] emission is not detected in the central 3.7 arcsec of 15 galaxies (see Table \ref{tabpropasy} in Appendix A). For the remaining galaxies, V$_{sys}^{H\alpha}$ and V$_{sys}^{[OIII]}$ are in good agreement (see Fig. \ref{vsys}b), $<$V$_{sys}^{H\alpha}$-V$_{sys}^{[OIII]}>$ = 2.2 $\pm$ 24.8 km s$^{-1}$. For nine objects, absolute differences range from 50 to 78 km s$^{-1}$. The poor signal-to-noise of [\ion{O}{iii}] emission lines in the central region of \object{NGC~0499} (V$_{sys}^{H\alpha}$-V$_{sys}^{[OIII]}\sim-63$ km s$^{-1}$), \object{NGC~6063} (V$_{sys}^{H\alpha}$-V$_{sys}^{[OIII]}\sim53$ km s$^{-1}$), \object{UGC~08234} (V$_{sys}^{H\alpha}$-V$_{sys}^{[OIII]}\sim-59$ km s$^{-1}$), and \object{UGC~08267} (V$_{sys}^{H\alpha}$-V$_{sys}^{[OIII]}\sim-67$ km s$^{-1}$) could explain these differences. The discrepancies for the other five objects (\object{NGC~6394}, \object{NGC~7466}, \object{UGC~03253}, \object{UGC~06036}, and \object{UGC~11717}) could be associated with nuclear activity, as in the case of a broad line region affecting the permitted (H$\alpha$) emission lines and/or the presence of strong outflows producing double peaked profiles. Indeed, \object{NGC~6394} and \object{NGC~7466} are classified as Seyfert 2 galaxies \citep{2010A&A...518A..10V,2009ApJ...707..787G}. We were unable to find published work on nuclear activity in \object{UGC~03253}, \object{UGC~06036}, or \object{UGC~11717}, although according to the emission-line diagnostic diagram for the most central spectrum (see Sect. \ref{sample}), \object{UGC~03253} is a star forming galaxy, while \object{UGC~06036}, and \object{UGC~11717} are LINERS. Indeed, \object{UGC~03253} and \object{UGC~11717} have clear evidence of asymmetric [\ion{O}{iii}] profiles in the central region  (see Sect. \ref{asym_results}), suggesting the presence of several gaseous systems.


Systemic velocities derived from [SII] (V$_{sys}^{[SII]}$) are also in good agreement with V$_{sys}^{H\alpha}$ values: $<$V$_{sys}^{H\alpha}$-V$_{sys}^{[SII]}>$ = -0.9 $\pm$ 23.4 km s$^{-1}$. We have omitted all objects with redshifts ranging from 5933 to 6588 km s$^{-1}$ in this comparison for which [SII] profiles are affected by a poor subtraction of the bright sky line at 6863.97 \AA . No [SII] emission is detected in the central region of 21 objects (see Table \ref{tabpropasy} in Appendix A).

With the adopted S/N thresholds and using the V1200 data cube from CALIFA, [\ion{O}{ii}]~$\lambda\lambda3726,3729$ emission is not detected in the central region of 12 galaxies (see Table \ref{tabpropasy} in Appendix A). For the remaining objects, we found $<$V$_{sys}^{H\alpha}$-V$_{sys}^{[OII]}> \sim-25 \pm$ 60 km s$^{-1}$. Different elements contribute to this discrepancy: (1) the relative strengths of the [\ion{O}{ii}]~$\lambda\lambda3726,3729$ emission lines, with ratios ranging from 0.35 -- for high electronic density regions -- to 1.5 -- for low electron density zones \citep[e.g.,][]{2006MNRAS.366L...6P}, and the strong blending of the two lines in almost all central spectra produce a large number of different [\ion{O}{ii}] observed profiles. ; (2) [\ion{O}{ii}] is more affected by dust obscuration than H$\alpha$+[\ion{N}{ii}]; and (3) the subtraction of the stellar continuum under the [\ion{O}{ii}] doublet is tricky because it is close to the blue border of the CALIFA V1200 spectral range.  All these factors complicate the definition of an adequate template for the [\ion{O}{ii}]~$\lambda\lambda3726,3729$ spectral range (see Sect. \ref{medir}), affecting the determination of the radial velocities.

\subsubsection{Velocity gradients: estimating the location of the kinematic center}
\label{k_results}

The 2D distribution of radial velocities derived from H$\alpha$+[\ion{N}{ii}] spectral range allows the derivation of a velocity gradient image for most of the galaxies in the analyzed sample. As we already noted, the gradient image should present a clear peak at the KC for a purely rotating galactic disk. With this idea in mind, we have divided our set of galaxies according to the structures of the velocity gradient images: multiple peaks/structures can be due to different factors, including the presence of dynamically distinct components (e.g., a bar), whose study is beyond the scope of this work. We refer as Multi velocity Gradient Peak (MGP hereafter) to those galaxies showing several velocity gradient peaks or clear structures in the velocity gradient map. Single velocity Gradient Peak (SGP hereafter) indicates galaxies with a conspicuous velocity gradient peak, sometimes surrounded by faint structures or secondary peaks of much lower intensities. Galaxies in the SGP class should correspond to systems dominated by rotation, while MGP galaxies to objects presenting circular and non-circular motions at the velocity resolution of the CALIFA data. We lack a reliable velocity gradient map for a subset of 26 galaxies: (1) for 14 objects\footnote{namely \object{NGC~0160}, \object{NGC~0499}, \object{NGC~3158}, \object{NGC~6146}, \object{NGC~6154}, \object{NGC~6166NED01}, \object{NGC~6338}, \object{NGC~7236}, \object{NGC7550}, \object{NGC~7671}, \object{UGC~00335NED02}, \object{UGC~08234}, \object{UGC~10695} and \object{UGC~11958}} H$\alpha$+[\ion{N}{ii}] emission is not extended enough in the central region to derive a velocity gradient image; for 7 edge-on galaxies\footnote{\object{IC~2095}, \object{MCG-01-54-016}, \object{NGC~6081}, \object{UGC~04722}, \object{UGC~6036}, \object{UGC~08250}, and \object{UGC~10650}} the information perpendicular to the galaxy plane/disk is not enough to identify peaks or structures; and 4 spiral galaxies\footnote{\object{NGC~5720}, \object{NGC~6032}, \object{UGC~01938}, and \object{UGC~11649}} show a patchy-distributed emission in the central ten arcsecs. We classify these objects as Unclear velocity Gradient Peak (UGP hereafter) galaxies. The optical nucleus is taken as reference to estimate the kinematic axes (when possible) for this subset of objects. Appendix A (Table \ref{tabpropkin}) includes the classification according to the structures in their H$\alpha$+[\ion{N}{ii}] velocity gradient maps (SGP, MGP or UGP) as result of combining the independent visual identification of several members of the CALIFA collaboration.

  \begin{figure}[h]
   \includegraphics[width=8.8cm]{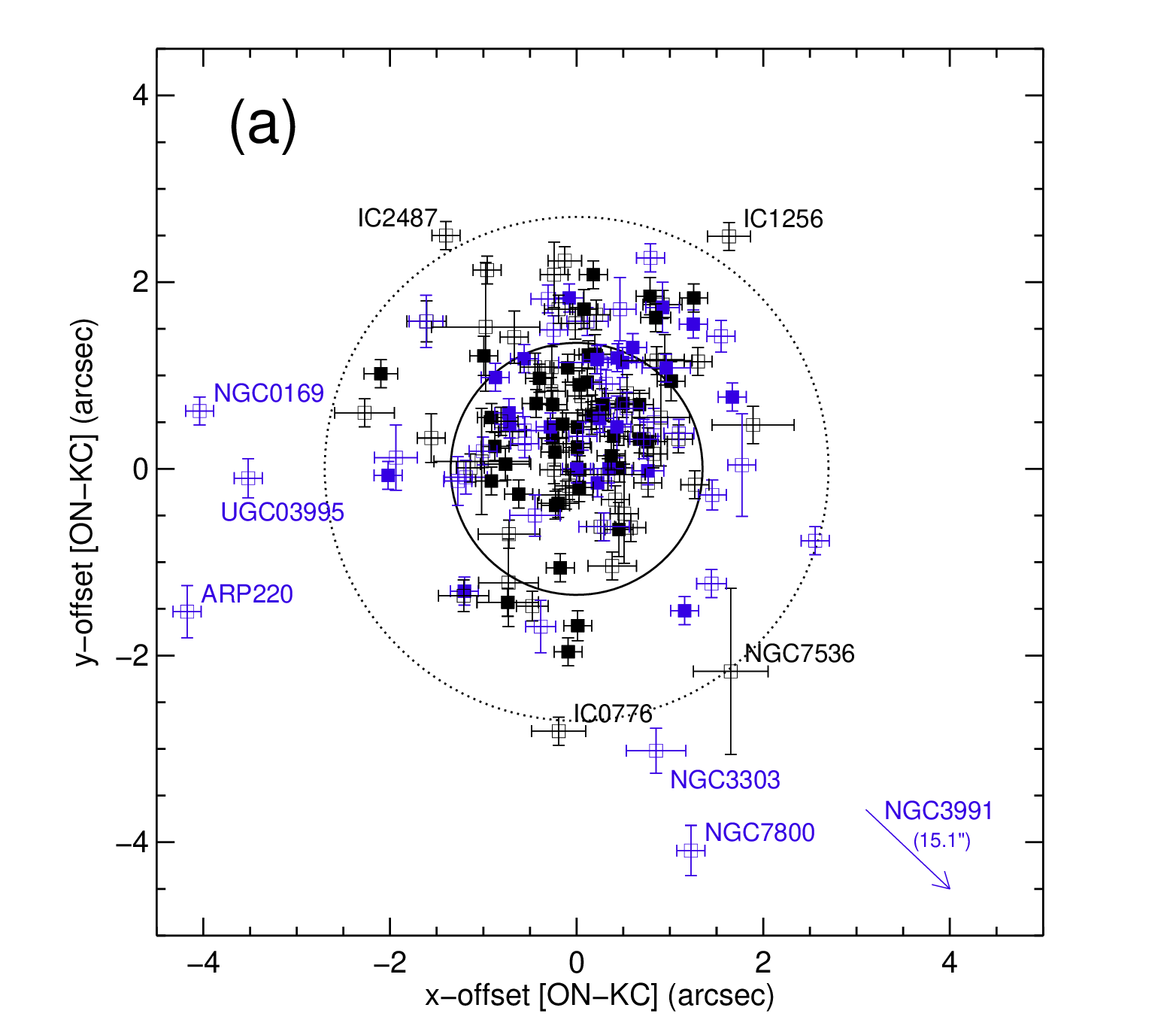} \\
   \includegraphics[width=8.4cm]{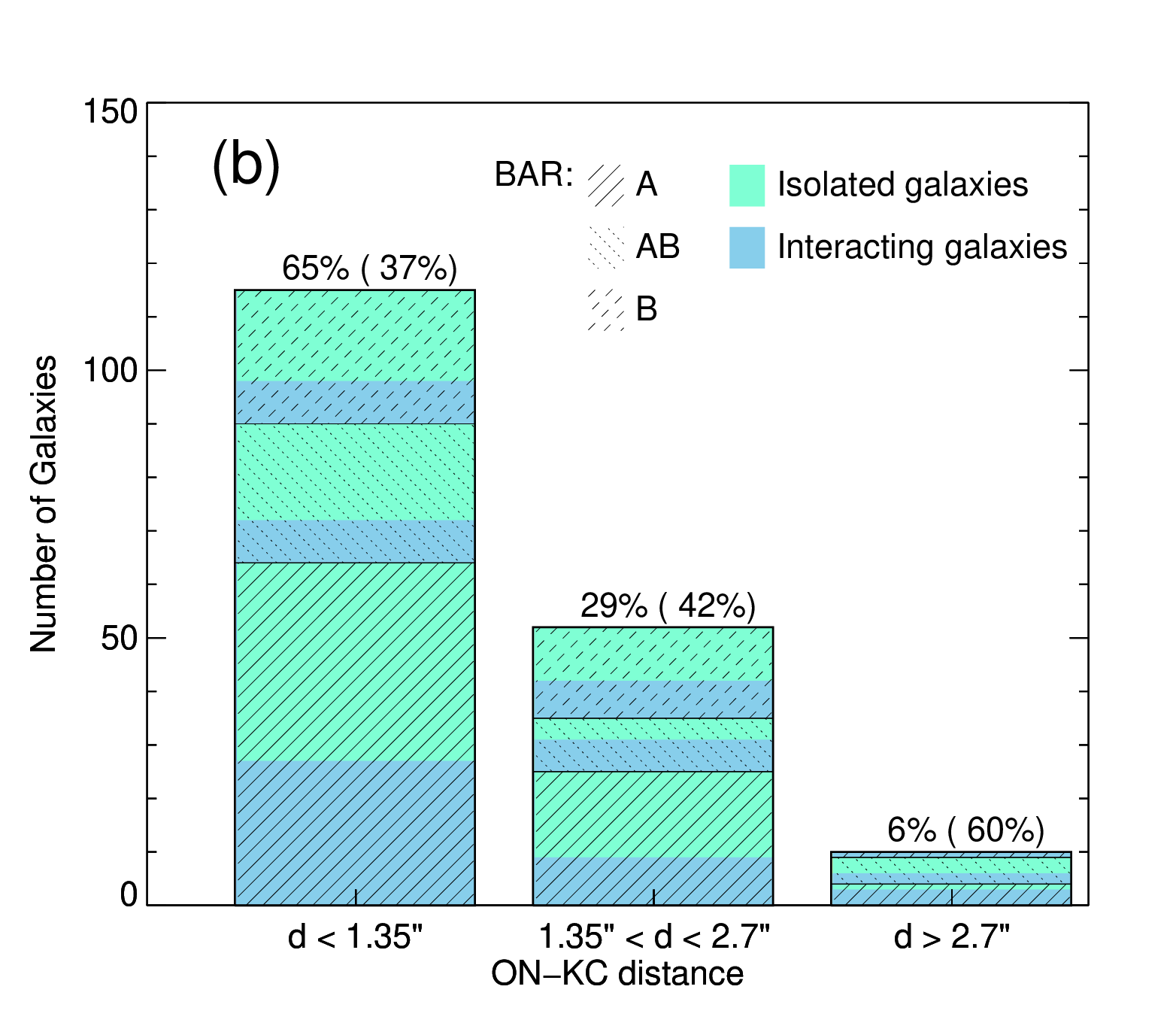}
   \caption{(a) Location of the estimated KC positions relative to the CALIFA data cube central spaxel (optical nucleus location, see H13) for all the galaxies in the sample. Filled squares correspond to SGP galaxies, while open squares to MGP objects (see text). Those objects identified as galaxies in interaction (see Table \ref{tabpropkin} in Appendix A) are in blue. Labels correspond to objects with ON-KC offsets $>2.7$ arscec. The arrow points to the location of the ON-OK for \object{NGC~3991}, indicating its ON-KC offset in parentheses. The solid-line circle marks the size of the original spatial sampling (fiber size) of the CALIFA observations centered on the coordinate origin (optical nucleus location for each object, see H13). The dashed-line circle corresponds to twice the size of the original spatial element in CALIFA, which delimits the region between an offset (out of the dashed-circle) or possible offset (annular region between the solid-circle and the dashed-circle). (b) Distribution of galaxies as a function of the distance (in arcsec) between ON and KC. Pattern styles and colors indicate bar and interacting status, respectively, as in Fig. 1. Numbers over each bin represent the fraction of objects of the sample in each bin. The relative fraction of interacting galaxies in each bin is indicated in parentheses. }
              \label{clear}%
    \end{figure}

The KC location was estimated from their velocity gradient images as explained in Sect. \ref{k_medir}. Figure \ref{clear}a shows the shifts between the derived KC position and the optical nucleus (ON hereafter) taken at the CALIFA data cube central spaxel (see H13). The ON was adopted as the KC for the UGP objects, and hence UGP galaxies are at the coordinate origin in Fig. \ref{clear}a. We adopted the original spatial fiber size of the CALIFA survey (2.7 arcsec, see Sect. \ref{observa}) as the minimum distance to report an offset between ON and KC. None of the SGP galaxies present ON-KC offsets larger than 2.7 arcsec (see Fig. \ref{clear}a). For 10 MGP galaxies (see Table \ref{tabpropkin} in Appendix A) KC and ON are shifted a distance larger than 2.7 arcsec. 6 of these 10 objects are marked interacting galaxies (see Table \ref{tabpropkin} in Appendix A). A possible ON-KC offset is found for a subset of 52 galaxies of the sample (20 SGP and 32 MGP), with an ON-KC distance in the range between 1.35 arcsec (half of the original fiber size) and 2.7 arcsec. 22 of these objects (9 SGP and 13 MGP) were also identified as interacting systems (see Fig. \ref{clear}b). At the CALIFA spatial resolution, ON and KC are in agreement for 47 SGP and 43 MGP galaxies. 

Offsets between ON and KC are reported for many different galaxies [e.g., for local tadpole galaxies \citep{2013ApJ...767...74S}; for bulgeless disk galaxies \citep{2011MNRAS.413.1875N}; for Wolf-Rayet galaxies \citep{2009A&A...508..615L}; for dwarf elliptical galaxies \citep{2000A&A...359..447B}; for AGN \citep{1993Natur.365..420M}]. Such offsets could be due to dust obscuration, which produce velocity fields and rotation curve gradients usually smoother than those for intermediate inclinations or almost face-on galaxies \citep{2008MNRAS.388..500E}. Offsets could be also due to actual displacement of a compact nucleus from the dynamical center \citep{1992ApJ...393..508M, 1998ApJ...496L..13L}. The ionized gas is only a small fraction of the total mass of a galaxy, and it can be quite sensitive to non-axisymmetric perturbations (such as interactions, bars or feedback from massive stars) that could drive large ON-KC offsets. Indeed, nine of the ten galaxies in our sample with ON-KC distance $>$ 2.7 arcsec have weak/strong bars and/or interactions (either or both) and only one (\object{IC~0776}) seems to be a single and non-barred galaxy. \object{IC~0776} is a peculiar late-type spiral galaxy displaying a large-scale asymmetry in its morphology. Its velocity field, at the CALIFA spectral and spatial resolutions, follows a general trend of receding and approaching velocities, with a quite distorted minor kinematic axis; its velocity gradient distribution (see Appendix C) is far from the expected single velocity gradient peak for a rotating system, suggesting that non-gravitational perturbations (may be warps and/or a minor merger) play a dominant role. For the galaxies (52 objects) with a possible ON-KC offset (ON-KC distance in the range 1.35-2.7 arcsec), 36 show signs of interaction and/or bars. The remaining 17 galaxies\footnote{\object{NGC~5633}, \object{NGC~5732}, \object{NGC~6063}, \object{NGC~6155}, \object{NGC~6301}, \object{UGC~00005}, \object{UGC~00148}, \object{UGC~00841}, \object{UGC~03899}, \object{UGC~03969}, \object{UGC~07145}, \object{UGC~09892}, \object{UGC~10972}, \object{UGC~11262}, \object{UGC~12054}, and \object{UGC~12816}} are apparently single late type spirals (Sbc type or later types) showing visual morphological lopsidedness, 8 of them with an ellipticity larger than 0.6. At the CALIFA spatial resolution, nuclear activity is unrelated to the origin of ON-KC displacements. Only 18 of the 62 objects with ON-KC offset larger than 1.35 arcsec are LINERS or AGN. 

The velocities derived for the KC (see Table \ref{tabpropkin} in Appendix A) are in good agreement with V$_{sys}^{H\alpha}$, with $<$V$_{sys}^{H\alpha}$-V$_{KC }>$ = 0 $\pm$ 23 km s$^{-1}$. Only three objects (namely, \object{NGC~0169}, \object{NGC~3991}, and \object{UGC~03995}) show a discrepancy larger than uncertainties. These galaxies present a large offset ($>3.5$ arcsec) between ON and KC.

\subsubsection{\bf Kinematic internal misalignment}
\label{pa_results}

\begin{figure*}
   \includegraphics[width=6cm,angle=90]{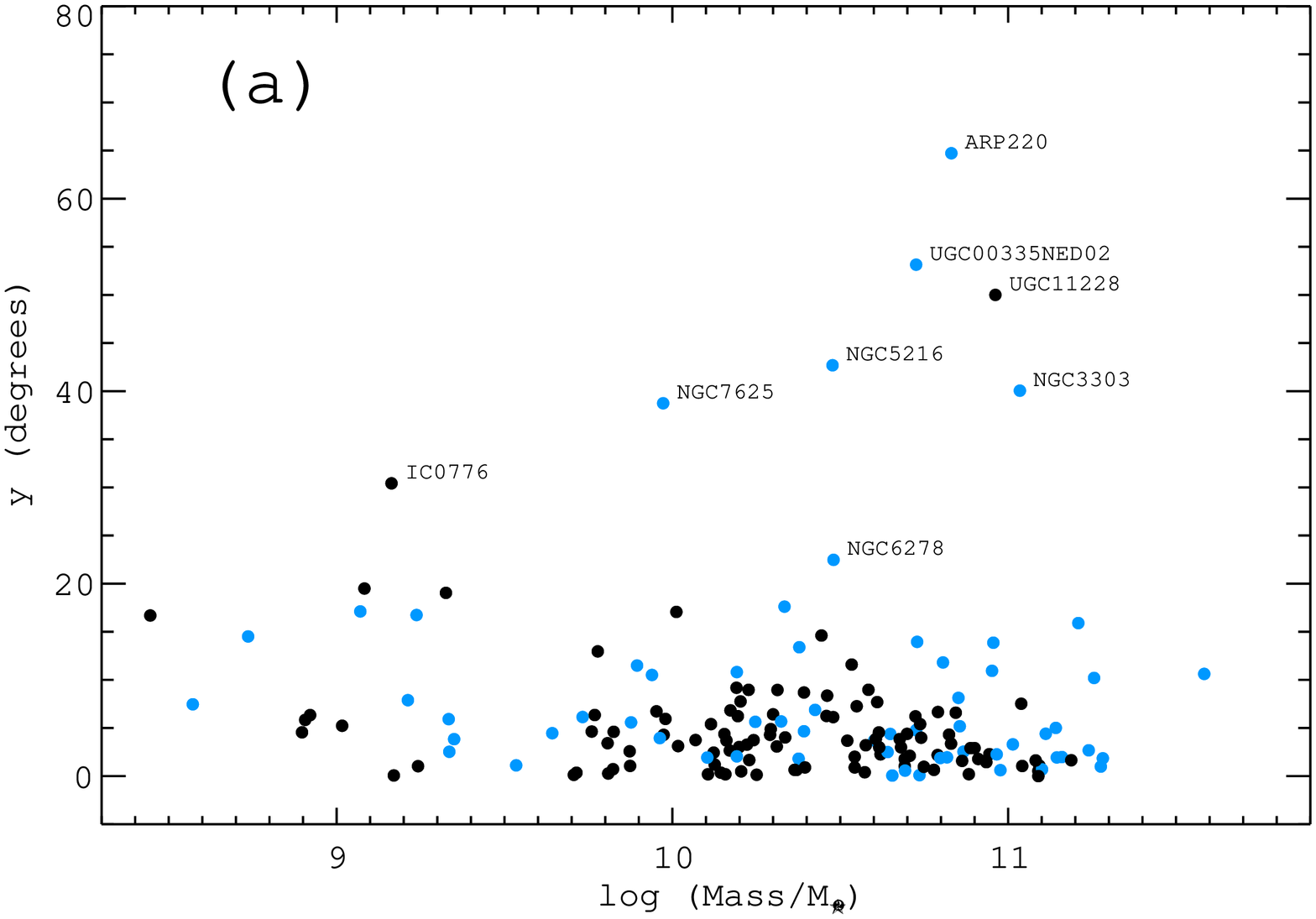}
   \includegraphics[width=6.2cm,angle=90]{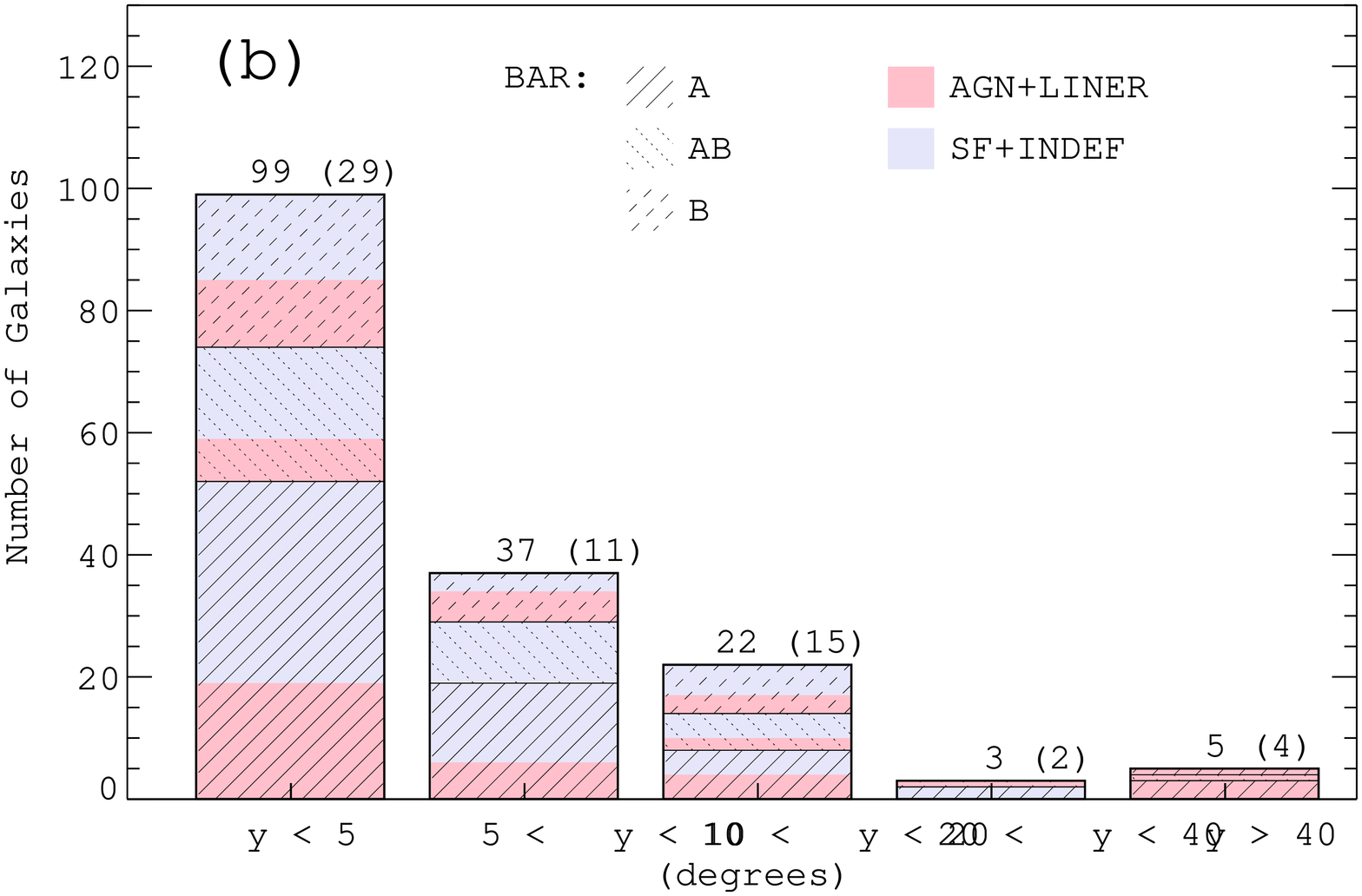}
   \includegraphics[width=6cm,angle=90]{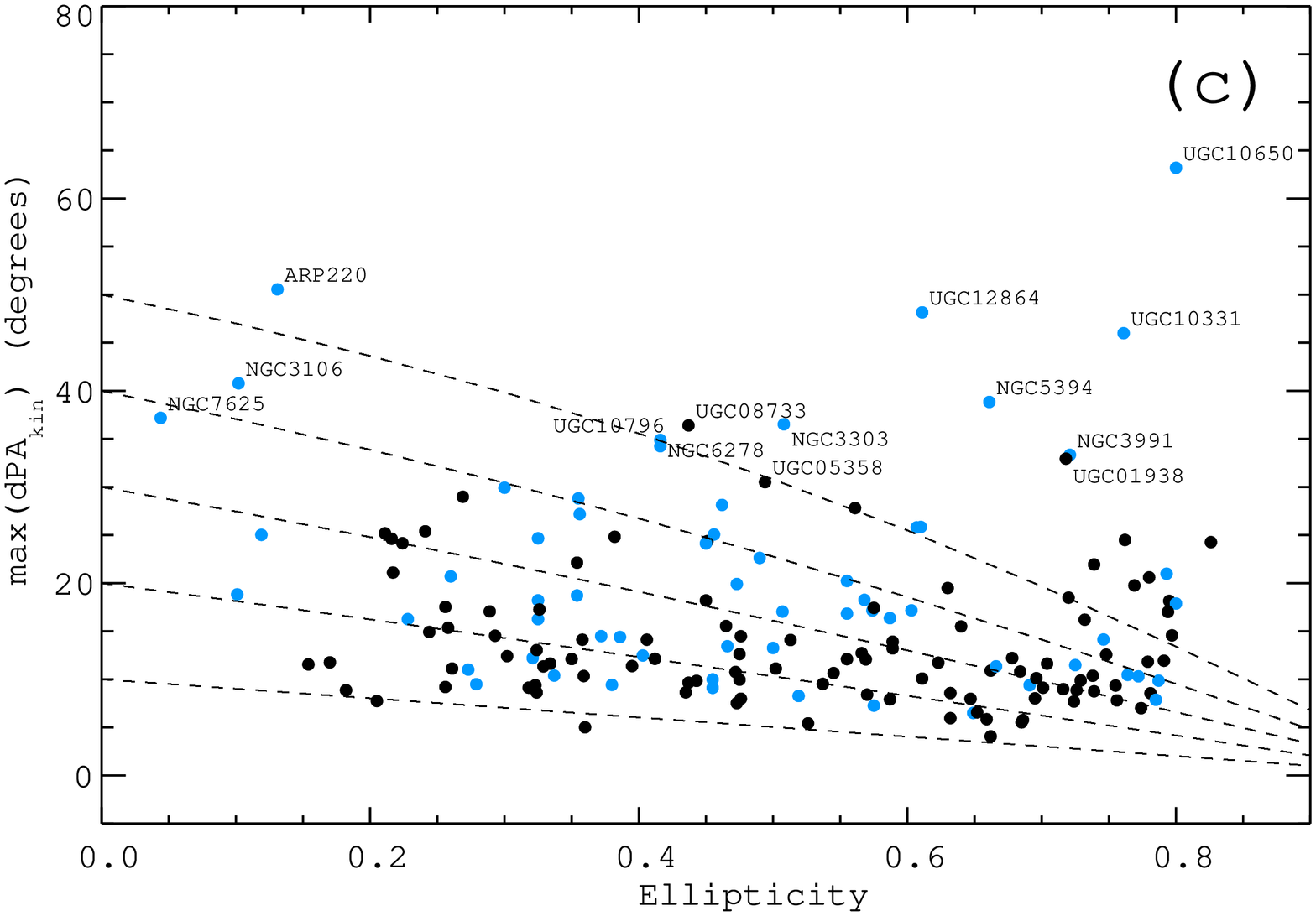}
   \includegraphics[width=6.2cm,angle=90]{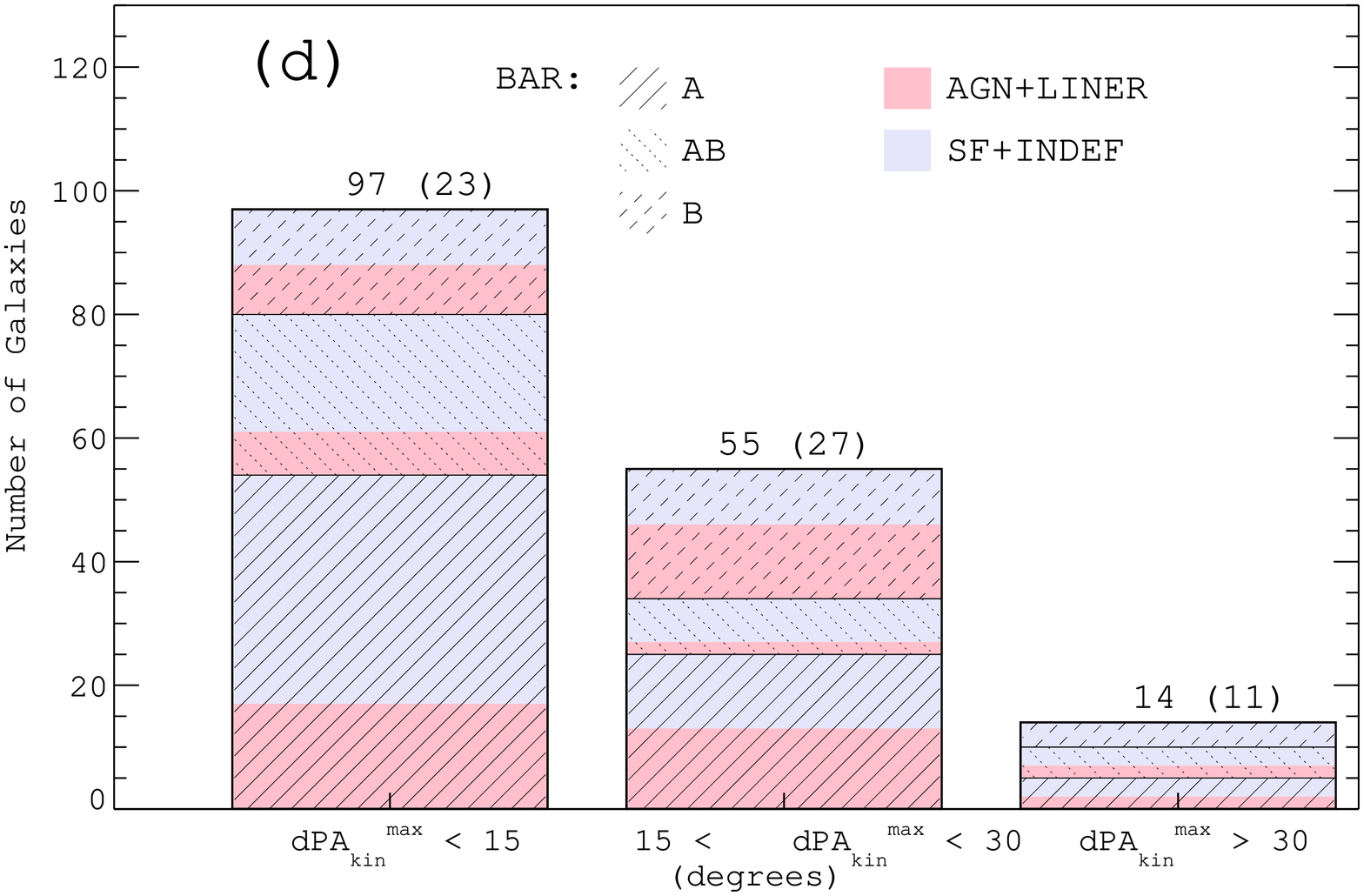}
   \caption{(a) Misalignment ($\psi$) of the major kinematic position angles estimated from the receding (PA$_{kin,rec}$) and approaching (PA$_{kin,app}$) sides of the velocity fields for each galaxy as a function of the integrated stellar masses for the CALIFA galaxies in the sample \citep[stellar masses are from][]{2014arXiv1407.2939W}. Labels indicate those objects with $\psi>20^{\circ}$ degrees. Blue dots correspond to objects identified as galaxies in interaction (see Sect. \ref{sample} and Table \ref{tabprop} in Appendix A). (b) Histogram of the kinematic misalignment of the major kinematic pseudo-axes estimated from the receding and approaching sides of the velocity fields. The fraction of non-barred and barred galaxies is indicated as in Fig. \ref{tipomorfo}. Colors indicate here the nuclear type of the galaxies (AGN+LINER or SF+INDEF). Numbers indicate the total number of objects in each bin, including the number of interacting galaxies in parentheses. (c) Largest standard deviation of the positions used to estimate the kinematic axes position angles (receding and approching sides) for each galaxy as a function of the ellipticity. Blue dots correspond to interacting galaxies. Labels indicate those objects with $\delta$PA$_{kin}>30^{\circ}$ degrees. Dashed lines draw the projection effects on $\delta$PA$_{kin}$. (d) Histogram of $\delta$PA$_{kin}$. Colors and filled lines are the same as in (b).   }
              \label{compara_PAs}%
    \end{figure*}

We have traced the kinematic line of nodes and derived their mean PA$_{kin,rec}$ and PA$_{kin,app}$ (see Sect. \ref{pa_medir}) for 
166 of the galaxies in our 
CALIFA subsample. The H$\alpha$+[\ion{N}{ii}] distribution extends $< 2.7$ arcsec (original spaxel diameter) or shows a patchy distribution in the 
remaining objects (see Table \ref{tabpropkin}). The difference between PA$_{kin,rec}$ and PA$_{kin,app}$ should be 180 degrees for a pure rotating disk system but, for many of the objects in the sample, this difference is far from this value. The misalignment of these two approaches of the mean position angle of the kinematic major axis is given by:
\begin{equation}
\label{psi}
sin \ \psi = | sin(PA_{kin}^{rec}-PA_{kin}^{app}) |
\end{equation}
with $\psi$ in the range between 0$^{\circ}$ and 90$^{\circ}$ degrees \citep{1991ApJ...383..112F}.  We adopt $\psi>10^{\circ}$ to define a kinematic lopsidedness in terms of the major kinematic axis. A similar limit was previously used to report misalignments between the photometric and kinematic axes in spiral \citep{2008A&A...488..117K} and early-type galaxies \citep{2011MNRAS.414.2923K}. Almost 82\% of the objects with $\psi$ estimation present internal kinematic misalignments smaller than 10$^{\circ}$ ($\sim77$\% of the objects in the full sample). Only 30 galaxies show $\psi>10^{\circ}$ along the major pseudo-axis from the receding to the approaching sides of the velocity fields, reaching a maximum value at around 65$^{\circ}$ for ARP~220 (see Fig. \ref{compara_PAs}a). 

During a major merger, complex kinematics could arise as a result of the tidal forces \citep[e.g.,][]{2007A&A...473..761K, 2005MNRAS.356.1177R}. The full analyzed sample (177 galaxies) includes 71 objects ($\sim40$\%) identified as interacting galaxies (see Sect. \ref{sample}). Only 21 of these systems present a clear internal asymmetry in the velocity fields ($\psi>$ 10$^{\circ}$). The rest of the objects showing $\psi> 10\,^{\circ}$ are apparently isolated galaxies. The degree of symmetry of a velocity field may also be affected by the presence of dust lanes, spiral arms, bars, warps, outflows/inflows, shocks or nuclear activity, minor mergers or enven interactions with difuse objects \citep[see, e.g.,][]{2005A&A...430...67F, 2005MNRAS.364..773F,2004ApJ...605..183W,2010A&A...521A..63L}. Their kinematics imprints could range from $\sim10$ km s$^{-1}$ \citep[for radial inflows, see]{2004ApJ...605..183W} to a few hundred of km s$^{-1}$ \citep[for outflows from an AGN, see, e.g.,][]{1996ApJ...463..509A}). Small-scale perturbations (in the spatial and/or velocity spaces) are smoothed or even undetectable depending on the spatial and spectral resolution of the observations. At the resolution of CALIFA V500 data cubes (see Sect. \ref{observa}), we find a similar proportion of galaxies with kinematic misalignments in barred and unbarred galaxies (see Fig. \ref{compara_PAs}b). The proportion of kinematic lopsided in terms of $\psi$ ($\psi> 10^ {\circ}$) seems to be slightly larger for interacting galaxies (21/61) than for AGN/LINER galaxies (15/63). However, the 5 objects with $\psi>40^{\circ}$ are AGN/LINERs, and 4 of them are galaxies in interaction. Moreover, it is important to highlight that 12 of the 15 AGN/LINER objects with $\psi>10^{\circ}$ are also classified as interacting/merger galaxies. This proportion is much larger than those AGN/LINER galaxies in interaction with $\psi<10^{\circ}$ (16/48). 

As already mentioned (see Sect. \ref{pa_medir}) the standard deviation ($\delta$PA$_{kin}$ hereafter) of the positions used to estimate the kinematic major axis indicates the degree of their alignment on the velocity field and their correspondence or not with a straight kinematic line of nodes (a negligible dependence on galactocentric distance of the major kinematic axis). The larger the $\delta$PA$_{kin}$ is the larger departure for pure rotation.  Fig. \ref{compara_PAs}c shows the largest $\delta$PA$_{kin}$ (maximum $\delta$PA$_{kin}$ at receding and approaching sides of the velocity field) for each galaxy as a function of morphological ellipticity. We note that we are estimating kinematic parameters on the plane of the sky plane (directly from observed radial velocities) and $\delta$PA$_{kin}$ should show a dependence like in equation \ref{inclination}. Figure \ref{compara_PAs}c includes $\delta$PA$_{kin}$ curves as a function of ellipticity for selected $\delta$PA$_{kin}$ at face-on. Many objects (mainly edge-on) present much larger $\delta$PA$_{kin}$ than expected, may due to the presence of dust lanes, warps, outflows/inflows inducing apparent or real vertical motions. \object{UGC~10650} is the object presenting the largest $\delta$PA$_{kin}$ (in its approaching side). \object{UGC~10650} seems to be an edge-on galaxy with a similar appearance to tadpole objects. At the spatial and spectral resolution of CALIFA, \object{UGC~10650} shows a quite chaotic velocity field, not compatible with simple rotation (see Appendix C). On the other side, the face-on galaxy \object{NGC~2347} presents the lowest $\delta$PA$_{kin}$ values (at the receding and approaching sides), suggesting that rotation is the dominant motion. Indeed, in terms of $\delta$PA$_{kin}$, \object{NGC~2347} seems to be the most symmetrical of the face-on galaxies in our sample. 

Almost 42\% of the studied objects present $\delta$PA$_{kin}$ (at the receding and/or approaching sides) larger than 15 degrees (see Fig. \ref{compara_PAs}d), while $\sim8$\% have $\delta$PA$_{kin}$s larger than 30 degrees. 45\% of the objects with $\delta$PA$_{kin}$ larger than 15 degrees are apparently isolated galaxies (31/69), but 22 of them present a bar or/and nuclear activity. Half of the remaining isolated galaxies with $\delta$PA$_{kin}$ $>15^{\circ}$ are edge-on and vertical motions in the disks and/or dust obscuration could be the responsible of the observed kinematic distortions. Poor gas content, the presence of hidden bars or a past interaction with a satellite object could explain the kinematic distortions of the other half non-barred isolated galaxies showing $\delta$PA$_{kin}$s larger than $15^{\circ}$. 

\begin{figure*}
   \includegraphics[width=6cm,angle=90]{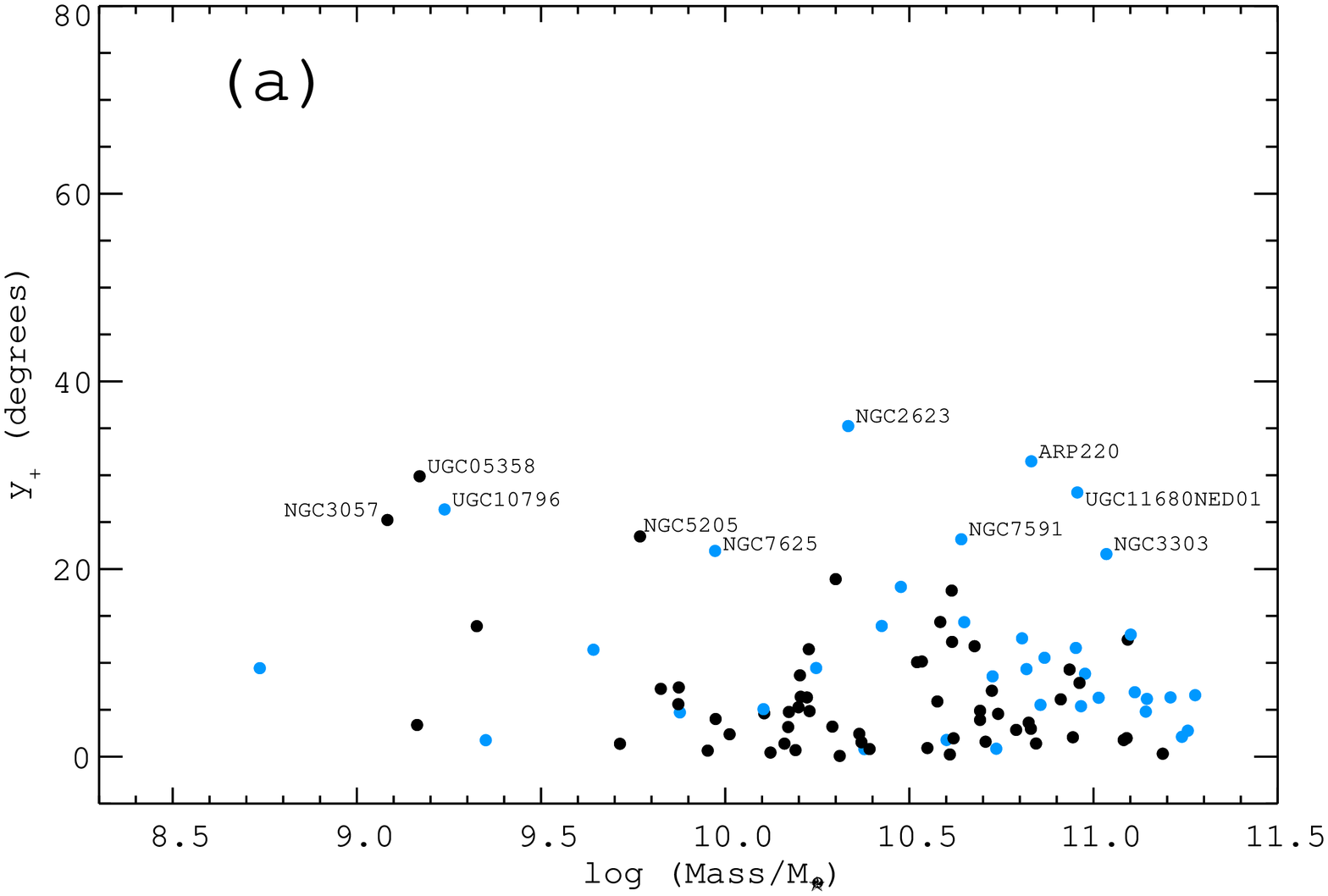}
   \includegraphics[width=6.2cm,angle=90]{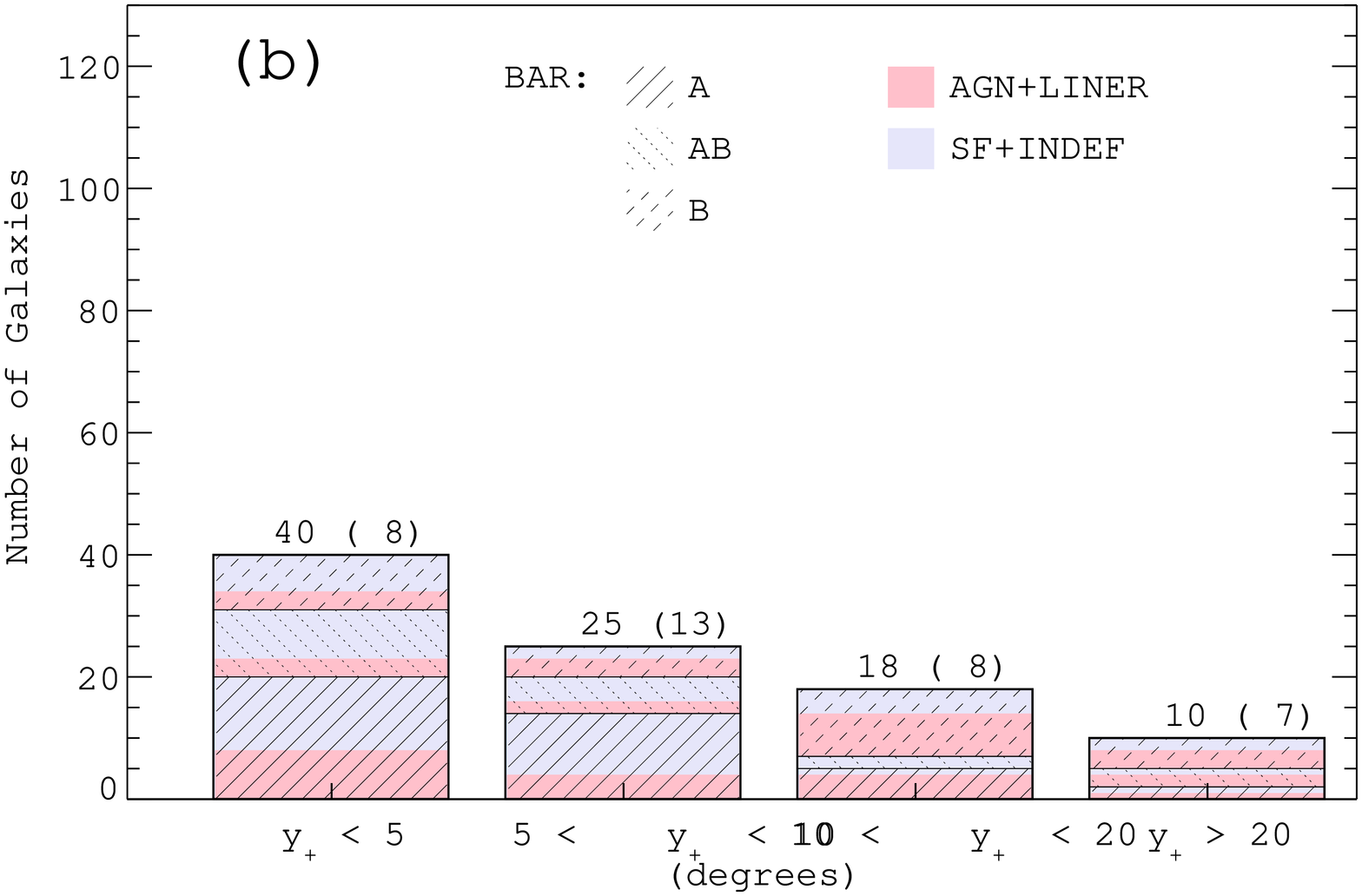}
   \includegraphics[width=6cm,angle=90]{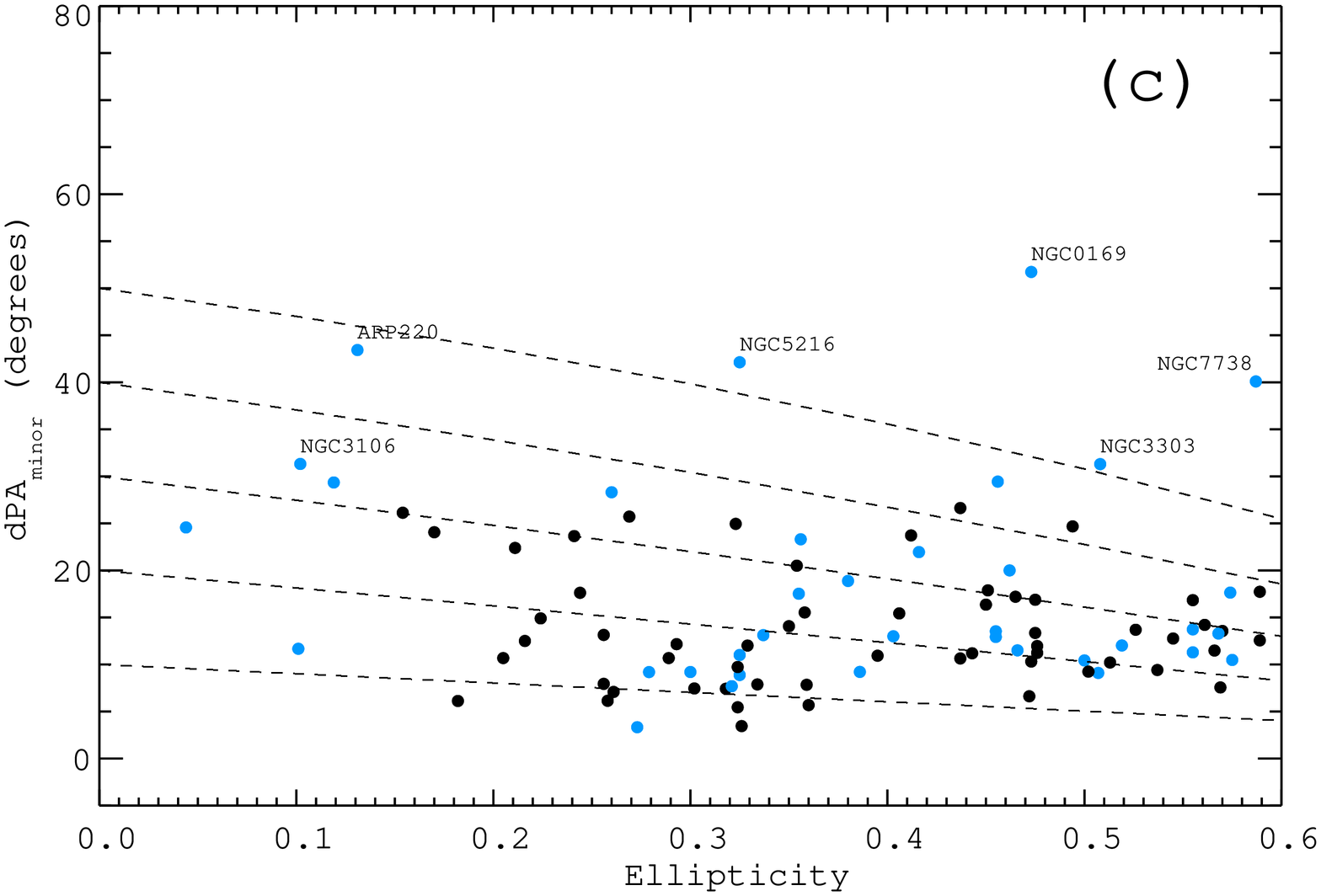}
   \includegraphics[width=6.2cm,angle=90]{pinta_dPA_minor_histo_new_Jul14.eps}
   \caption{(a) Misalignment respect to the normal ($\psi_+$) of the minor and major kinematic angles estimated for each galaxy directly from the measured radial velocities as a function of the integrated stellar masses for the CALIFA galaxies in the sample \citep{2014arXiv1407.2939W}. Labels indicate those objects with $\psi_+>20^{\circ}$ degrees. Blue circles correpond to objects identified as galaxies in interaction (see Sect. \ref{sample} and Table \ref{tabprop} in Appendix A). (b) Histogram of misalignment of the minor and major kinematic pseudo-axes respect to perpendicularity. The fraction of non-barred and barred galaxies is indicated as in Fig. \ref{tipomorfo}. Colors indicate here the faction of galaxies in each bin and bar strength of the nuclear type of the galaxies (AGN+LINER or SF+INDEF). Numbers indicate the total number of objects in each bin, including the number of interacting galaxies in parentheses. (c) Standard deviation of the positions used to estimate position angle of the minor kinematic axis for each galaxy as a function of the ellipticity. Blue dots correspond to interacting galaxies. Labels indicate those objects with $\delta$PA$_{minor}$$>30^{\circ}$ degrees.(d) Histogram of $\delta$PA$_{minor}$. Colors and filled lines indicate the same than in (b).}
              \label{compara_PAminor}%
    \end{figure*}

The orthogonality of the kinematic pseudo-axes also provides an approach to the distortions in a velocity field. Figure \ref{compara_PAminor}a shows the difference respect to normal ($\psi_+$ hereafter) calculated from the average kinematic minor and major PA through:

\begin{equation}
\label{psi_m}
\psi_+ = 90 - arcsin(| sin(PA_{kin}-PA_{minor}) |)
\end{equation}

We took the average of PA$_{kin,rec}$ and PA$_{kin,app}$ as PA$_{kin}$ in this equation. We only estimated the kinematic minor axis for a reduced number of objects (93 galaxies) because of the limited extent of the ionized gas along this direction (mainly galaxies with ellipticity larger than 0.6). $\sim70$\% of the galaxies (65 of 93) in this subsample have $\psi_+<10^{\circ}$, while only 10 (of 93) deviate from normality more than 20$^{\circ}$ (see Fig. \ref{compara_PAminor}a and b). 75\% (21 of 28) of the objects with $\psi_+>10^{\circ}$ are barred galaxies, 12 of them also have an AGN/LINER type nuclear spectrum and even 9 of them show signatures of interactions. Only \object{NGC~7819} shows $\psi_+>10^{\circ}$ being an apparently isolated and unbarred galaxy with a nuclear spectrum compatible with star formation (see H13). An additional indicator of large distortions in the minor kinematic axis is the standard deviation of the angles averaged to estimate PA$_{minor}   $ ($\delta$PA$_{minor}$ hereafter). $\sim62$\% of the objects (58 of 93) present $\delta$PA$_{minor} < 15^{\circ}$. Only 6 of 93 galaxies have $\delta$PA$_{minor}$$ > 30^{\circ}$, identified the 6 as interacting galaxies with nuclear activity (AGN/LINER).

\subsubsection{\bf Photometric to kinematic pseudo-axes mislignment}
\label{pa_results}
  \begin{figure*}
   \includegraphics[width=5.8cm,angle=90]{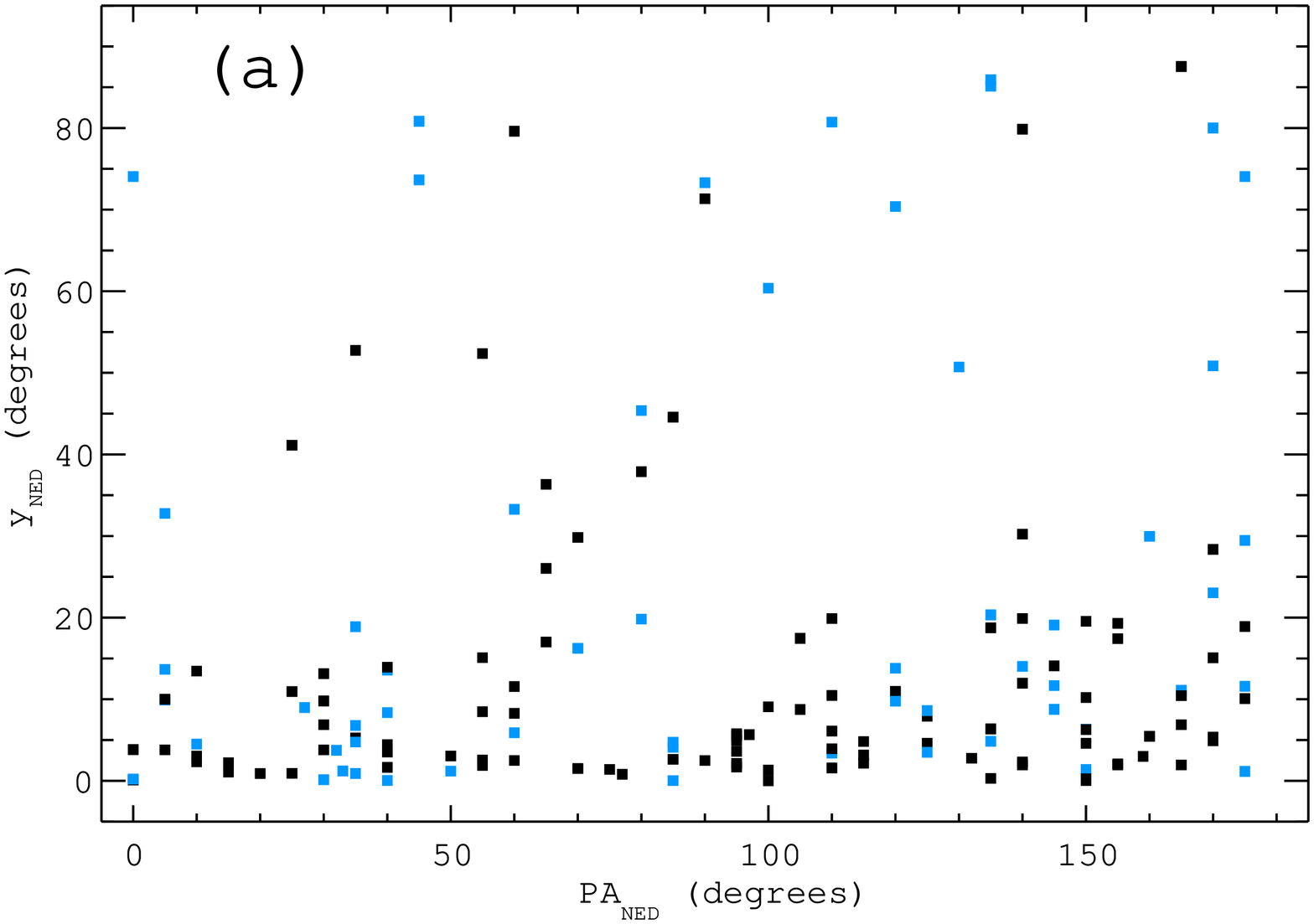}
   \includegraphics[width=6cm,angle=90]{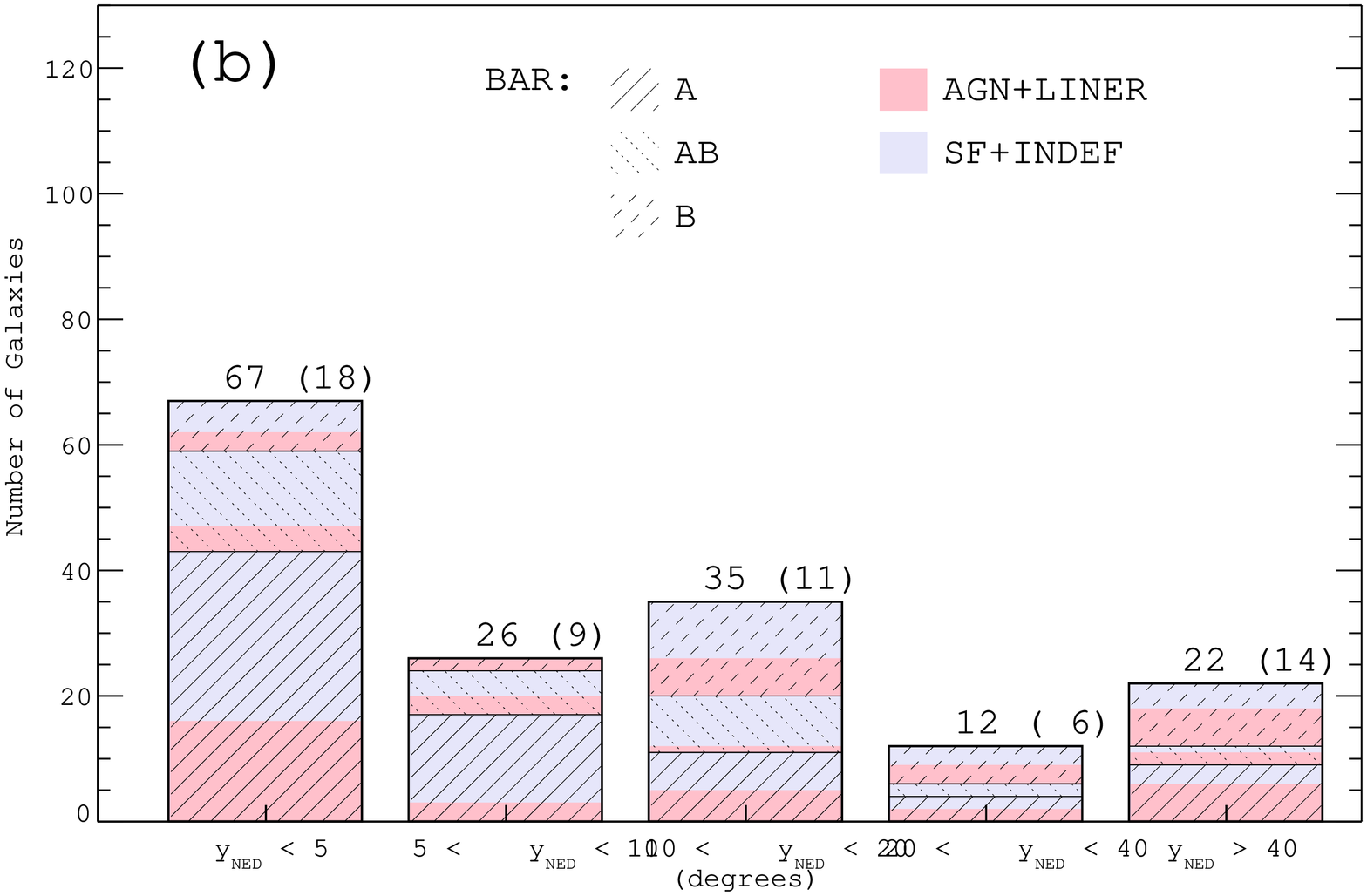}
   \includegraphics[width=5.8cm,angle=90]{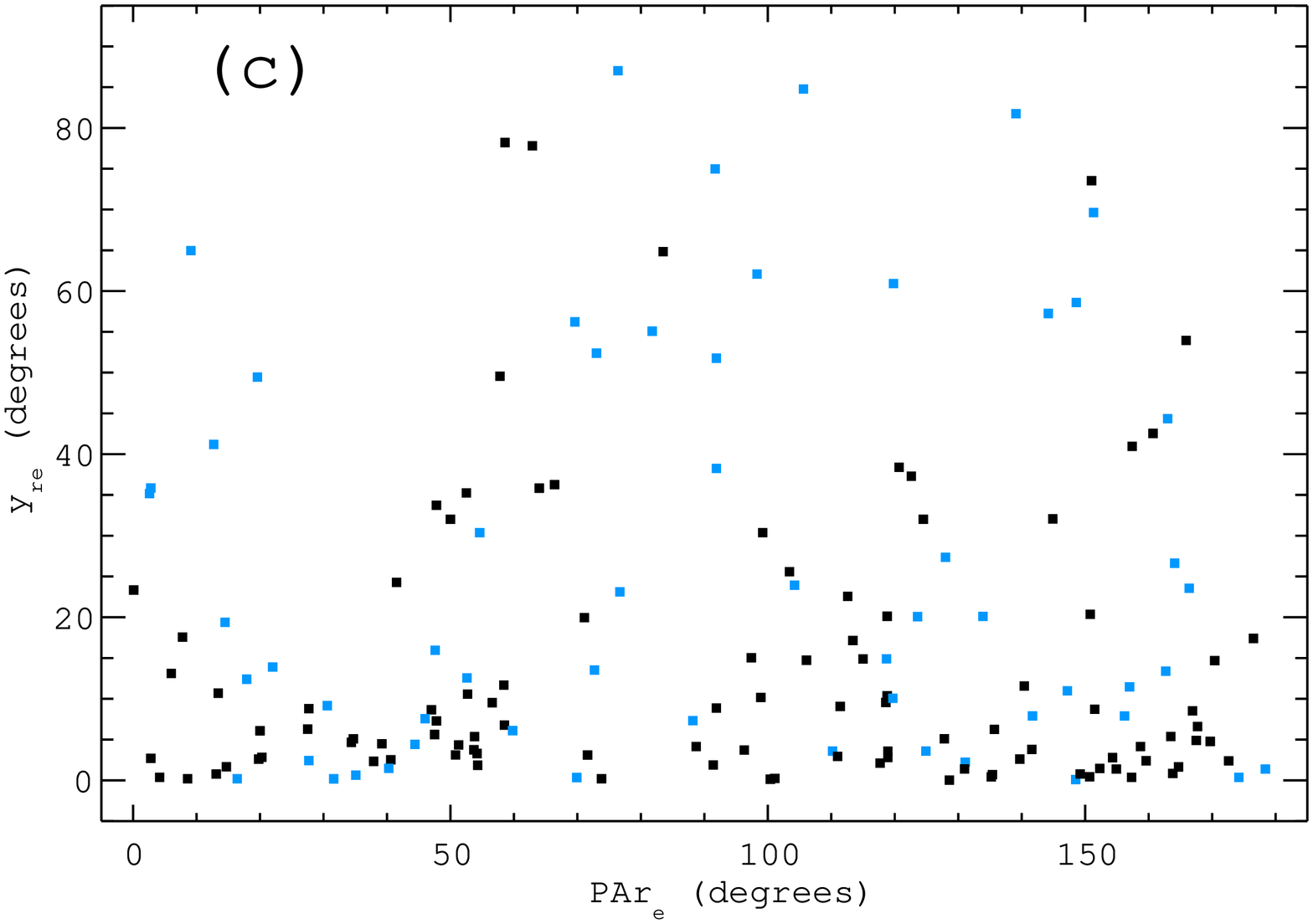}
   \includegraphics[width=6cm,angle=90]{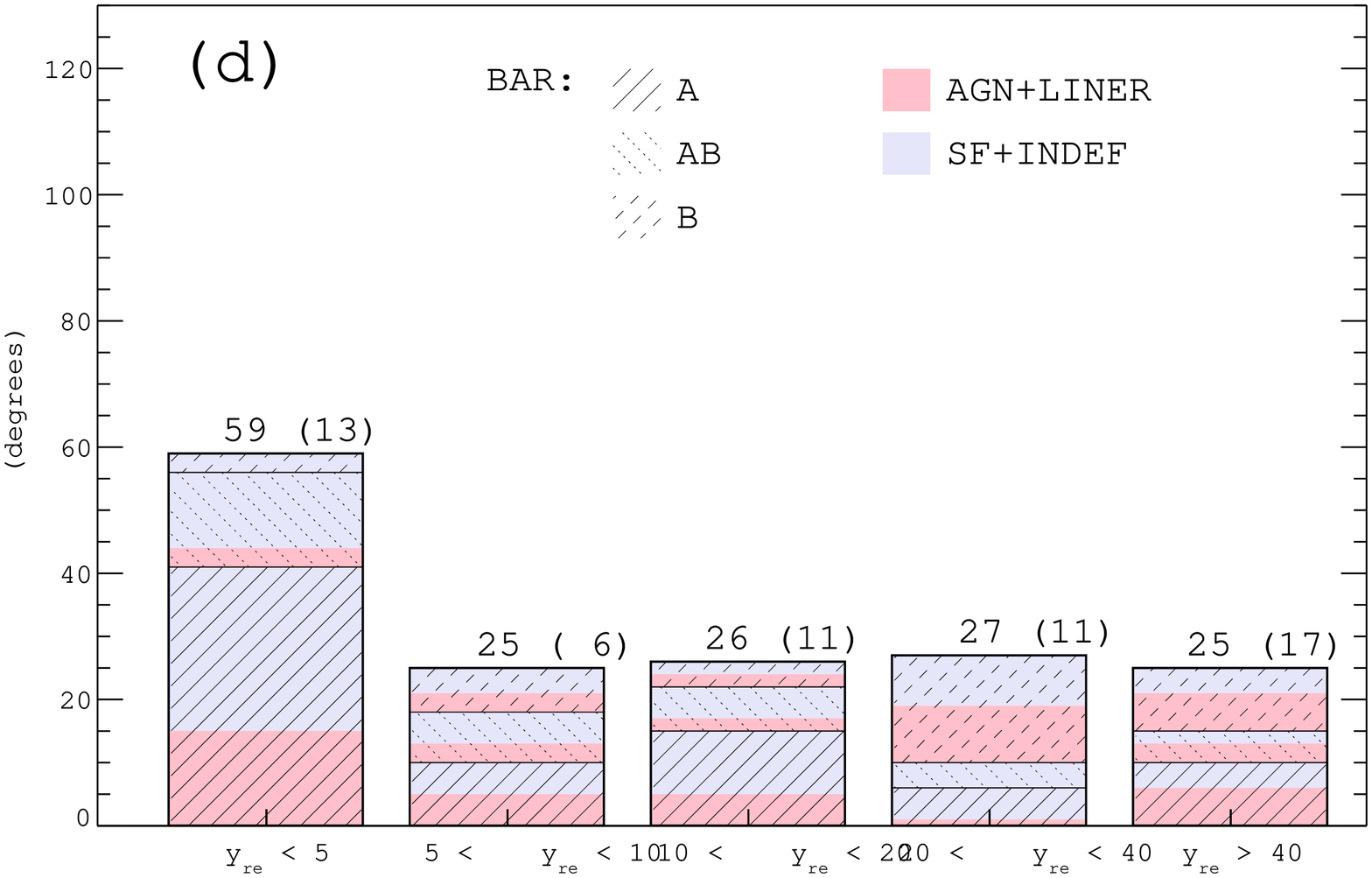}
   \includegraphics[width=5.8cm,angle=90]{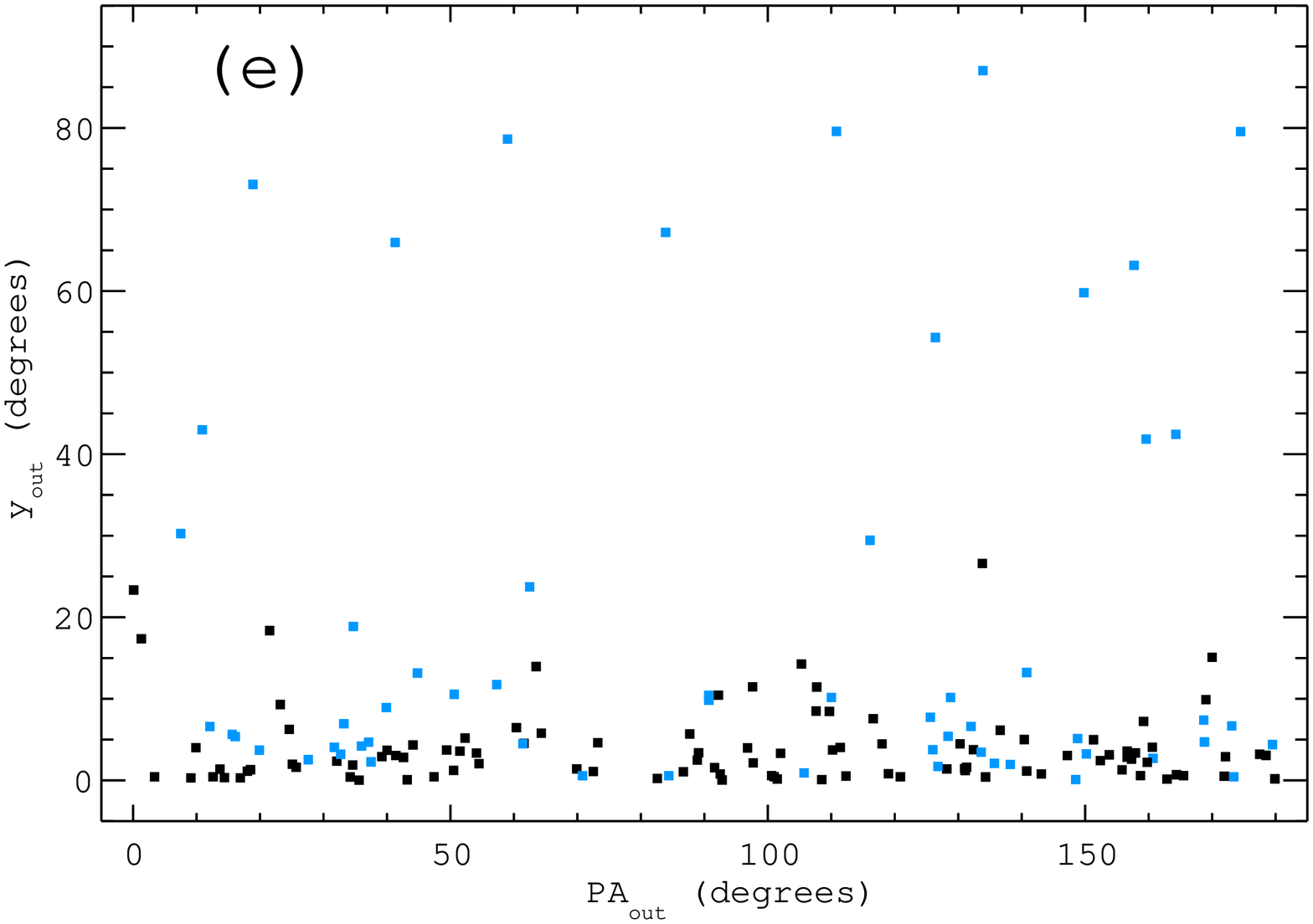}
   \includegraphics[width=6cm,angle=90]{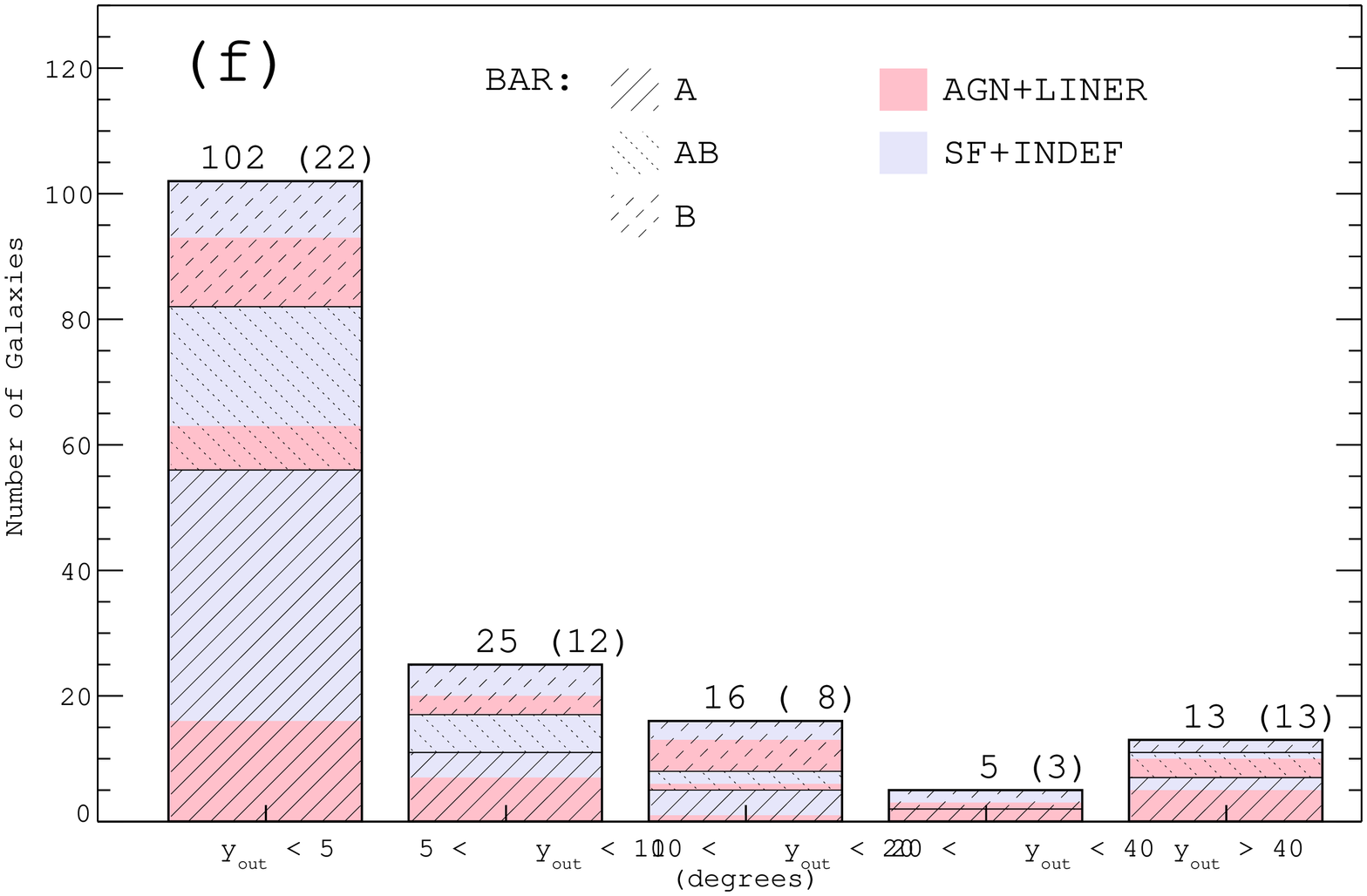}
   \caption{(a) Misalignment ($\psi_{NED}$) of the photometric position angles (from NED) and the major kinematic position angles (estimated from the average of the estimations at the receding (PA$_{kin,rec}$) and approaching (PA$_{kin,app}$) sides of each galaxy). (b) Histogram of the morpho-kinematic misalignment $\psi_{NED}$. The fraction of non-barred and barred galaxies is indicated as in Fig. \ref{tipomorfo}. (c) Misalignment ($\psi_{r_{e}}$) of the photometric position angles at one effective radius and the major kinematic position angles (PA$_{kin}$). (d) Histogram of the morpho-kinematic misalignment $\psi_{r_{e}}$. (d) Misalignment ($\psi_{out}$) of the photometric position angles at the external isophotes and the major kinematic position angles (PA$_{kin}$). (e) Histogram of the morpho-kinematic misalignment $\psi_{out}$. In (a), (c), and (e), blue circles correpond to objects identified as galaxies in interaction (see Sect. \ref{sample} and Table \ref{tabprop} in Appendix A). In (b), (d), and (f), colors indicate the fraction of galaxies in each bin and bar strength of the nuclear type of the galaxies (AGN+LINER or SF+INDEF). Numbers indicate the total number of objects in each bin, including the number of interacting galaxies in parentheses.}
              \label{compara_PA_phot}%
    \end{figure*}

The photometric position angles of the galaxies in the sample were obtained from NED (PA$_{NED}$ hereafter). Most of PA$_{NED}$ correspond to estimations from K${_s}$ images from 2MASS, but some of the  PA$_{NED}$ comes from other sources (see NED). The average of PA$_{kin,rec}$ and PA$_{kin,app}$ (see Sect. \ref{pa_results}) is taken as the orientation of the velocity fields (PA$_{kin}$). Following equation \ref{psi}, we estimated the misalignment ($\psi_{NED}$ hereafter) between photometric (PA$_{NED}$) and kinematics (PA$_{kin}$) maps orientations. Figure \ref{compara_PA_phot}b shows the histogram of $\psi_{NED}$ for 162 objects in our sample (those with both PA$_{NED}$ and PA$_{kin}$ values). $\sim$57\% of the objects (93/162) are in the first two bins ($\psi_{NED} < 10^{\circ}$), while 21\% of the galaxies (34/162) present $\psi_{NED} > 20^{\circ}$, with the maximum misalignment reaching $\sim$87$^{\circ}$ for \object{UGC~11649}. The visual inspection of \object{UGC~11649} broad band images reveals a strong bar crossing the galaxy and almost perpendicular to the apparent rotation axis of its velocity field (see appendix C). Indeed, 45 of the 69 objects with $\psi_{NED} > 10^{\circ}$ are barred galaxies, and half of the objects with $\psi_{NED} > 20^{\circ}$ have strong bars (see Fig. \ref{compara_PA_phot}b). We find an excess of early-type galaxies with photometric-kinematic misalignments, with 9 of the 10 ellipticals and S0 galaxies in this subsample showing $\psi_{NED} > 10^{\circ}$ (for 6 of them $\psi_{NED} > 20^{\circ}$). The largest misaligment for the early-type galaxies corresponds to \object{NGC~7671} ($\psi_{NED}\sim86^{\circ}$), morphologically classified as an S0 galaxy and forming a pair with \object{NGC~7672} (placed at $\sim98$ kpc from \object{NGC~7671} and with a difference in systemic velocity of $\sim-118$ km/s). It is important to note that 8 of the 10 early-types (E+S0) galaxies in this subsample of 163 objects are involved in dynamical interactions (according to the established criteria in Sect. \ref{sample}), and 9 show a LINER type nuclear spectrum. It is also important to point out that most of the early-type galaxies in the sample have a limited extent of ionized gas ($<20$ arcsec). 

For a homogeneous dataset of photometric position angles, we fitted the isophotes of the SDSS r-band images of the galaxies in the sample by ellipses using the standard IRAF task {\it ellipse} \citep{1987MNRAS.226..747J}, deriving the radial variation of photometric position angles and ellipticities. We defined an external photometric position angle (PA$_{out}$ hereafter) by averaging the outer isophotes. PA$_{out}$ represents the global stellar structure but could be affected by close companions. We also calculated the morphological position angle at one effective radius for each galaxy (PA$_{r_{e}}$ hereafter), which can account for internal morphological structures such as bars. PA$_{out}$ and PA$_{r_{e}}$ are listed in Appendix A (Table \ref{tabprop}). These two approaches of the photometric position angles are also used in Falc\'on-Barroso et al. (in preparation) for the statistical analysis of the stellar kinematics of a sample of CALIFA galaxies and in Barrera-Ballesteros et al. (in preparation) for the comparison of stellar and ionized gas kinematic for a sample of interacting galaxies following a merging sequence. Figure \ref{compara_PA_phot} shows the comparison of these photometric position angles with the orientation of the velocity fields following equation \ref{psi}.

Similar statistical results are obtained when comparing PA$_{r_{e}}$ and PA$_{kin}$ ($\psi_{re}$ hereafter) to those obtained for $\psi_{NED}$. We have an excess of barred galaxies (48 of 78) with $\psi_{re} > 10^{\circ}$ and also an excess of early-type galaxies with large misaligments (8 of 10). Moreover, 18 of the 28 objects in interaction with nuclear activity (AGN or LINER) present inner morpho-kinematic misaliments ($\psi_{re} > 10^{\circ}$). On the other hand, the misalignment between kinematics and PA$_{out}$ ($\psi_{out}$ hereafter) is quite small for a large number of objects in the sample (see Fig. \ref{compara_PA_phot}e and \ref{compara_PA_phot}f): 129 of 162 objects ($\sim80$ \%) have $\psi_{out} < 10^{\circ}$, and in total $\sim$89\% of galaxies present $\psi_{out} < 15^{\circ}$ (in agreement with results in \citet{2011MNRAS.414.2923K} for early-type galaxies). $\psi_{out}$ is larger than $10^{\circ}$ for only 14 of the galaxies with a weak/strong bar ($\sim24$\% of the barred galaxies in the sample). Otherwise, 19 of the 33 galaxies with $\psi_{out} > 10^{\circ}$ present some degree of nuclear activity ($\sim$31\% of the AGN+LINER objects in the sample), while 23  of the objects with $\psi_{out} > 10^{\circ}$ are identified as interactions ($\sim40$\% of the interacting galaxies in the sample), 14 of them having also nuclear activity. Indeed, half of the galaxies with nuclear activity and in interaction show $\psi_{out} > 10^{\circ}$. Again, the largest excess of galaxies showing photometric and kinematic misalignments correspond to early-type objects (E+S0), with 7 of the 10 early-type galaxies in the sample having $\psi_{out} > 10^{\circ}$. However, we note that all the objects with strong morpho-kinematic misaligments ($\psi_{out} > 40^{\circ}$) are interacting galaxies (see Fig. \ref{compara_PA_phot}f). Barrera-Ballesteros et al. (in preparation) study the morpho-kinematic misaligments (as well as the stellar-ionized gas kinematic misaligments) for a sample of interacting galaxies.


\subsection{Presence of kinematically distinct gaseous components}
\label{asym_results}

The presence of asymmetries in the observed spectra can be studied for 122 galaxies, $\sim69$\% of the galaxies in the analyzed sample, those satisfying the signal-to-noise threshold (S/N$\ge40$ in [\ion{O}{iii}]~$\lambda5007$, see Sect. \ref{asym_medir}). Obviously, the total number of spectra with estimated S/N larger than 40 varies from object to object, from a minimum of three to more than thousand spatial elements of the CALIFA data cubes. We assume three contiguous spectra as the minimum number of spaxels to define a region. 

  \begin{figure}
   \resizebox{\hsize}{!}{\includegraphics[width=9cm]{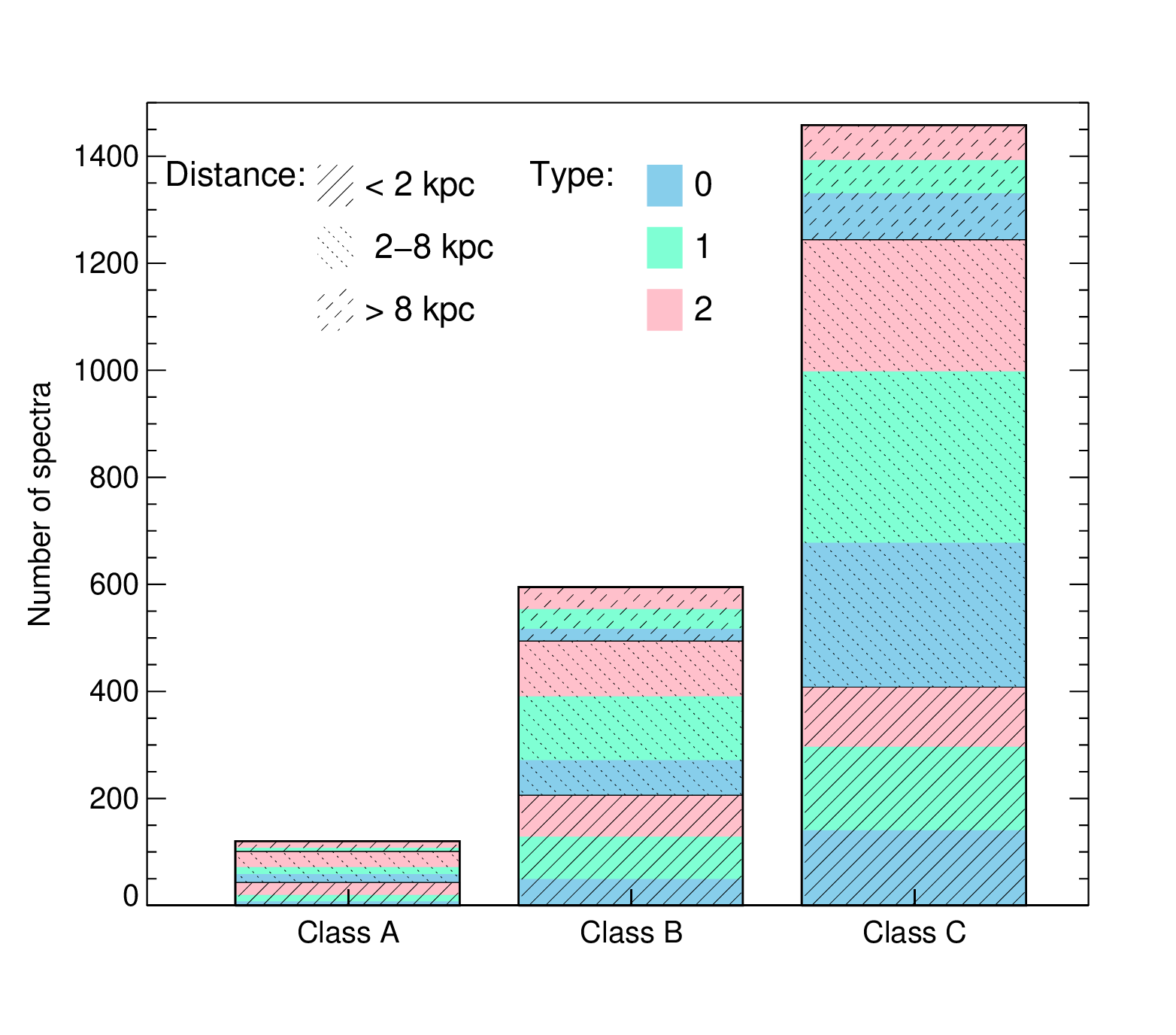}}
   \caption{Distribution of asymmetries detected in the [\ion{O}{iii}] profiles according to the classes defined. We indicate the fraction of spectra located at different distances from the galaxy center (as coded in the plot). Colors indicate the proportion of profiles with detected asymmetric types according to bisector deviation from a Gaussian (as coded in the plot) in each division.}
              \label{histo_class}%
    \end{figure}

Asymmetries are detected  in 117 objects at different bisector levels with absolute velocity shifts respect to a Gaussian bisector ($\mid$$\Delta$V$_{b}$$\mid$) larger than the limits estimated in Appendix B (F$_{\Delta V_{b}}$(S/N), equations \ref{eqc1} and \ref{eqc4}) for noise induced asymmetries in CALIFA profiles. We classified the detected asymmetries into three categories according to the number of bisector levels at which the profiles appear asymmetric:

\begin{itemize}
\item[$\bullet$] Class A: $\mid$$\Delta$V$_{b}$$\mid$ is larger than F$_{\Delta V_{b}}$(S/N) in more than five bisector levels. In general, class A profiles correspond to asymmetries first detected over a 30\% of the peak intensity level. \\

\item[$\bullet$] Class B: $\mid$$\Delta$V$_{b}$$\mid$ is larger than F$_{\Delta V_{b}}$(S/N) in a number of bisector levels between 3 and 5. Frequently, class B profiles correspond to asymmetries first detected at intensity levels between 20\% and 30\% of the intensity peak. \\

\item[$\bullet$] Class C: $\mid$$\Delta$V$_{b}$$\mid$ is larger than F$_{\Delta V_{b}}$(S/N) only in  two bisector levels. Commonly, class C corresponds to profiles with asymmetries in the 10\% and 15\% bisector levels. \\
\end{itemize}

Moreover, we can classify the detected asymmetries according to the maximum deviation of the bisector from a Gaussian ($\Delta$V$_{b}$), adopting the following types:
\begin{itemize}
\item[$\bullet$] Type 0: max($\mid \Delta$V$_{b} \mid) - $F$_{\Delta V_{b}}$(S/N) $\le$ F$_{\Delta V_{b}}$(S/N) km/s
\item[$\bullet$] Type 1: F$_{\Delta V_{b}}$(S/N) $<$ max($\mid \Delta$V$_{b} \mid) - $F$_{\Delta V_{b}}$(S/N) $\le 2\times$ F$_{\Delta V_{b}}$(S/N)
\item[$\bullet$] Type 2: max($\mid \Delta$V$_{b} \mid) - $F$_{\Delta V_{b}}$(S/N) $> 2\times$ F$_{\Delta V_{b}}$(S/N) km/s
\end{itemize}

 Following these divisions, we have nine different categories of asymmetries in the profiles (namely, A0, A1, A2, B0, B1, B2, and C0, C1, and C2) depending on the bisector velocity deviation from a Gaussian and bisector level at which the asymmetry is first detected. In appendix B, we explored the parameter space of two kinematically distinct gaseous components producing asymmetric emission profiles, founding that classes and types result from a complex combination of the parameters (velocity, velocity dispersion and intensity) of each gaseous component contributing to a particular emission line profile.

Class C profiles are found in the spectra of 108 objects. In $\sim$79\% of the objects (92 of 117), we detect class B profiles, while only 25 galaxies have class A profiles. Obviously, a single galaxy can present spectra of different asymmetry classes depending on the kinematics of the gaseous systems from region to region. Table \ref{tabpropasy} in Appendix A indicates the classes and types of asymmetric profiles detected for the different galaxies. The presence of several gaseous components in the rest of the objects of our sample and/or at additional spectra cannot be ruled out, but we are not able to detect them at the resolution and 
depth of the CALIFA data cubes. These two facts could also be masking any trend on the presence of multiple gaseous systems with galaxy types. Indeed, asymmetric emission line profiles are found in spiral galaxies but also in elliptical galaxies. We stress that most 
elliptical galaxies in the sample do not satisfy the signal-to-noise threshold to look for asymmetries. Neither the interaction with a nearby galaxy nor the presence or absence of a bar (strong or weak) seems to be associated with the detection of asymmetries in the emission line profiles of the galaxies in the sample.

 Asymmetries in the [\ion{O}{iii}] profiles are found in regions around the nuclear zone (up to 2 kpc), in compact regions out of the nuclear zone but also in dispersed regions all over the galaxies. Figure \ref{histo_class} presents the distribution of profiles according to the detected asymmetry class, indicating the fraction of them located at different distances from the galaxy center. Although all categories of asymmetries in the profiles are found at different scales in the analyzed objects, in general asymmetries in the emission line profiles for spectra coming from the central regions suggest brighter secondary components with respect to the dominant than the secondary features for spectra in the outer regions of the galaxies. Moreover, the difference in velocity between the dominant and the secondary components seems to be also larger for spectra in the central region than those in the outer parts. However, from a model of two Gaussian (see Appendix B) we found that many combinations of the parameters characterizing each component can result in an specific class and/or type asymmetric profile. Commonly, asymmetric emission line profiles detected out of the circumnuclear region are generally in clouds surrounding bright emission knots and/or at bright emission regions. At the spatial resolution of the CALIFA survey many of the emission knots could lumps of unresolved emission knots. 
 

Complex emission line profiles in the central region of galaxies have been commonly interpreted as bi-polar outflows driven by starbursts or AGN radiation pressure \citep[see, e.g.,]{2011ApJ...735...48S, 2010A&A...517A..27M, 1981ApJ...247..403H}. Multicomponent emission line profiles are also observed in many luminous \ion{H}{ii} regions associated with the expansion of bubbles produced by stellar winds from massive stars \citep[e.g.,][]{2007A&A...467..603R,2007ApJ...656..168L}. Luminous \ion{H}{ii} regions emission line profiles are characterized by a central peak and one or two high velocity features that could be associated with the asymmetries detected in the CALIFA [\ion{O}{iii}]~$\lambda\lambda4959,5007$ profiles according to the observed velocity shifts \citep[see, e.g.,][]{2005A&A...431..235R}. Multiple massive star-forming clumps have been also suggested as the origin of complex profiles in dynamically young host galaxies \citep{2012ApJ...754L..22A}. The presence of a nearby galaxy or a small satellite companion (in interaction or not) could also explain the observation of complex emission line profiles depending on the orientation respect to our line of sight.  Even in absence of kinematically distinct gaseous components, complex emission line profiles could result from beam-smearing of the velocity gradients within the observation aperture. A detailed analysis of the origin of asymmetric emission line profiles for each galaxy in our sample is far from the scope of this work but could be the issue of a future dedicated work. The main intention here is to indicate the presence of multiple gaseous systems in the objects to users of the CALIFA database. As we already mention, the presence of secondary gaseous systems could give rise to complex emission lines, and such deviation from a Gaussian profile could have an impact on the calculation of parameters from emission line fluxes (see Appendix B).

\section{Summary and conclusions}
\label{summary}

In this paper we present a basic analysis of the ionized gas kinematics of the galaxies in CALIFA. Our main results and conclusions are summarized by the following items.

\begin{itemize}
\item[$\bullet$] At the spatial and spectral resolution of CALIFA, the ionized gas velocity fields of the galaxies in the sample present, in general, the typical pattern of receding and approaching velocities.

\item[$\bullet$] Systemic velocities derived from different emission lines are in good agreement and they are compatible (within uncertainties) to values in NED.

\item[$\bullet$] Almost half of the galaxies in the sample have clear structures in the velocity gradient maps, indicating clear departures from rotation at the resolution of CALIFA.

\item[$\bullet$] We find evidence of displacements between the photometric and kinematic centers for 35\% of the objects in the sample. The largest offsets mainly correspond to galaxies in interaction.

\item[$\bullet$] The major kinematic position angles estimated from the receding and approaching sides of the velocity fields suggest a kinematic lopsided in 17\% of the galaxies in the sample. A significant fraction of these galaxies correspond to interacting objects with nuclear activity (AGN or LINER).

\item[$\bullet$] Deviations larger than 15$^{\circ}$ in tracing the major kinematic axes are found in almost 40\% of the analyzed objects, indicating clear departures from pure rotation.

\item[$\bullet$] Deviations ($>10^{\circ}$) from the normal between the minor and major kinematic axes are mainly associated with the presence of a bar in the objects .

\item[$\bullet$] We find an excess of early-type galaxies (E+S0) showing photometric-kinematic misaligments.

\item[$\bullet$] Evidence of the presence of kinematically distinct gaseous systems are found in 69\% of the galaxies in the sample.

\end{itemize}

\begin{acknowledgements}
This study makes uses of the data provided by the Calar Alto Legacy Integral Field
Area (CALIFA) survey (http://www.califa.caha.es). Based on observations collected at the Centro Astron\'omico Hispano Alem\'an (CAHA) at Calar Alto,
operated jointly by the Max-Planck-Institut f\"{u}r Astronomie and the Instituto
de Astrof\'{\i}sica de Andaluc\'{\i}a (CSIC). CALIFA is the first legacy survey being performed at Calar Alto. The CALIFA collaboration would like to thank
the IAA-CSIC and MPIA-MPG as major partners of the observatory, and
CAHA itself, for the unique access to telescope time and support in manpower and infrastructures. The CALIFA collaboration thanks also the CAHA
staff for the dedication to this project. We thank the Viabilidad, Dise\~no ,
Acceso y Mejora funding program (ICTS-2009-10) for supporting the initial developement of this project. B.G-L and J.B-B thank the support from the Plan Nacional de I+D+i (PNAYA) funding programs (AYA2012- 39408-C02-02) of the Spanish Ministerio de Econom\'{\i}a y Competitividad (MINECO). I.M., J.M. and A. d.O. acknowledge financial support from the Spanish grant AYA2010-15169 and Junta de Andalucia TIC114 and Excellence Project P08-TIC-03531. S.F.S. and D. Mast also thank the support given to this project from the PNAYA of the MINECO under grant AYA2012-31935. S.F.S thanks the the Ram\'on y Cajal project (RyC-2011-07590) of the Spanish MINECO, for the support giving to this project. JMA acknowledges support from the European Research Council Starting Grant (SEDmorph; P.I. V. Wild). We acknowledge financial support for the ESTALLIDOS collaboration by the Spanish MINECO under grant AYA2010- 21887-C04-03. J. F.-B. acknowledges financial support from the Ram\'on y Cajal Program and grant AYA2010-21322-C03-02 from the MINECO, as well as to the DAGAL network from the People Programme (Marie Curie Actions) of the European Union’s Seventh Framework Programme FP7/2007-2013/ under REA grant agreement number PITN-GA-2011-289313. KS acknowledges support from the National Sciences and Engineering Research Council of Canada. A. M.-I. acknowledges support from Agence Nationale de la Recherche through the STILISM
project (ANR-12-BS05-0016-02) and from BMBF through the Erasmus-F project (grant number 05 A12BA1).
P.P. is supported by Ciencia 2008 Contract, funded by FCT/MCTES (Portugal) and POPH/FSE (EC), and JMG by a Post-Doctoral grant, funded by FCT/MCTES (Portugal) and POPH/FSE (EC). PP\&JMG acknowledge support by the Funda\c{c}\~{a}o para a Ci\^{e}ncia e a Tecnologia (FCT) under project FCOMP-01-0124-FEDER-029170 (Reference FCT PTDC/FIS-AST/3214/2012), funded by FCT-MEC (PIDDAC) and FEDER (COMPETE). R.A. Marino was also funded by the Spanish programme of International Campus of Excellence Moncloa (CEI). Finally, we are grateful to the referee, Matt Bershady, for a careful reading of the paper and his several comments that helped to improve this paper.
\end{acknowledgements}

\bibliographystyle{aa} 
\bibpunct{(}{)}{;}{a}{}{,}
\bibliography{CALIFA_ionized_gas_astroph.bib} 

\begin{appendix}


\section{TABLES}
In this appendix, we summaryze the main photometric parameters of the galaxies in the sample (Table \ref{tabprop}) obtained (within the CALIFA collaboration) from SDSS r-band images of the galaxies in the sample (see Walcher et al. 2013, in preparation for details). In Table \ref{tabprop} columns correspond to:
\begin{itemize}
\item Columns [1] and [2]: Object and CALIFA unique ID number for the galaxy, respectively.
\item Columns [3] and [4]: Systemic velocity and position angle of the apparent major axis obtained from NED.
\item  Columns [5]: Effective radius in arcsec of the disk estimated as detailed in \citet{2014A&A...563A..49S} .
\item  Columns [6]: Radial distance (in units of the effective radii) used to estimate the large-scale photometric position angles (in column 8) and ellipticities (in column 9).
\item Columns [7] and [8]: Morphological position angles at one effective radius (PA$_{r_{e}}$) and at the largest scale of the SDSS images (PA$_{out}$). Both measurements were inferred from the SDSS r-band image of the galaxy using the {\it IRAF} task {\it ellipse}.
\item Column [9]: Ellipticity of the outer isophotes of the SDSS r-band image obtained using the IRAF task {\it ellipse}.
\item Column [10]: Identification of the galaxy as isolated (I G), interacting/merging (IoM G), or group of galaxies (GoG) (see Section 2.3 for criteria on this division). Here, we divide the interacting sample analyse in the work (pair of galaxies, small groups of galaxies and mergers with tidal features) in IoM G and GoG just for reference to future works.
\item Column [11]: Morphological type from visual classification performed by the CALIFA collaboration (see H13 and Walcher at al. 2013, in preparation for details).
\item Column [12]: Bar strength of the galaxy as an additional outcome of the CALIFA visual classification (see H13 and Walcher at al. 2013, in preparation for details). We divided the galaxies into non-barred (A), weakly barred (AB) and strongly barred (B).
\item Column [13]: Nuclear type of the object indicating the main ionization mechanisms in the central region determined through diagnostic diagrams (see Section 2.3). SF, LINERS and AGN indicate pure star formation, low-ionization nuclear
emission-line regions and active galactic nuclei, respectively. INDEF indicates that the nuclear type could not be inferred (see Section 2.3).
\end{itemize}

In Table \ref{tabpropkin} we include the kinematic parameters directly derived from the measured radial velocities of the H$\alpha$+[\ion{N}{ii}] emission lines for each galaxy. Each column corresponds to:

\begin{itemize}
\item Column [1]: CALIFA ID number for the galaxy.
\item Column [2]: Classification of the galaxy according to the structures in the velocity gradient map obtained from the H$\alpha$+[\ion{N}{ii}] velocity field. SGP, MGP, and UGP indicate Single, Multi and Unclear velocity Gradient Peak (see Section \ref{k_results}).
\item Columns [3] and [4]: Position (in arcsec) of the kinematic center (right ascension [4], and declination [5]) relative to the central spaxel of the CALIFA data cube (see Table 4 in H13 for keyword) (see Section \ref{k_medir}).
\item Column [5]:  Average velocity (in km/s) in an aperture of 3.7 arcsec in radius centered at the location of the kinematic center. Errors correspond to the standard deviation of the average radial velocities.
\item Columns [6] and [7]: Position angle of the major kinematic pseudo-axis estimated from the receding ([6]) and approaching ([7]) sides of the velocity field and taking the reference position at the kinematic center. Errors correspond to the standard deviation of the polar coordinates of the spaxels tracing these axes (see Section \ref{pa_medir}).
\item Column [8]: Position angle of the minor kinematic pseudo-axis.  Errors correspond to the standard deviation of the polar coordinates tracing this axis (see Section \ref{pa_medir}).
\end{itemize}

Table \ref{tabpropasy} includes the systemic velocities derived from different emission lines through a 3.5 arcsec aperture in radius on the zero reference spaxel (galactic nucleus) of each CALIFA datacube (see Sections \ref{vsys_medir} and \ref{vsys_results}). Table \ref{tabpropasy} also indicates the class and type of asymmetries detected in the [\ion{O}{iii}] emission line profiles in each object (see Sections \ref{asym_medir} and \ref{asym_results}). Each column in Table \ref{tabpropasy} corresponds to:
\begin{itemize}
\item Column [1]: Object
\item Columns [2]-[5]: Sytemic velocities derived from: [2] [\ion{O}{ii}] (V$_{sys}^{[OII]}$); [3] [\ion{O}{iii}] (V$_{sys}^{[OIII]}$); [4] H$\alpha$+[\ion{N}{ii}] (V$_{sys}^{H\alpha}$); and [5] [SII] (V$_{sys}^{[SII]}$) emission lines.
\item Columns [6]-[8]: Class and types of asymmetries detected in the [\ion{O}{iii}] profiles (see Sections \ref{asym_medir} and \ref{asym_results}).

\end{itemize}  

 \begin{landscape}
   \begin{table*}
      \caption[]{Morphological parameters of the sample of CALIFA galaxies analized in this work.}
         \label{tabprop}
     $$ 
\begin{array}{llccccccccccl}
            \hline
            \noalign{\smallskip}
NAME & CAL & V_{sys}^{NED} & PA_{NED} & r_{e} & r_{out} & PA_{r_{e}}\pm\sigma_{PA_{r_{e}}} & PA_{out}\pm\sigma_{PA_{out}} & \epsilon\pm\sigma_{\epsilon} & Stage & Morph. & Bar & Nuclear \\
     & ID  & (km s^{-1}) &  (degrees) & (arcsec) & (r_{e}) & (degrees) & (degrees) &     type      & Strength &  Type  \\
            \noalign{\smallskip}
            \hline
            \noalign{\smallskip}
ARP220  &    802  &   5434  &    170  &     21  &    2.5  &  139.1  \pm    6.6  &  157.7  \pm    6.6  &   0.13  \pm   0.02  &IoM G  &         S         d  &         A  &
LINER  \\
IC0540  &    274  &   2035  &    170  &     15  &    1.0  &  170.4  \pm    0.1  &  170.0  \pm    0.1  &   0.63  \pm   0.00  &  I G  &         S        ab  &        AB  &
AGN  \\
IC0776  &    528  &   2468  &     97  &     27  &    1.5  &   66.4  \pm    0.4  &   92.2  \pm    0.4  &   0.45  \pm   0.02  &  I G  &         S        dm  &         A  &
SF  \\
IC0944  &    663  &   6997  &    110  &     20  &    2.8  &  110.2  \pm    0.5  &  105.7  \pm    0.5  &   0.65  \pm   0.00  &IoM G  &         S        ab  &         A  &
SF  \\
IC1199  &    824  &   4731  &    140  &     24  &    2.2  &  122.6  \pm    2.1  &  157.3  \pm    2.1  &   0.57  \pm   0.02  &  I G  &         S         b  &        AB  &
SF  \\
IC1256  &    856  &   4730  &     90  &     17  &    3.1  &  124.5  \pm    1.5  &   89.1  \pm    1.5  &   0.36  \pm   0.02  &  I G  &         S         b  &        AB  &
SF  \\
\hline
         \end{array}
     $$ 
\end{table*}
   \begin{table*}
    \addtocounter{table}{-1} 
      \caption[]{Continued}
         \label{tabprop}
     $$ 
         \begin{array}{llccccccccccl}
            \hline
            \noalign{\smallskip}
NAME & CAL & V_{sys}^{NED} & PA_{NED} & r_{e} & r_{out} & PA_{r_{e}}\pm\sigma_{PA_{r_{e}}} & PA_{out}\pm\sigma_{PA_{out}} & \epsilon\pm\sigma_{\epsilon} & Stage & Morph. & Bar & Nuclear \\
     & ID  & (km s^{-1}) &  (degrees) & (arcsec) & (r_{e}) & (degrees) & (degrees) &     type      & Strength &  Type  \\
            \noalign{\smallskip}
            \hline
            \noalign{\smallskip}
NGC3057  &    312  &   1524  &    175  &     35  &    1.1  &  176.5  \pm    0.7  &   23.2  \pm    0.7  &   0.27  \pm   0.01  &  I G  &         S        dm  &         B  &
SF  \\
NGC3106  &    311  &   6205  &    175  &     25  &    2.4  &  157.0  \pm    6.9  &  116.1  \pm    6.9  &   0.10  \pm   0.02  &IoM G  &         S        ab  &         A  &
LINER  \\
NGC3158  &    318  &   6865  &    150  &     33  &    2.8  &  151.8  \pm    3.0  &  165.7  \pm    3.0  &   0.19  \pm   0.02  &  GoG  &         E         3  &         A  &
INDEF  \\
NGC3160  &    319  &   6920  &    140  &     15  &    2.5  &  135.7  \pm    0.5  &  140.8  \pm    0.5  &   0.76  \pm   0.00  &  I G  &         S        ab  &        AB  &
AGN  \\
NGC3303  &    340  &   6281  & ---  &     22  &    2.2  &  144.2  \pm    1.2  &  159.6  \pm    1.2  &   0.51  \pm   0.01  &IoM G  &         S        0a  &        AB  &     LINER  \\
NGC3614  &    388  &   2333  &     65  &     42  &    1.9  &  101.1  \pm    0.7  &  101.5  \pm    0.7  &   0.41  \pm   0.01  &  I G  &         S        bc  &        AB  &
SF  \\
\hline
         \end{array}
     $$ 
\end{table*}
   \begin{table*}
    \addtocounter{table}{-1} 
      \caption[]{Continued}
         \label{tabprop}
     $$ 
         \begin{array}{llccccccccccl}
            \hline
            \noalign{\smallskip}
NAME & CAL & V_{sys}^{NED} & PA_{NED} & r_{e} & r_{out} & PA_{r_{e}}\pm\sigma_{PA_{r_{e}}} & PA_{out}\pm\sigma_{PA_{out}} & \epsilon\pm\sigma_{\epsilon} & Stage & Morph. & Bar & Nuclear \\
     & ID  & (km s^{-1}) & (degress) & (arcsec) & (r_{e}) & (degrees) & (degrees) &     type      & Strength &  Type  \\
            \noalign{\smallskip}
            \hline
            \noalign{\smallskip}
NGC6310  &    848  &   3419  &     70  &     25  &    2.2  &   71.6  \pm    0.3  &   69.9  \pm    0.3  &   0.72  \pm   0.01  &  I G  &         S         b  &         A  &
LINER  \\
NGC6314  &    850  &   6633  &    175  &     15  &    2.7  &  174.2  \pm    0.7  &  173.4  \pm    0.7  &   0.47  \pm   0.01  &IoM G  &         S        ab  &         A  &
INDEF  \\
NGC6338  &    851  &   8222  &     15  &     30  &    1.7  &   15.0  \pm    0.8  &   15.1  \pm    0.8  &   0.38  \pm   0.01  &IoM G  &         E         5  &         A  &
LINER  \\
NGC6394  &    857  &   8486  &     30  &     15  &    2.4  &   34.7  \pm    0.1  &   42.6  \pm    0.1  &   0.64  \pm   0.01  &  I G  &         S        bc  &         B  &
AGN  \\
NGC6478  &    862  &   6776  &     30  &     25  &    1.8  &   27.5  \pm    0.2  &   34.2  \pm    0.2  &   0.63  \pm   0.01  &  I G  &         S         c  &         A  &
SF  \\
NGC6497  &    863  &   6162  &    115  &     14  &    3.2  &  144.9  \pm    0.3  &  112.3  \pm    0.3  &   0.51  \pm   0.01  &  I G  &         S        ab  &         B  &
INDEF  \\
\hline
         \end{array}
     $$ 
\end{table*}
   \begin{table*}
    \addtocounter{table}{-1} 
      \caption[]{Continued}
         \label{tabprop}
     $$ 
         \begin{array}{llccccccccccl}
            \hline
            \noalign{\smallskip}
NAME & CAL & V_{sys}^{NED} & PA_{NED} & r_{e} & r_{out} & PA_{r_{e}}\pm\sigma_{PA_{r_{e}}} & PA_{out}\pm\sigma_{PA_{out}} & \epsilon\pm\sigma_{\epsilon} & Stage & Morph. & Bar & Nuclear \\
     & ID  & (km s^{-1}) &  (degrees) & (arcsec) & (r_{e}) & (degrees) & (degrees) &     type      & Strength &  Type  \\
            \noalign{\smallskip}
            \hline
            \noalign{\smallskip}
UGC03253  &    146  &   4130  &     65  &     17  &    2.8  &   50.0  \pm    0.0  &   87.7  \pm    0.0  &   0.47  \pm   0.00  &  I G  &         S         b  &         B  &
SF  \\
UGC03899  &    150  &  3884.  &    55.  &  ---  & ---  & ---  & --- & --- &  I G&         S         d  &         A  &        SF  \\
UGC03944  &    152  &   3895  &    110  &     18  &    2.4  &  111.4  \pm    1.0  &  120.9  \pm    1.0  &   0.57  \pm   0.03  &  I G  &         S        bc  &        AB  &
SF  \\
UGC03969  &    153  &   8130  &    135  &     16  &    2.3  &  135.4  \pm    0.1  &  134.3  \pm    0.1  &   0.78  \pm   0.00  &  I G  &         S         b  &         A  &
SF  \\
UGC03973  &    154  &   6652  &     60  &     18  &    2.5  &   58.6  \pm    5.1  &  133.8  \pm    5.1  &   0.22  \pm   0.01  &  I G  &         S        bc  &         B  &
AGN  \\
UGC03995  &    155  &   4750  &     85  &     28  &    2.8  &  104.2  \pm    0.9  &   90.7  \pm    0.9  &   0.56  \pm   0.02  &IoM G  &         S         b  &         B  &
AGN  \\
\hline
         \end{array}
     $$ 
\end{table*}
   \begin{table*}
    \addtocounter{table}{-1} 
      \caption[]{Continued}
         \label{tabprop}
     $$ 
         \begin{array}{llccccccccccl}
            \hline
            \noalign{\smallskip}
NAME & CAL & V_{sys}^{NED} & PA_{NED} & r_{e} & r_{out} & PA_{r_{e}}\pm\sigma_{PA_{r_{e}}} & PA_{out}\pm\sigma_{PA_{out}} & \epsilon\pm\sigma_{\epsilon} & Stage & Morph. & Bar & Nuclear \\
     & ID  & (km s^{-1})   & (degrees)  & (arcsec) & (r_{e}) & (degrees) & (degrees) &     type      & Strength &  Type  \\
            \noalign{\smallskip}
            \hline
            \noalign{\smallskip}
UGC11228  &    865  &   5771  &    165  &     13  &    3.3  &  166.9  \pm    1.6  &  178.5  \pm    1.6  &   0.33  \pm   0.03  &  I G  &         S         0  &         B  &
LINER  \\
UGC11262  &    866  &   5606  &     55  &     20  &    1.8  &   47.5  \pm    0.7  &   49.4  \pm    0.7  &   0.65  \pm   0.00  &  I G  &         S         c  &         A  &
SF  \\
UGC11649  &    872  &   3798  &    165  &     19  &    2.5  &  151.0  \pm    1.1  &   63.5  \pm    1.1  &   0.22  \pm   0.02  &  I G  &         S        ab  &         B  &
LINER  \\
UGC11680NED01  &    873  &   7791  &    175  &     20  &    3.5  &   19.6  \pm    1.8  &   57.3  \pm    1.8  &   0.46  \pm   0.02  &IoM G  &         S         b  &         B  &
AGN  \\
UGC11717  &    877  &   6303  &     35  &     19  &    1.8  &   40.3  \pm    0.5  &   37.1  \pm    0.5  &   0.61  \pm   0.00  &IoM G  &         S        ab  &         A  &
LINER  \\
UGC11792  &    880  &   4812  &    159  &     17  &    2.1  &  159.6  \pm    0.1  &  159.8  \pm    0.1  &   0.80  \pm   0.00  &  I G  &         S        bc  &         A  &
SF  \\
\hline
         \end{array}
     $$ 
\end{table*}
 
\end{landscape}
\newpage

   \begin{table*}
      \caption[]{Kinematic parameters estimated directly from the radial velocity derived for each spectra of the CALIFA datacubes with a S/N$>$6 (in both the stellar and ionized gas components) through the application of the cross-correlation technique in the H$\alpha$+[\ion{N}{ii}] spectral range. }
         \label{tabpropkin}
     $$ 
         \begin{array}{llcccccc }
            \hline
            \noalign{\smallskip}
CAL & Grad. & \Delta\alpha_{KC}\pm\sigma_{\Delta\alpha_{KC}} & \Delta\delta_{KC}\pm\sigma_{\Delta\delta_{KC}} & V_{KC} \pm \sigma_{KC} & PA_{kin,rec}\pm \delta PA_{kin,rec} &  PA_{kin,app}\pm \delta PA_{kin,app} &  PA_{minor}\pm \delta PA_{minor} \\
  ID   & type     &   (arcsec)                          &    (arcsec)                              &  (km s^{-1})          &  (degrees)                      & (degrees)                        & (degrees) \\
            \noalign{\smallskip}
            \hline
            \noalign{\smallskip}

   802 &    MGP &     -4.2 \pm      0.1 &     -1.5 \pm      0.3 &   5339.6 \pm     84.6 &      8.5 \pm     21.5 &    253.2 \pm     50.5 &     99.4 \pm     43.4 \\ 
   274 &    MGP &     -0.3 \pm      0.1 &      1.1 \pm      0.2 &   2054.6 \pm     32.6 &    191.6 \pm     13.1 &     -1.4 \pm     19.5 &     87.8 \pm     40.9 \\ 
   528 &    MGP &     -0.2 \pm      0.3 &     -2.8 \pm      0.1 &   2448.4 \pm     12.6 &    267.5 \pm     15.0 &    117.9 \pm     24.4 &     16.0 \pm     17.9 \\ 
   663 &    SGP &     -1.2 \pm      0.1 &     -1.3 \pm      0.1 &   6913.1 \pm     83.1 &    287.6 \pm      3.4 &    105.6 \pm      6.5 &     13.6 \pm     14.2 \\ 
   824 &    MGP &      0.4 \pm      0.1 &     -0.3 \pm      0.1 &   4710.1 \pm     52.6 &    156.0 \pm      6.3 &    343.8 \pm      8.4 &     61.2 \pm     13.6 \\ 
   856 &    MGP &      1.6 \pm      0.2 &      2.5 \pm      0.1 &   4720.9 \pm     40.0 &    274.0 \pm      6.0 &     91.0 \pm     14.1 &     -2.8 \pm     15.5 \\
\hline
         \end{array}
     $$ 

\end{table*}

  \begin{table*}
    \addtocounter{table}{-1} 
      \caption[]{Continued.}
         \label{tabpropkin}
     $$ 
         \begin{array}{llcccccc }
            \hline
            \noalign{\smallskip}
CAL & Grad. & \Delta\alpha_{KC}\pm\sigma_{\Delta\alpha_{KC}} & \Delta\delta_{KC}\pm\sigma_{\Delta\delta_{KC}} & V_{KC} \pm \sigma_{KC} & PA_{kin,rec}\pm \delta PA_{kin,rec} &  PA_{kin,app}\pm \delta PA_{kin,app} &  PA_{minor}\pm \delta PA_{minor} \\
  ID   & type     &   (arcsec)                          &    (arcsec)                              &  (km s^{-1})          &  (degrees)                      & (degrees)                        & (degrees) \\
            \noalign{\smallskip}
            \hline
            \noalign{\smallskip}
   633 &    MGP &     -1.3 \pm      0.1 &     -0.1 \pm      0.3 &   2907.5 \pm     45.6 &    241.5 \pm     24.4 &     18.8 \pm     24.7 &    112.0 \pm     42.1 \\ 
   676 &    MGP &     -1.1 \pm      0.1 &      0.0 \pm      0.1 &   2970.6 \pm     61.9 &     85.9 \pm     24.6 &    269.6 \pm     19.0 &      7.8 \pm     12.5 \\ 
   680 &    SGP &     -0.3 \pm      0.1 &      0.4 \pm      0.1 &   3435.1 \pm     10.9 &     25.8 \pm     38.8 &    207.6 \pm     24.0 &    115.3 \pm     18.9 \\ 
   684 &    MGP &      0.2 \pm      0.1 &      0.6 \pm      0.1 &   5424.8 \pm     53.6 &    287.9 \pm     14.5 &    106.8 \pm     10.8 &     29.8 \pm     12.2 \\ 
   708 &    SGP &      0.4 \pm      0.1 &      0.4 \pm      0.3 &   1917.7 \pm     85.0 &    251.9 \pm     12.2 &     76.2 \pm      9.3 &     -9.1 \pm      7.7 \\ 
   748 &    SGP &      0.0 \pm      0.1 &     -1.7 \pm      0.2 &   2335.0 \pm     31.1 &    199.1 \pm      9.2 &     15.4 \pm      6.9 &    105.8 \pm     13.1 \\
\hline
         \end{array}
     $$ 

\end{table*}

  \begin{table*}
    \addtocounter{table}{-1} 
      \caption[]{Continued.}
         \label{tabpropkin}
     $$ 
         \begin{array}{llcccccc }
            \hline
            \noalign{\smallskip}
CAL & Grad. & \Delta\alpha_{KC}\pm\sigma_{\Delta\alpha_{KC}} & \Delta\delta_{KC}\pm\sigma_{\Delta\delta_{KC}} & V_{KC} \pm \sigma_{KC} & PA_{kin,rec}\pm \delta PA_{kin,rec} &  PA_{kin,app}\pm \delta PA_{kin,app} &  PA_{minor}\pm \delta PA_{minor} \\
  ID   & type     &   (arcsec)                          &    (arcsec)                              &  (km s^{-1})          &  (degrees)                      & (degrees)                        & (degrees) \\
            \noalign{\smallskip}
            \hline
            \noalign{\smallskip}

    17 &    UGP & 0 & 0 &   5392.8 \pm     59.7 &    243.4 \pm     13.0 &    116.6 \pm     22.6 &      0.0 \pm      0.0 \\ 
    41 &    MGP &     -0.1 \pm      0.2 &      2.2 \pm      0.1 &   5527.7 \pm     26.3 &    232.5 \pm      7.0 &     52.4 \pm      5.3 &    142.5 \pm     11.9 \\ 
    53 &    MGP &     -0.1 \pm      0.3 &      0.3 \pm      0.3 &   6317.2 \pm     34.3 &    152.7 \pm      5.6 &    327.3 \pm      5.8 &     65.9 \pm     14.2 \\ 
    88 &    UGP & 0 & 0 &   6393.8 \pm     32.0 &    339.0 \pm     29.9 &    155.2 \pm     32.9 &     80.5 \pm     33.1 \\ 
   116 &    SGP &     -0.4 \pm      0.1 &      1.0 \pm      0.1 &   7703.2 \pm     56.1 &    165.4 \pm      5.8 &    344.5 \pm      5.8 &     68.4 \pm     11.0 \\ 
   133 &    SGP &     -0.8 \pm      0.3 &      0.1 \pm      0.2 &   8346.2 \pm     49.4 &     76.7 \pm      5.1 &    250.5 \pm      7.7 &    167.2 \pm     10.2 \\
\hline
         \end{array}
     $$ 

\end{table*}

  \begin{table*}
    \addtocounter{table}{-1} 
      \caption[]{Continued.}
         \label{tabpropkin}
     $$ 
         \begin{array}{llcccccc }
           \hline
            \noalign{\smallskip}
CAL & Grad. & \Delta\alpha_{KC}\pm\sigma_{\Delta\alpha_{KC}} & \Delta\delta_{KC}\pm\sigma_{\Delta\delta_{KC}} & V_{KC} \pm \sigma_{KC} & PA_{kin,rec}\pm \delta PA_{kin,rec} &  PA_{kin,app}\pm \delta PA_{kin,app} &  PA_{minor}\pm \delta PA_{minor} \\
  ID   & type     &   (arcsec)                          &    (arcsec)                              &  (km s^{-1})          &  (degrees)                      & (degrees)                        & (degrees) \\
            \noalign{\smallskip}
            \hline
            \noalign{\smallskip}
   894 &    MGP &      0.3 \pm      0.4 &     -0.1 \pm      0.2 &   7621.1 \pm     56.1 &    322.8 \pm     12.2 &    141.8 \pm     10.8 &     60.7 \pm     28.5 \\ 
   895 &    MGP &      0.5 \pm      0.1 &     -0.5 \pm      0.5 &   2214.1 \pm     13.8 &    305.4 \pm      8.9 &    119.5 \pm     17.0 &     45.2 \pm     38.7 \\ 
   905 &    MGP &     -1.9 \pm      0.2 &      0.1 \pm      0.3 &   4150.6 \pm     14.0 &     38.5 \pm      8.3 &    221.0 \pm     11.4 &    135.5 \pm     10.4 \\ 
   909 &    MGP &     -0.7 \pm      0.3 &     -1.2 \pm      0.5 &   4388.2 \pm     21.9 &    156.4 \pm      9.1 &    332.7 \pm      6.4 &     66.1 \pm     10.5 \\ 
   930 &    SGP &      0.1 \pm      0.1 &      1.7 \pm      0.2 &   5281.5 \pm     34.4 &    325.2 \pm     11.0 &    145.6 \pm     11.1 &     56.8 \pm      9.2 \\ 
   935 &    MGP &      0.3 \pm      0.3 &     -0.6 \pm      0.1 &   4673.3 \pm     11.2 &    285.1 \pm     18.5 &     94.6 \pm     48.2 &     75.0 \pm     37.1 \\ 
   892 &    MGP &      0.3 \pm      0.1 &      0.9 \pm      0.0 &   4916.2 \pm     23.5 &     69.2 \pm      5.7 &    251.3 \pm     10.3 &     -9.4 \pm     16.2 \\ 
 \hline
         \end{array}
     $$ 

\end{table*}

   \begin{table*}
      \caption[]{Types and classes of asymmetries detected from the [\ion{O}{iii}] emission line profiles. We also include the systemic velocity derived from different emission lines ([\ion{O}{ii}],[\ion{O}{iii}], H$\alpha$+[\ion{N}{ii}] and [SII].}
         \label{tabpropasy}
     $$ 
         \begin{array}{lcccccccc}
            \hline
            \noalign{\smallskip}
NAME &   V_{sys}^{[OII]}\pm\sigma_{V_{sys}^{[OII]}} &  V_{sys}^{[OIII]}\pm\sigma_{V_{sys}^{[OIII]}} &  V_{sys}^{H\alpha}\pm\sigma_{V_{sys}^{H\alpha}} &  V_{sys}^{[SII]}\pm\sigma_{V_{sys}^{[SII]}} &   & Asymmetries &  \\ 
   & (km s^{-1}) & (km s^{-1})  & (km s^{-1})  & (km s^{-1})  &  Class A & Class B & Class C \\ 
    &                                     &                                      &                                         &                                     & (types) & (types) & (types) \\ 
            \noalign{\smallskip}
            \hline
            \noalign{\smallskip}
\object{ARP220} &     5455 \pm       42 &     5370 \pm       42 &     5350 \pm       53 &     5354 \pm       59  &   A0  -   A2  &   -   B1  -   &   C0  -   C2 \\ 
\object{IC0540} &     2159 \pm       44 &     2103 \pm       33 &     2067 \pm       24 &     2083 \pm       27  &   -   -   -   &   -   -   -   &   C0  C1  -  \\ 
\object{IC0776} &     2510 \pm       15 &     2449 \pm       10 &     2454 \pm       10 &     2457 \pm       12  &   -   -   -   &   B0  B1  B2  &   C0  C1  C2 \\ 
\object{IC0944} &     6947 \pm       48 &     6961 \pm       68 &     6961 \pm       81 &     7023 \pm       42  &   -   -   -   &   -   -   -   &   -   -   -  \\ 
\object{IC1199} &     4782 \pm       52 &     4681 \pm       44 &     4710 \pm       51 &     4702 \pm       58  &   -   -   -   &   -   -   -   &   -   -   -  \\
\hline
         \end{array}
     $$ 

\end{table*}

   \begin{table*}
    \addtocounter{table}{-1} 
      \caption[]{Continued.}
         \label{tabpropasy}
     $$ 
         \begin{array}{lcccccccc}
            \hline
            \noalign{\smallskip}
NAME &   V_{sys}^{[OII]}\pm\sigma_{V_{sys}^{[OII]}} &  V_{sys}^{[OIII]}\pm\sigma_{V_{sys}^{[OIII]}} &  V_{sys}^{H\alpha}\pm\sigma_{V_{sys}^{H\alpha}} &  V_{sys}^{[SII]}\pm\sigma_{V_{sys}^{[SII]}} &   & Asymmetries &  \\ 
  & (km s^{-1}) & (km s^{-1})  & (km s^{-1})  & (km s^{-1})  &  Class A & Class B & Class C \\ 
       &                                     &                                      &                                         &                                     & (types) & (types) & (types) \\ 
            \noalign{\smallskip}
            \hline
            \noalign{\smallskip}
\object{NGC5216} &     2960 \pm       56 &     2918 \pm       42 &     2910 \pm       45 &     2915 \pm       43  &   -   -   -   &   -   B1  -   &   C0  -   -  \\ 
\object{NGC5378} &     3091 \pm       55 &     2943 \pm       47 &     2938 \pm       57 &     2957 \pm       69  &   -   -   A2  &   -   B1  B2  &   -   C1  -  \\ 
\object{NGC5394} &     3555 \pm       18 &     3440 \pm       11 &     3433 \pm       10 &     3436 \pm       11  &   -   -   -   &   B0  -   B2  &   C0  C1  C2 \\ 
\object{NGC5406} &     5476 \pm       45 &     5365 \pm       57 &     5413 \pm       53 &       ---  &   -   -   -   &   -   -   -   &   -   -   -  \\ 
\object{NGC5485} &     2044 \pm       54 &     1923 \pm       54 &     1898 \pm       84 &     1908 \pm      102  &   -   -   -   &   -   -   -   &   -   -   -  \\
\hline
         \end{array}
     $$ 

\end{table*}
   \begin{table*}
    \addtocounter{table}{-1} 
      \caption[]{Continued.}
         \label{tabpropasy}
     $$ 
         \begin{array}{lcccccccc}
            \hline
            \noalign{\smallskip}
NAME &   V_{sys}^{[OII]}\pm\sigma_{V_{sys}^{[OII]}} &  V_{sys}^{[OIII]}\pm\sigma_{V_{sys}^{[OIII]}} &  V_{sys}^{H\alpha}\pm\sigma_{V_{sys}^{H\alpha}} &  V_{sys}^{[SII]}\pm\sigma_{V_{sys}^{[SII]}} &   & Asymmetries &  \\ 
  & (km s^{-1}) & (km s^{-1})  & (km s^{-1})  & (km s^{-1})  &  Class A & Class B & Class C \\ 
   &                                     &                                      &                                         &                                     & (types) & (types) & (types) \\ 
            \noalign{\smallskip}
            \hline
            \noalign{\smallskip}
\object{UGC00335NED02} &     5541 \pm       67 &     5391 \pm       43 &     5392 \pm       59 &     5359 \pm        4  &   -   -   -   &   -   -   -   &   -   -   -  \\ 
\object{UGC00841} &     5604 \pm       32 &     5540 \pm       30 &     5546 \pm       24 &     5550 \pm       24  &   -   -   -   &   -   -   -   &   -   -   -  \\ 
\object{UGC01057} &     6422 \pm       28 &     6325 \pm       47 &     6319 \pm       34 &     6328 \pm       26  &   -   A1  A2  &   B0  B1  B2  &   C0  C1  C2 \\ 
\object{UGC01938} &     6384 \pm       30 &     6405 \pm       50 &     6393 \pm       32 &     6398 \pm       30  &   -   -   -   &   -   B1  -   &   C0  C1  C2 \\ 
\object{UGC02405} &     7755 \pm       46 &       --- &     7725 \pm       58 &     7714 \pm       49  &   -   -   -   &   -   -   -   &   -   -   -  \\
\hline
         \end{array}
     $$ 

\end{table*}
   \begin{table*}
    \addtocounter{table}{-1} 
      \caption[]{Continued.}
         \label{tabpropasy}
     $$ 
         \begin{array}{lcccccccc}
            \hline
            \noalign{\smallskip}
NAME &   V_{sys}^{[OII]}\pm\sigma_{V_{sys}^{[OII]}} &  V_{sys}^{[OIII]}\pm\sigma_{V_{sys}^{[OIII]}} &  V_{sys}^{H\alpha}\pm\sigma_{V_{sys}^{H\alpha}} &  V_{sys}^{[SII]}\pm\sigma_{V_{sys}^{[SII]}} &   & Asymmetries &  \\ 
  & (km s^{-1}) & (km s^{-1})  & (km s^{-1})  & (km s^{-1})  &  Class A & Class B & Class C \\ 
   &                                     &                                      &                                         &                                     & (types) & (types) & (types) \\ 
            \noalign{\smallskip}
            \hline
            \noalign{\smallskip}
\object{UGC12274} &     7655 \pm       44 &     7623 \pm       56 &     7613 \pm       59 &       ---  &   -   -   -   &   -   -   -   &   -   -   -  \\ 
\object{UGC12308} &     2259 \pm       24 &     2214 \pm       13 &     2213 \pm       13 &     2218 \pm       11  &   A0  -   -   &   B0  B1  -   &   C0  C1  C2 \\ 
\object{UGC12494} &     4289 \pm       12 &     4144 \pm       14 &     4141 \pm       13 &     4142 \pm       12  &   -   -   -   &   B0  B1  B2  &   C0  C1  C2 \\ 
\object{UGC12519} &     4417 \pm       39 &     4383 \pm       23 &     4374 \pm       22 &     4372 \pm       21  &   -   -   -   &   -   B1  B2  &   C0  C1  C2 \\ 
\object{UGC12816} &     5348 \pm       32 &     5245 \pm       29 &     5258 \pm       31 &     5259 \pm       31  &   -   -   -   &   -   -   B2  &   C0  C1  C2 \\
\hline
         \end{array}
     $$ 

\end{table*}
\newpage




\section{Presence of double/multiple gaseous components: detection limits for CALIFA V500 spectra.}

The capability of detecting double-peaked/multi-component emission-line profiles in the spectra of galaxies strongly depends on the spectral resolution as well as on its signal-to-noise. In this appendix we analize the capabilities and limitations of the CALIFA V500 data for such detection. 

  \begin{figure*}
   \resizebox{\hsize}{!}{\includegraphics[angle=0]{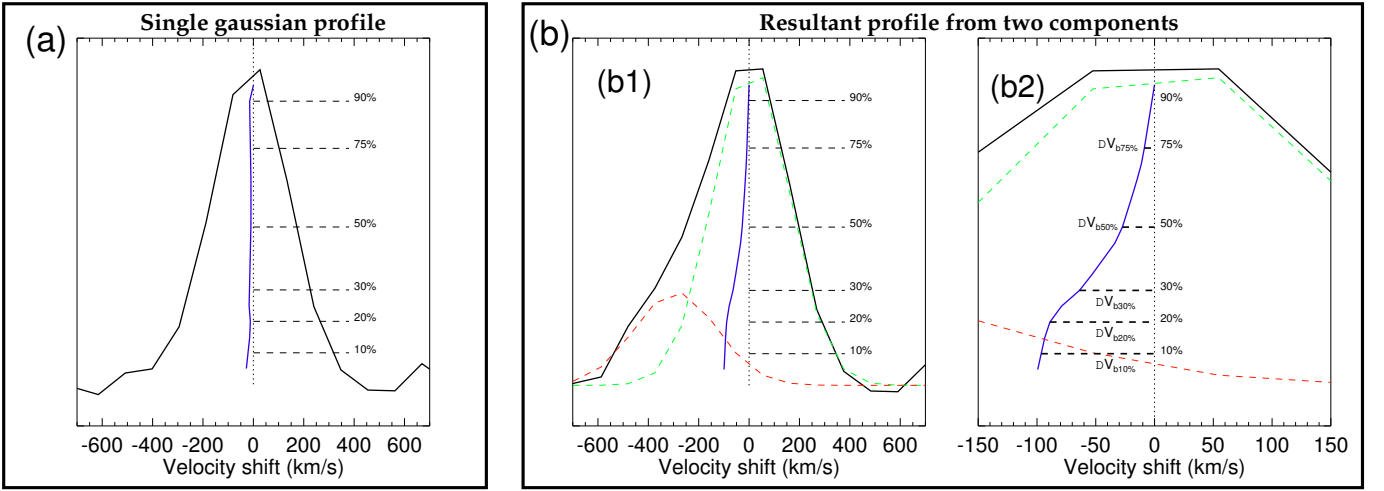}}
   \caption{(a) Single-Gaussian model (including random noise) of an single emission line (S/N$\sim$45) for CALIFA V500. The bisector of a symmetric profile should remain at constant velocity (or wavelength) for all intensity levels, dividing it into two equal parts; any existing asymmetry between the base and the peak of the line will remain reflected in the shape of the bisector. The dotted vertical line corresponds to the bisector of an ideal symmetric Gaussian profile, while the blue line is the actual bisector of the modeled profile, which deviates from the ideal bisector due to noise and velocity sampling. Dashed lines only indicate some intensity levels (in percentage of the intensity peak) that are refered to as ''bisector levels''. (b)-(b1) Resultant profile (black line) from the combination of two Gaussian components. Both Gaussians profiles have similar velocity dispersions, a flux ratio of 0.3 and a velocity shift of 300 km/s (red and green dashed curves). The blue line is the bisector of the resultant profile and the dotted vertical line is the same as in (a). (b)-(b2) Zoom of the profile in (b1). Horizontal dashed lines indicate the difference in velocity between the central velocity (traced by the dotted line) and the velocity at different bisector levels ($\Delta$V$_{b}$).}
  \label{bisector}
    \end{figure*}

\subsection{Noise influence}

For an ideal Gaussian profile, the difference in velocity between the peak and the bisector velocity at any intensity level ($\Delta$V$_{b}$ hereafter) is zero, but the presence of noise and the limited spectral resolution of the data would be perturbing this behavior. Figure \ref{bisector} shows two examples of a Gaussian model for a single and a double-component emission line profiles for the same spectral resolution than CALIFA V500, including the shape of their bisectors. We note the difference between the actual bisector (blue line in Fig. \ref{bisector}) and the ideal bisector of a single non-perturbed Gaussian (dotted line in Fig. \ref{bisector}). In order to determine the limits on the detection of double/multiple gaseous components in the CALIFA V500 spectra, an isolated emission line (e.g. [\ion{O}{iii}]~$\lambda5007$) has been modeled through a single Gaussian profile using the same spectral sampling than the CALIFA V500 configuration. The width of this Gaussian profile was selected to be the instrumental FWHM (6 \AA, see H13) and a set of central wavelengths were selected to cover the redshift range of CALIFA (from 1000 to 9000 km s$^{-1}$ in steps of 500 km s$^{-1}$). Each of these single-emission line profiles were perturbed by normally-distributed, pseudo-random noise with a mean of zero and a standard deviation of one, obtaining a set of single-Gaussian profiles of signal-to-noise (S/N) ranging from 5 to 110. The S/N was estimated from the ratio of the flux at the peak of the perturbed Gaussian profile and the standard deviation in the continuum. Simulations assume a negligible impact on the line profile of the underlaying stellar continuum subtraction.

\begin{figure*}
   \resizebox{\hsize}{!}{\includegraphics[angle=90]{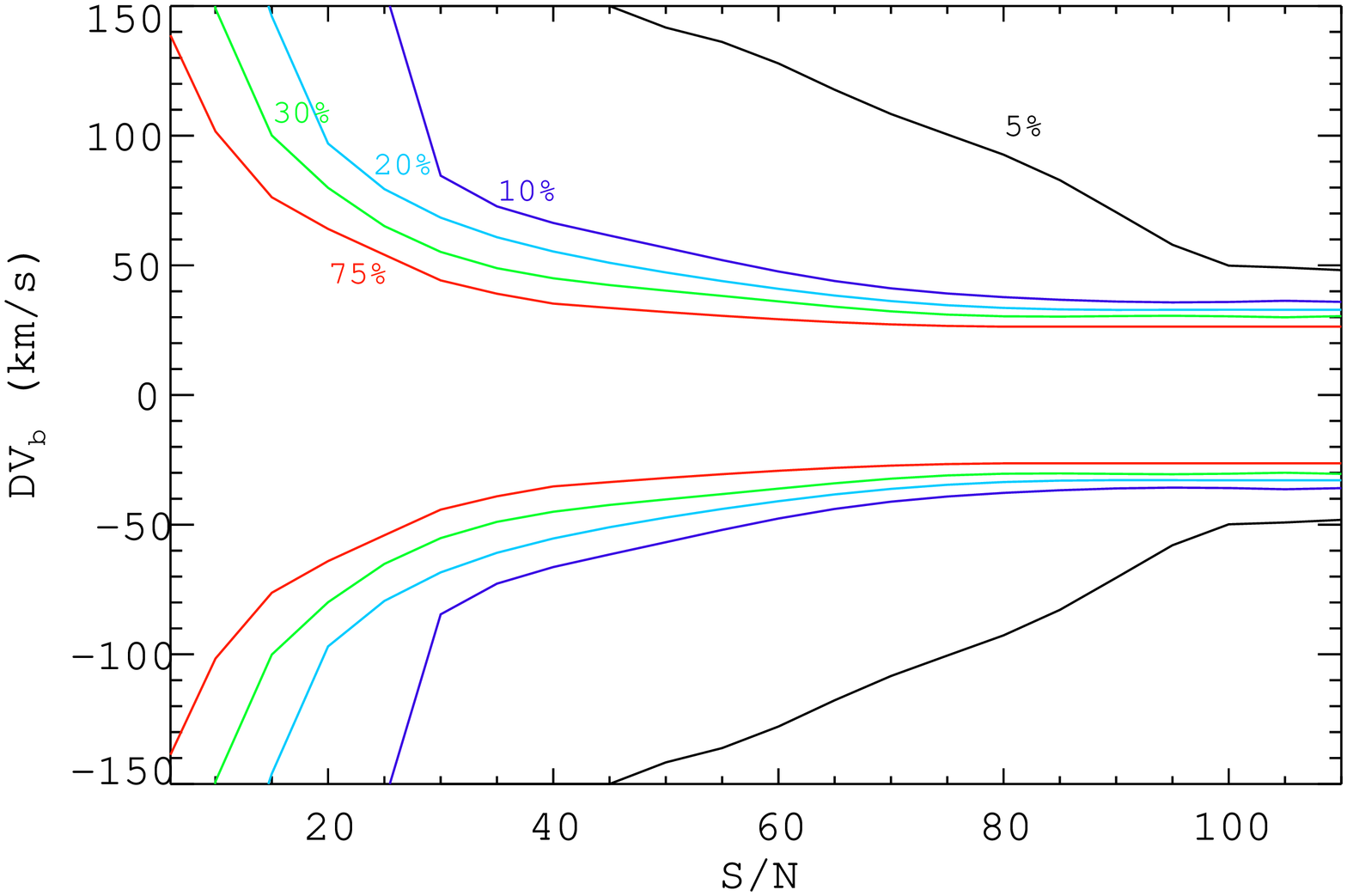}}
   \caption{Noise-induced asymmetries limits, as a function of the S/N, of the difference in velocity between the peak and selected bisector levels (in percentage of the intensity peak, as indicated in the plot) for V500 CALIFA spectra derived from a large number of modeled single-Gaussian profiles. The $\Delta$V$_{b}$ obtained for each selected bisector level are within the curves for that bisector. Any $\Delta$V$_{b}$ beyond these curves should be indicating the presence of secondary gaseous components.}
  \label{singleperturbed}
    \end{figure*}

The bisectors of each profile (1.5x10$^{6}$ in total) were traced at twenty intensity levels (from 5\% to 100\% of the intensity peak in steps of 5\%), providing the shift in velocity at each intensity level respect to an unperturbed Gaussian bisector ($\Delta$V$_{b}$ hereafter, see Fig. \ref{bisector}). Figure \ref{singleperturbed} shows the dependence of the measured $\Delta$V$_{b}$ on S/N for the set of single-Gaussian profiles generated. It is clear that a single Gaussian profile at the resolution of CALIFA V500 data could appear as an asymmetric profile at some degree depending on its S/N and the intensity level at which the bisector velocity is measured. Curves drawn in the Fig. \ref{singleperturbed} at positive and negative $\Delta$V$_{b}$ delimit the region of $\Delta$V$_{b}$ derived from the different spectra in these single-Gaussian experiments. That is, all the $\Delta$V$_{b}$ at a selected bisector level (for a single-Gaussian emission line) are within the region defined by the curves in Fig. \ref{singleperturbed} associated with that bisector level. Therefore, if the bisector of an unknown emission line profile is traced and $\Delta$V$_{b}$ at a particular bisector level is in the region between the curves associated with that bisector level in Fig. \ref{singleperturbed}, we cannot say anything about the presence or not of double components. However, we are confident that an observed profile is the result of at least two gaseous components if a $\Delta$V$_{b}$ is measured at a particular bisector level beyond the curves associated with that level in Fig. \ref{singleperturbed}. For example, if we have an emission line profile of S/N$\sim$60 and we have measured a $\Delta$V$_{b}$ of 50 km s$^{-1}$ at the intensity level of 10\% of the peak (10\% bisector level), we do not know if we actually have a single or a double component profile, but if we measured a  $\Delta$V$_{b}$ of 50 km/s in the same emission line profile at the intensity level of 75\% of the peak (75\% bisector level) we are confident that we have a double/multiple components.


 For S/N between 40 and 90, $\Delta$V$_{b}$ curves can be approched at first order by a linear function of S/N (F$_{\Delta V_{b}}$(S/N) hereafter), with the linear coefficients parametrized in terms of bisector levels. Then, for practical applications of searching kinematically distinct gaseous components in V500 CALIFA spectra through unblended emission lines, the F$_{\Delta V_{b}}$(S/N) limits at a particular bisector level can be approached through

\begin{equation}
F_{\Delta V_{b}}(S/N) =  A(l) + B(l) \times S/N \hspace{0.25cm}  for \ 40 \lesssim S/N \lesssim 90,
\label{eqc1}
\end{equation}

\noindent where, l denotes the bisector level, and the polynomial functions of the second degree A(l) and B(l) are

\begin{equation}
A(l) = 107.6 - 2.12 \times l + 0.0135 \times l^2
\label{eqc2}
\end{equation}
\begin{equation}
B(l) = -0.816 + 0.0204 \times l - 1.26e-4 \times l^2.
\label{eqc3}
\end{equation}

\noindent The $\Delta$V$_{b}$ curves in Fig. \ref{singleperturbed} are almost constant for S/N$\gtrsim90$, constant value that is approximately the F$_{\Delta V_{b}}$(S/N) at S/N=90 for each bisector level. Then:

\begin{equation}
F_{\Delta V_{b}}(S/N) \sim F_{\Delta V_{b}}(90) \hspace{1cm} for \ S/N \gtrsim 90,
\label{eqc4}
\end{equation}

\noindent With this simple approach of the $\Delta$V$_{b}$ curves in Fig. \ref{singleperturbed}, we have a probability smaller than 0.6\% of having a $\Delta$V$_{b}$ associated with a single Gaussian profile beyond the limit provided by F$_{\Delta V_{b}}$(S/N) at any bisector level and any S/N$\ge$40.

\subsection{Parameter space for asymmetries in CALIFA emission line profiles}

If we could isolate and observe a single gaseous system in a galaxy, we could derive its parameters (flux, velocity and velocity dispertion) from the ideal Gaussian profiles of its spectrum. The presence of two or more gaseous systems with distinct kinematics along the line of sight would result in complex emission lines in the spectra of the observed object. These profiles can show blueshifted or/and redshifted wings, shoulders and/or double peaks, shapes that are a complex functions of the parameters characterizing each gaseous component.

\begin{figure*}
   \resizebox{\hsize}{!}{\includegraphics[angle=90]{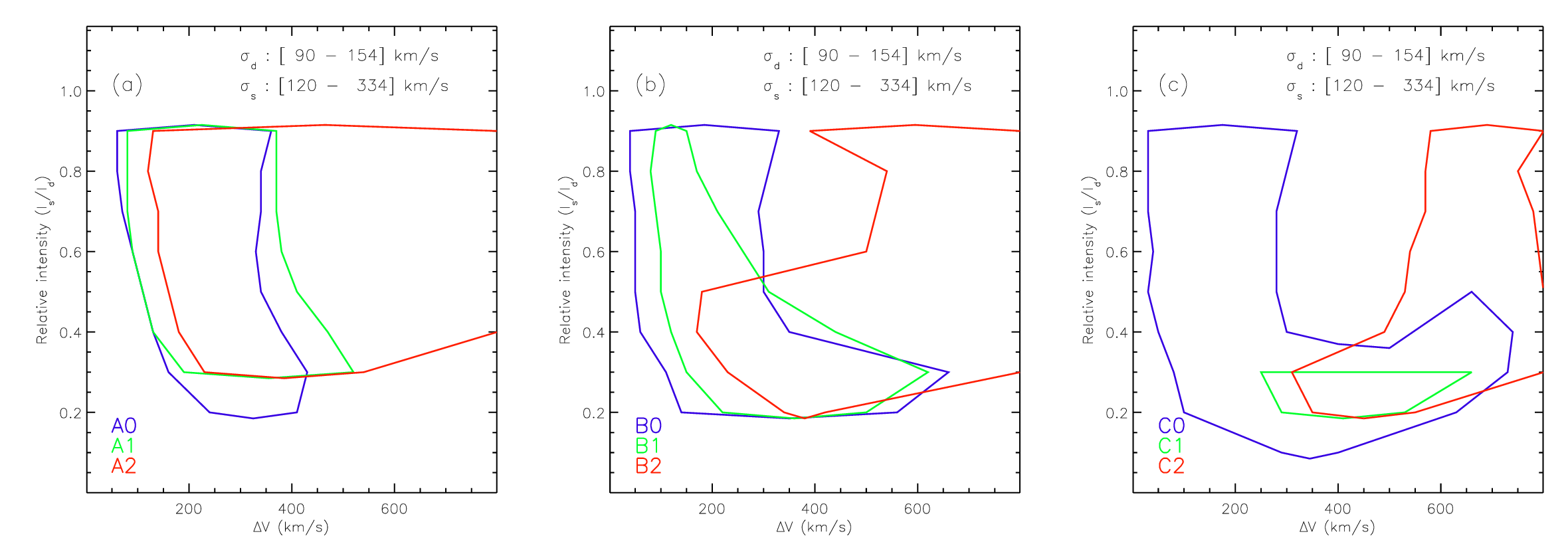}}
   \caption{Examples of the parameter space for two Gaussians combined to model asymmetric emission line profiles of the types and classes defined in this work (see section \ref{asym_results}). Lines encicle the region of possible values of the $\mid\Delta$V$\mid$ and I$_{s}$/I$_{d}$ parameters resulting in an asymmetric profile of the classes (a) A0, A1, and A2, (b) B0, B1, and B2, and (c) C0, C1, and C2 indicated by colors. Each plot includes the velocity dispersion ranges for the secondary ($\sigma_{s}\sim[120-334]$ km/s) and dominant ($\sigma_s\sim[90-154]$ km/s) components used to model the profiles. }
  \label{modelo_two_gauss}
    \end{figure*}

In this section we explore the parameter space (relative intensity, velocity and velocity dispersion) of two gaseous systems associated with the different profiles classes and types defined in section \ref{asym_results}. With this aim, we have used a two Gaussians functions to model [\ion{O}{iii}] emission line profiles from two gaseous components.We performed a set of experiments playing with the input parameters: (1) The central wavelength of the main component, covering the CALIFA redshift range (from $\sim1000$ to $\sim9000$ km/s) in steps of 100 km/s; (2) the FWHM of the main component, changing from 60\% to 100\% in steps of 10\% of the CALIFA V500 instrumental FWHM (6 \AA, see H13); (3) the relative intensity ratio of the secondary (I$_{s}$) and dominant (I$_{d}$) components, varying from 0.1 to 0.9 in steps of 0.1; (4) The relative difference in velocity ($\Delta$V) between both components, ranging from 0 to 800 km/s in steps of 10 km/s; and (5) the FWHM difference (in \AA) of the secondary and dominant components, spaning from 0 to 3 in steps of 0.5. We only generated redshifted asymmetries, assuming that blueshifted profile will follow a similar parameter space but with negative difference in velocity between the main and the secondary components. The resultant profiles were perturbed by normally-distributed, pseudo-random noise with a mean of zero and a standard deviation of one, obtaining S/N values from $\sim$40 to $\sim$100. We classified the simulated profiles showing asymmetries according to the classes and types defined in section \ref{asym_results}. Asymmetries classes/types result from many combination of the parameters of both Gaussian components. Figure \ref{modelo_two_gauss} shows some examples of the possible combinations for relative velocity ($\mid\Delta$V$\mid$) and intensity (I$_{s}$/I$_{d}$) providing asymmetric profiles for a range of velocity dispersions for the dominant and secondary components.

From the modeled profiles we can also estimate the uncertanties when fitting an asymmetric profile with a single Gaussian model. With this aim, we fit a single Gaussian to all the asymmetric profiles and we compared its flux with the total flux of the two Gaussians forming the modeled profile. For simplicity, noise was not included in these tests. For A0, B0, C0 and C1 profiles, uncertainties in the estimated flux induced using a single Gaussian instead of two Gaussians components are smaller than 0.5\% for any combination of the two Gaussians parameters. For A1, B1, and C3 profiles, uncertainties can reach 4\%, being smaller than 0.5\% in average. Uncertainties larger than 10\% in flux can be induced fitting a single Gaussian to B2 and A2 profiles when the absolute difference in velocity between the two components is larger than 300 km/s. For B2 and A2 profiles coming from two Gaussian with $\mid\Delta$V$\mid\leq300$ km/s, flux uncertainties are smaller than 3\% when approching the profile by a single Gaussian. 


\section{tlas of ionized gas velocity fields for CALIFA galaxies}
In this appendix we present the kinematic maps derived from the CALIFA datacubes for objects in the analyzed sample. In each plot we include narrow band recovered images from the datacubes for [\ion{O}{ii}]~$\lambda\lambda3726,3728$, [\ion{O}{iii}]~$\lambda\lambda4959,5007$, H$\alpha$+[\ion{N}{ii}]~$\lambda\lambda6548,6584$, and [SII]$\lambda\lambda6716,6730$ emission lines and their velocity fields. We also show the gradient velocity maps obtained from the H$\alpha$+[\ion{N}{ii}] velocity fields, the pseudo-rotation curves and the location of spectra showing asymmetric profiles, indicative of the presence of kinematically distinct gaseous systems. Panels in each figure correspond to:

\begin{itemize}
\item[(a)] Narrow band images (stellar continuum subtracted) of (1) [OII], (2) [OIII], (3) H$\alpha$+[\ion{N}{ii}], and (4) [SII] obtained by integrating the signal in selected spectral bands (see section \S\ref{medir}). Maps are normalized to the flux of their brightest spaxel. Contours correspond to the stellar component distribution as traced by a continuum map made from the total flux of the best stellar fit in the spectral range 3800-7000. \AA in logarithmic scle (minimum value is -1.65 dex, and step between countours is 0.15 dex).
\item[(b)] Velocity field obtained by averaging the obtained radial velocities from Monte Carlo simulations . Radial velocities were inferred from cross-correlation in the (1) [OII], (2) [OIII], (3) H$\alpha$+[\ion{N}{ii}], and (4) [SII] spectral ranges (see section \S\ref{medir}). 
\item[(c)] Standard deviation of the radial velocities for each spaxel from Monte Carlo simulations for (1) [OII], (2) [OIII], (3) H$\alpha$+[\ion{N}{ii}], and (4) [SII] emission lines  (see section \S\ref{medir}).
\item[(d)] Sloan Digital Sky Survey r-band image of the object. The hexagonal area drawn is the CALIFA field of view.
\item[(e)] Velocity gradient map obtained from the H$\alpha$+[\ion{N}{ii}] velocity field in (b3). Green cross indicate the location of the kinematic center (see section \S\ref{k_medir}). 
\item[(f)] Distance to the kinematic center (in arcsec) versus the radial velocity for each spectra of the CALIFA data cube with a S/N larger than the threshold. Blue open squares trace the pseudo-rotation curve. Green circles correspond to those spaxels with the lowest difference in velocity to the KC selected to estimate the minor kinematic axis (see section \S\ref{pa_medir}). 
\item[(g)] Location (black squares) of the spectra tracing the pseudo-rotation curve in (f) on the H$\alpha$+[\ion{N}{ii}] velocity field and tracing the major kinematic pseudo-axis (the same than those marked in blue open squares in (f)). Filled green circles correspond to those spaxels with a similar velocity than the KC (the same than green open circles in (f)) and tracing the minor kinematic axis. 
\item[(h)] Location on the H$\alpha$+[\ion{N}{ii}] narrow-band image of the spectra showing asymmetric [\ion{O}{iii}]~$\lambda\lambda4959,5007$ profiles. Green, yellow and black squares correpond to profile classes A, B, and C, respectively.
\end{itemize} 

Maps for emission lines not reaching the detection/content threshold (see section \S\ref{medir}) are missing in the panels.
Only a sample of these figures is included here; the full version will be available by request.

   \begin{figure*}
\centering
      \includegraphics[width=12.3cm,angle=90]{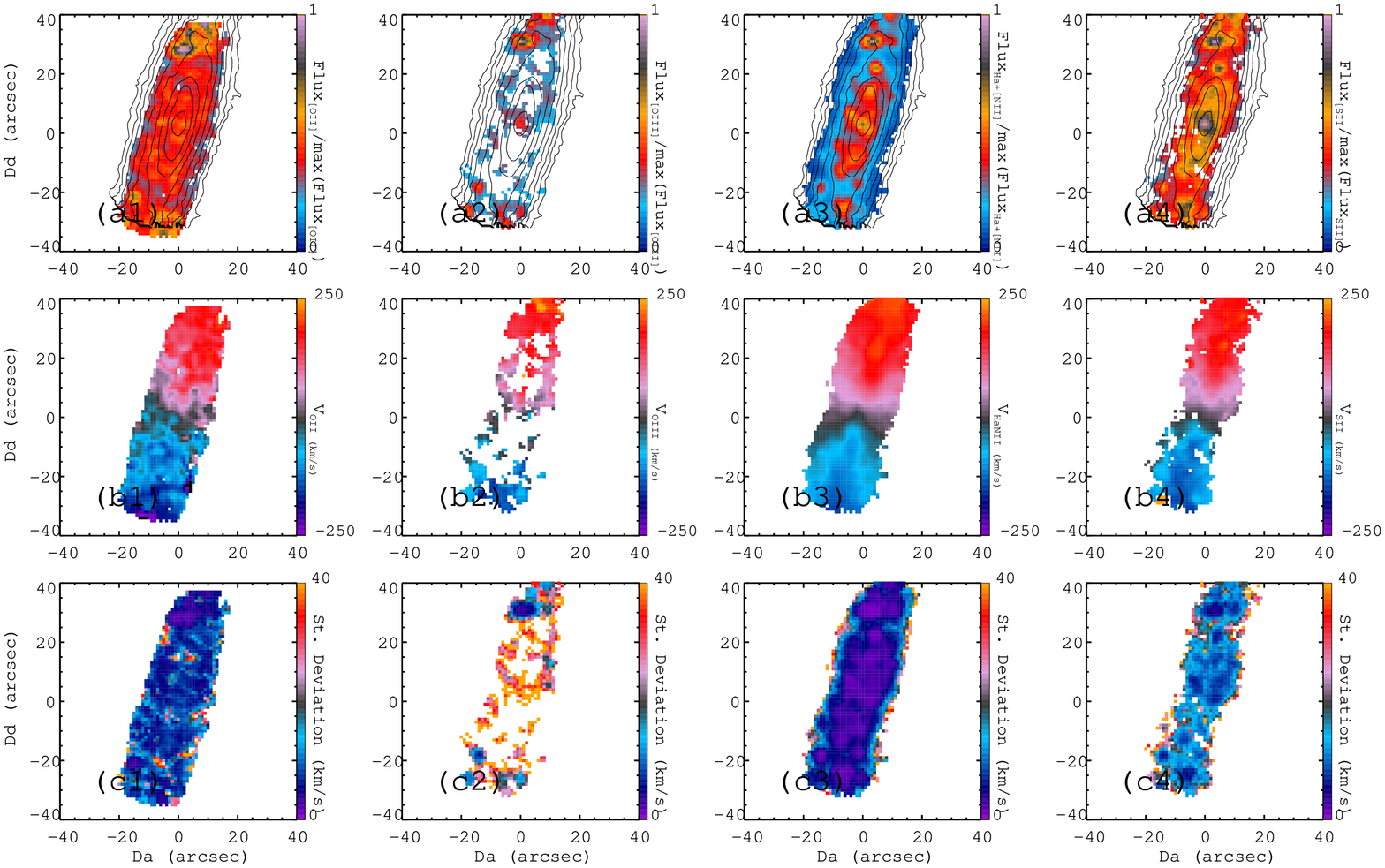}
\hspace*{-0.25cm}     \includegraphics[width=12cm,angle=90]{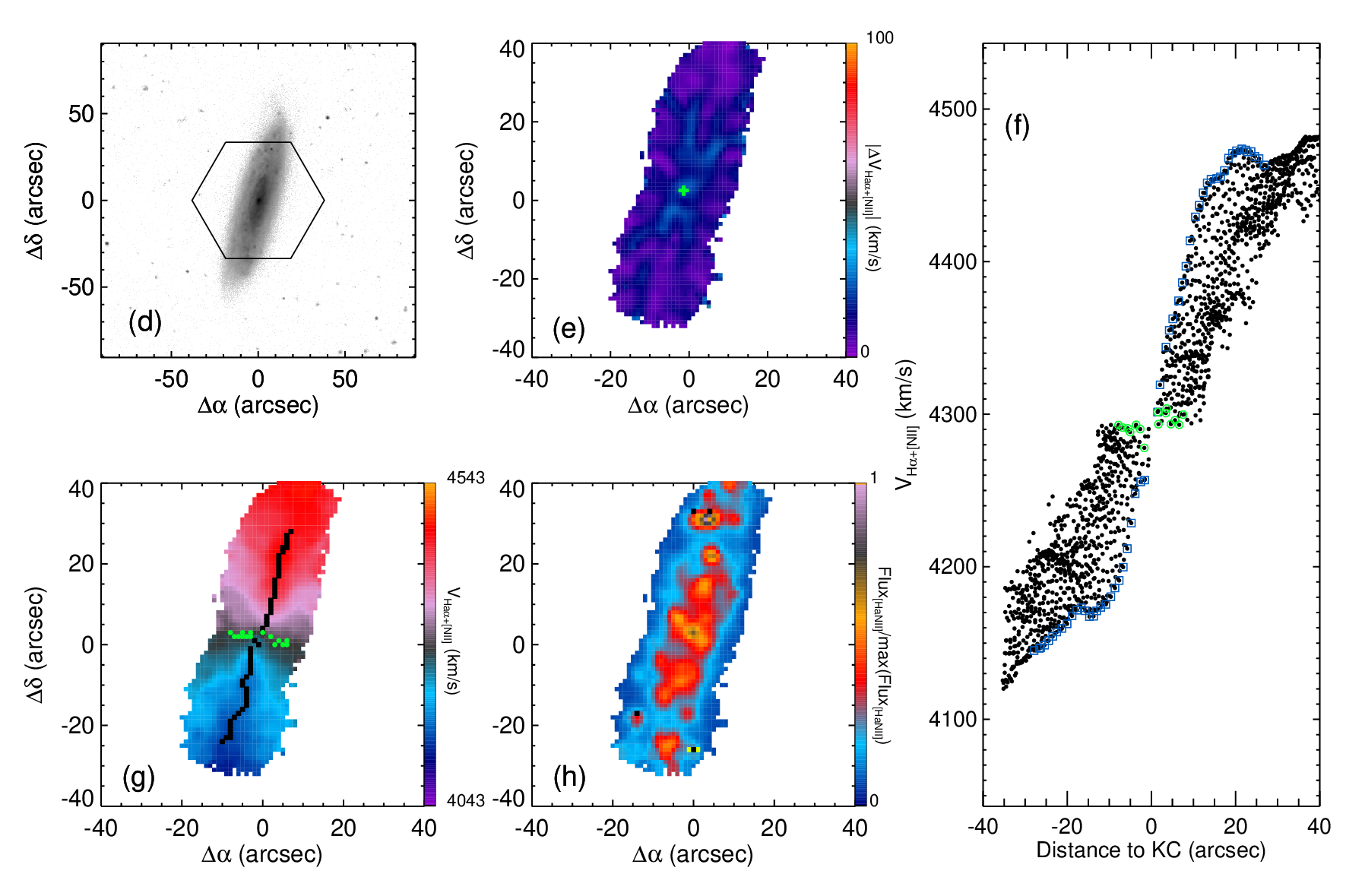}
   \caption{Summarizing results for IC2487.}
   \label{ARP220_all}
    \end{figure*}

\end{appendix}

\end{document}